# Non-Transferable Utility Coalitional Games
# via Mixed-Integer Linear Constraints

**Gianluigi Greco**                                    GGRECO@MAT.UNICAL.IT
*Dipartimento di Matematica*
*Università della Calabria*
*I-87036 Rende, Italy*

**Enrico Malizia**                                     EMALIZIA@DEIS.UNICAL.IT
**Luigi Palopoli**                                     PALOPOLI@DEIS.UNICAL.IT
**Francesco Scarcello**                                SCARCELLO@DEIS.UNICAL.IT
*DEIS*
*Università della Calabria*
*I-87036 Rende, Italy*

## Abstract

Coalitional games serve the purpose of modeling payoff distribution problems in scenarios where agents can collaborate by forming coalitions in order to obtain higher worths than by acting in isolation. In the classical Transferable Utility (TU) setting, coalition worths can be freely distributed amongst agents. However, in several application scenarios, this is not the case and the Non-Transferable Utility setting (NTU) must be considered, where additional application-oriented *constraints* are imposed on the possible worth distributions.

In this paper, an approach to define NTU games is proposed which is based on describing allowed distributions via a set of mixed-integer linear constraints applied to an underlying TU game. It is shown that such games allow non-transferable conditions on worth distributions to be specified in a natural and succinct way. The properties and the relationships among the most prominent solution concepts for NTU games that hold when they are applied on (mixed-integer) constrained games are investigated. Finally, a thorough analysis is carried out to assess the impact of issuing constraints on the computational complexity of some of these solution concepts.

## 1. Introduction

Cooperative game theory provides—under the concept of *coalitional games*—an elegant framework for modeling multi-agent systems where agents might collaborate with other agents, by forming *coalitions* in order to guarantee themselves some advantage. Within this framework, each coalition $S \subseteq N$ (where $N$ is the set of all the players, also called the grand-coalition), is assigned a certain worth $v(S) \in \mathbb{R}$, and the outcome of the game is a vector of real payoffs $(x_i)_{i \in N}$ that is meant to specify the distribution of the worth granted to the players as the result of the game.

Coalitional games are very often classified according to the mechanisms underlying payoff distributions. The best known and most widely studied class therein is that of coalitional games *with transferable utility* (or TU games) (Osborne & Rubinstein, 1994), where no constraints whatsoever are imposed over the way coalitional worths can be distributed amongst coalition members. In this context, several outcomes might be associated with a given game, and hence a relevant question is to understand which outcomes most properly capture the rational behavior of the players. This matter has been extensively studied in economics and social sciences (Aumann & Hart, 2002). In fact, various *solution concepts* have been proposed in the literature to identify worth distributions that embody some rational concept of stability, i.e., that are somehow "immune" to deviations caused by groups of players who may decide to leave the grand-coalition and to form sub-coalitions in order to claim for higher worths.

There are cases, however, where players cannot freely distribute the coalition worth so that a pure TU framework is not appropriate for such modeling purposes (Aumann & Peleg, 1960). To deal with those scenarios, coalitional games *without transferable utility* (or NTU games) have been introduced in the literature, where the worth function is defined as to return all those allowed worth distributions (called *consequences*, in this setting) associated with any given coalition, rather than associating just one real value with it. In fact, it is easily understood that NTU games are more general than TU ones, since any game of the latter kind can be expressed as an NTU game where any possible worth distribution among the members of a coalition $S$ is a consequence for $S$.

## 1.1 Modeling NTU Specifications via Mixed-Integer Linear Constraints on TU Games

Enhancing TU games with application-oriented *constraints* over the set of all possible outcomes is an approach that has been exploited in the literature in order to model non-transferable scenarios. The first occurrence of the name "constrained games" goes back to the seventies, and is due to Aumann and Dreze (1974), who considered games with *coalition structures*, where players are partitioned in groups $\mathcal{S}_1, \ldots, \mathcal{S}_k$, and where any outcome $(x_i)_{i \in N}$ must allocate the total payoff $v(\mathcal{S}_j)$ exactly amongst the members of each group $\mathcal{S}_j$, that is, so as to satisfy the equalities $\sum_{i \in \mathcal{S}_j} x_i = v(\mathcal{S}_j)$, for $1 \leq j \leq k$. However, Aumann and Dreze noticed in their turn that considering constraints over TU games was not a novel idea, since the *core* and the *nucleolus* (which are two prominent solution concepts for TU games) were indeed defined by Gillies (1959) and Schmeidler (1969), respectively, on games with outcomes restricted to subsets of $\mathbb{R}^{|N|}$.

Recently, constrained games have been reconsidered under the pragmatic perspective of modeling some relevant application scenarios, such as price formation (Byford, 2007) and autonomic wireless networks (Jiang & Baras, 2007). As a matter of fact, however, they received considerably little attention over the years. In particular, no general framework was proposed in the literature and no systematic study of the (analytical, as well as computational) properties of this kind of approaches was conducted so far.

In this paper, we embark on a systematic formalization of constrained games, and we investigate a framework allowing to succinctly specify non-transferable conditions on the outcomes of an underlying TU game, via a set of constraints expressed as *mixed-integer linear (in)equalities*.[1] Note that such constrained games are defined on top of an underlying TU specification, and hence they are expected to retain some of the nice properties of the transferable setting. However, their ability of restricting the set of possible outcomes makes them fit the more general framework of NTU games, from which they smoothly inherit the solution concepts that we shall use.

---

1. A good source for basic notions and results on mixed-integer linear programming is the book by Nemhauser and Wolsey (1988).

By allowing integer variables, the constrained games studied in this paper will improve the expressiveness of classical NTU formalizations, in that admissible outcomes might be possibly restricted over *non-convex* and *non-comprehensive* regions (definitions for these properties are recalled in Section 3.2). Indeed, those NTU games that attracted much attention in earlier literature do not allow to specify arbitrary consequences (Aumann & Hart, 2002). Rather, according to the classical definition due to Aumann and Peleg (1960), an NTU game is actually a game that must satisfy additional conditions such as, in particular, convexity and comprehensiveness. This view features several nice properties under a mathematical perspective (Weber, 1994; McLean, 2002), and it influenced several other proposals for defining NTU games which appeared in the literature, where further additional conditions are often considered. However, this view is not appropriate to model application scenarios where those required properties do not naturally hold.

In fact, the framework of constrained games proposed in this paper can be viewed as a framework to (succinctly) define NTU games where convexity and comprehensiveness do not necessarily hold. This is an important peculiarity of our approach from a knowledge representation perspective. An intuitive exemplification of those scenarios where this peculiarity might be very useful is illustrated below.

**Example 1.1.** Three brothers, Tim, John and Jim, aged 10, 8 and 5, resp., have collected into a piggy money-box all the small Euro coins (values 1, 2, 5, and 10 cents) that Mom every week has given to each of them since the age of four. Now, the time has come to break the money-box and divide its content. In order to avoid quarrels among the kids, Mom decides that the distribution has to go with their ages, so that Tim will deserve at least 10/8 the money John will get and John, in its turn, will receive at least 8/5 of Jim's money share. (Jim is not very happy with that, but agrees to comply with Mom's rule). The money-box gets broken and the little treasure of seven Euros and ninety Euro cents, as resulting from the available coin set including one-hundred 1-cent coins, seventy 2-cent coins, fifty 5-cents coins, and thirty 10-cent coins, can then be divided amongst the kids.

Note that this scenario is based on the non-transferable condition that the treasure cannot freely be distributed amongst the brothers. The specific distribution rule, however, does not fit the classical NTU formalization by Aumann and Peleg (1960).

On the other hand, it is easily seen that the scenario can be modeled by means of a set of linear (in)equalities, with a few variables taking values from the set $\mathbb{Z}$ of integer numbers. In this example, admissible outcomes can indeed be identified as the solutions to the following set of mixed-integer linear (in)equalities (the three brothers Tim, John and Jim are denoted by using the indexes 1, 2 and 3, respectively):

$$
\begin{cases}
x_i = 1 \times \alpha_1^i + 2 \times \alpha_2^i + 5 \times \alpha_5^i + 10 \times \alpha_{10}^i, \forall 1 \le i \le 3 \\
\alpha_1^1 + \alpha_1^2 + \alpha_1^3 = 100 \\
\alpha_2^1 + \alpha_2^2 + \alpha_2^3 = 70 \\
\alpha_5^1 + \alpha_5^2 + \alpha_5^3 = 50 \\
\alpha_{10}^1 + \alpha_{10}^2 + \alpha_{10}^3 = 30 \\
x_1 \ge 10/8 \times x_2 \\
x_2 \ge 8/5 \times x_3 \\
\alpha_1^i, \alpha_2^i, \alpha_5^i, \alpha_{10}^i \ge 0, \forall 1 \le i \le 3 \\
x_i \in \mathbb{R}, \ \alpha_1^i, \alpha_2^i, \alpha_5^i, \alpha_{10}^i \in \mathbb{Z}, \forall 1 \le i \le 3
\end{cases}
$$

Note that the auxiliary variables $\alpha_j^i$ denote the number of coins of value $j$ taken by player $i$, the first five equalities encode restrictions on the domains of the variables as defined by the available coin set, and the subsequent two inequalities encode Mom's rule (which can be seen, for instance, as playing the role of a central market regulation authority). ◁

## 1.2 Contribution and Organization

Despite the intuitiveness of the modeling approach adopted in Example 1.1, there is no reference framework in the literature accounting for it, both because of the specificity of Mom's rule and because money distribution is constrained by the available coin set, so that those allowed outcomes do not form a convex set.

Proposing and investigating a framework that may serve to model such kinds of scenarios is the main contribution of this paper. In more detail:

▶ We define a formal framework for NTU games based on mixed-integer linear constraints applied to an underlying TU game, we discuss its modeling capabilities, and we show how main solution concepts for NTU games—in particular, *core*, *bargaining set*, *kernel*, *nucleolus*, and *Shapley value*—specialize within this novel framework.

▶ We analyze the impact of constraints on the basic properties of such solution concepts. Moreover, we highlight similarities and differences featured by constrained games as opposed to TU games, by investigating in particular whether an outcome that is stable (under these concepts) in a TU game remains stable if constraints are issued.

▶ We assess the impact of adding constraints on the computational complexity underlying some of these concepts. In particular, we consider games in *characteristic function form* (von Neumann & Morgenstern, 1944) within a setting where worths are given as oracles. In this context, we discuss both the intrinsic difficulty of checking whether a given worth distribution is in the core or in the bargaining set, and of deciding the non-emptiness of these solutions. Complexity results for constrained games are also compared with those characterizing TU games.

The rest of the paper is organized as follows. In Section 2, we overview some basic notions of cooperative game theory. The formal framework of constrained games is defined in Section 3. The properties of this novel framework are illustrated in Section 4, and its analysis from the computational viewpoint is carried out in Section 5. A discussion and a few concluding remarks are reported in Section 6.

## 2. Coalitional Games

An important issue in cooperative game theory is to determine payoff distributions for the agents in scenarios where they can collaborate by forming coalitions, in order to obtain higher worths than by acting in isolation. In this context, one usually does not take care of other relevant problems emerging in the coalition formation process, such as the *coalition value calculation* and the *coalition structure generation* problems—an excellent overview of these problems and a state-of-the-art algorithm facing the latter one can be found in the work by Rahwan, Ramchurn, Jennings, and Giovannucci (2009).

Coalitional games have been introduced by von Neumann and Morgenstern (1944) in order to model payoff distributions problems in scenarios where utility can be freely transferred among players. In these cases, coalitional games can be described by associating a payoff with each possible coalition. Thus, a *coalitional game with transferable utility* (TU) is a pair $\langle N, v \rangle$, where $N$ is a finite set of players, and $v$ is a function ($v \colon 2^N \mapsto \mathbb{R}$) that associates with every coalition $S \subseteq N$ a real number $v(S)$, called the *worth* of $S$.

Scenarios where utility cannot be freely transferred among players were first formalized by Aumann and Peleg (1960). In these scenarios, games have to be described by specifying all the possible payoff distributions for the players in each coalition, rather than by just one (global) payoff. For any coalition $S \subseteq N$, let $|S|$ denote the cardinality of $S$, and let $\mathbb{R}^S$ be the $|S|$-dimensional real coordinate space, whose coordinates are labeled with the members of $S$; in particular, given a *payoff vector* $x \in \mathbb{R}^S$, $x_i$ denotes the component associated with the player $i \in S$. Then, a *coalitional game without transferable utility* (NTU) is a pair $\langle N, V \rangle$, where $N$ is a finite set of players, and $V$ is a function associating with every coalition $S \subseteq N$ a set of payoff vectors $V(S) \subseteq \mathbb{R}^S$, also called *consequences*.

Note that NTU games are generalizations of TU games. In particular, according to the standard encoding[2] discussed, e.g., in the handbook edited by Aumann and Hart (2002) and in the book by Peleg and Sudhölter (2007), the TU game $\langle N, v \rangle$ will be viewed throughout this paper as the NTU game $\langle N, V_v \rangle$, where:

$$V_v(S) = \left\{ x \in \mathbb{R}^S \Big| \sum_{i \in S} x_i \leq v(S) \right\}, \ \forall \ S \subseteq N. \tag{1}$$

Let $\mathcal{G} = \langle N, V \rangle$ be an NTU game. A consequence $x \in V(N)$ is an *imputation* of $\mathcal{G}$ if the following two properties hold (see, e.g., Peleg, 1963; Peleg & Sudhölter, 2007):

(1) *Efficiency:* for each $y \in V(N)$, there is a player $i \in N$ such that $x_i \geq y_i$—this property is also known as *weak Pareto optimality*; and

(2) *Individual Rationality:* for each player $i \in N$, $x_i \geq \max\{ y_i \mid y_i \in V(\{i\}) \}$.

The set of all imputations of an NTU game $\mathcal{G}$ is denoted by $X(\mathcal{G})$. If $\mathcal{G}$ is actually a TU game, i.e., $\mathcal{G} = \langle N, v \rangle$ (or, equivalently, $\mathcal{G} = \langle N, V_v \rangle$), it is immediate to check that:

$$X(\mathcal{G}) = \left\{ x \in \mathbb{R}^N \Big| \sum_{i \in N} x_i = v(N) \text{ and } x_j \geq v(\{j\}), \ \forall j \in N \right\}.$$

---

2. Indeed, this encoding allows most of the solution concepts originally defined for TU games to be smoothly generalized to the NTU framework—as we shall discuss later in the section.

In particular, note that $x_j \geq v(\{j\})$ encodes the individual rationality of player $j$ at $x$.

An *outcome* for $\mathcal{G}$ is an imputation taken from $X(\mathcal{G})$ specifying a payoff distribution for all the players of the game. This outcome should represent a kind of agreement amongst players, which has to be "stable" with respect to the possibility that subsets of players get an incentive to deviate from it, by forming coalitions on their own. Depending on the criterium adopted to define this concept of stability, various *(solution) concepts* for coalitional games can be defined. The most relevant solution concepts for coalitional games—such as the *core*, the *bargaining set*, the *nucleolus*, the *kernel*, and the *Shapley value*—have been originally defined within the TU framework (see, e.g., Osborne & Rubinstein, 1994). Several efforts have been subsequently paid to apply these concepts within the more general NTU framework (see, e.g., Aumann & Hart, 2002). Natural extensions have been defined in some cases, while natural counterparts are still missing and looked for in others.

In the following, we shall provide an overview of the definitions of the basic solution concepts for TU games and their canonical extensions to NTU games.

## 2.1 Core

The concept of the *core* goes back to the work by Edgeworth (1881). In the TU framework, it has been formalized by Gillies (1959), and it was first extended to the NTU framework by Aumann (1961). In fact, this is the solution concept that enjoys the most canonical extension to the NTU case, which is the one presented next.

Let $\mathcal{G} = \langle N, V \rangle$ be an NTU game. For any coalition $S \subseteq N$, a vector $y \in \mathbb{R}^S$ of real numbers is called *S-feasible* if $y \in V(S)$. Let $x$ be a consequence in $\mathbb{R}^N$. Then, the pair $(y, S)$ is an *objection to* $x$ if $y$ is an $S$-feasible payoff vector such that $y_k > x_k$ for all $k \in S$—in this case, the coalition $S$ is also said to *block* $x$ *via* $y$.

**Definition 2.1.** The *core* $\mathscr{C}(\mathcal{G})$ of an NTU game $\mathcal{G} = \langle N, V \rangle$ is the set of all imputations $x$ to which there is no objection; that is,

$$\mathscr{C}(\mathcal{G}) = \{x \in X(\mathcal{G}) \mid \nexists S \subseteq N, y \in V(S) \text{ such that } y_k > x_k, \forall k \in S\}.$$  □

Thus, an imputation $x$ in the core is "stable" because there is no coalition whose members will receive a higher payoff than in $x$ by leaving the grand-coalition.

The application of Definition 2.1 over TU games exactly coincides with the original formulation by Gillies (1959). Moreover, it is easily seen that, when applied over TU games, Definition 2.1 can be equivalently restated as illustrated next (see, e.g., Osborne & Rubinstein, 1994). For a coalition $S \subseteq N$ and a payoff vector $x \in \mathbb{R}^N$, we define $x(S)$ as the value of the expression $\sum_{i \in S} x_i$.

**Definition 2.2.** The *core* $\mathscr{C}(\mathcal{G})$ of a TU game $\mathcal{G} = \langle N, v \rangle$ is the set of all imputations $x \in X(\langle N, V_v \rangle)$ such that, for each coalition $S \subseteq N$, $x(S) \geq v(S)$.  □

Thus, the core of a coalitional game with transferable utility and $|N|$ players is defined by a set of inequalities over $|N|$ variables and, in fact, it is a polytope in $\mathbb{R}^N$.

## 2.2 Bargaining Set

The concept of bargaining set was defined by Aumann and Maschler (1964), and has many variants even within the TU context (see, e.g., Maschler, 1992). A natural extension to the NTU framework was given by Peleg (1963), which is discussed next.

Let $\mathcal{G} = \langle N, V \rangle$ be an NTU game, and $x$ be a consequence in $V(N)$. Let $S \subseteq N$ be a coalition, and $y$ be an $S$-feasible payoff vector (i.e., $y \in V(S)$). The pair $(y, S)$ is an *objection of player $i$ against player $j$ to $x$* if $i \in S$, $j \notin S$, and $y_k > x_k$ for all $k \in S$.

A *counterobjection to the objection* $(y, S)$ *of $i$ against $j$ to $x$* is a pair $(z, T)$ where $j \in T$, $i \notin T$, and $z$ is a $T$-feasible payoff vector such that $z_k \geq x_k$ for all $k \in T \setminus S$ and $z_k \geq y_k$ for all $k \in T \cap S$. If there does not exist any counterobjection to $(y, S)$, we say that $(y, S)$ is a *justified objection*.

**Definition 2.3.** The *bargaining set* $\mathscr{B}(\mathcal{G})$ of an NTU game $\mathcal{G}$ is the set of all imputations $x$ to which there is no justified objection. □

Note that the above definitions straightforwardly apply to TU games, and coincide for them with the one originally proposed by Aumann and Maschler (1964). For the sake of completeness, we just recall here that $y$ (resp., $z$) is an $S$-feasible (resp., $T$-feasible) payoff vector in the TU game $\langle N, v \rangle$ if $y \in V_v(S)$ (resp., $z \in V_v(T)$) holds; that is, if $y(S) \leq v(S)$ (resp., $z(T) \leq v(T)$)—recall that $y(S) = \sum_{i \in S} y_i$ (resp., $z(T) = \sum_{i \in T} z_i$).

## 2.3 Nucleolus

The *nucleolus* is a solution concept introduced by Schmeidler (1969). Its definition is based on the notion of excess $e(S, x, V)$ of a coalition $S$ at an imputation $x$, which is a measure of the dissatisfaction of $S$ at $x$.

In the case of TU games (where $v$ denotes the worth function), it is widely accepted that the canonical definition of the excess is $e(S, x, V_v) = v(S) - x(S)$. Then, for each vector $x \in \mathbb{R}^N$, let us define $\theta(x)$ as the vector where the excesses associated with all coalitions (but the empty one) are arranged in non-increasing order:

$$\theta(x) = (e(S_1, x, V), e(S_2, x, V), \ldots, e(S_{2^{|N|}-1}, x, V)).$$

Let $\theta(x)[i]$ denote the $i$-th element of $\theta(x)$. For a pair of imputations $x$ and $y$, we say that $\theta(x)$ is *lexicographically smaller* than $\theta(y)$, denoted by $\theta(x) \prec \theta(y)$, if there exists a positive integer $q$ such that $\theta(x)[i] = \theta(y)[i]$ for all $i < q$ and $\theta(x)[q] < \theta(y)[q]$.

Since the excess is a measure of dissatisfaction, the imputations lexicographically minimizing the vector of the excesses are very natural candidates to be the "stable" outcomes for the game. This is indeed the idea underlying the definition of the nucleolus, as it was defined by Schmeidler (1969) for TU games.

**Definition 2.4.** The *nucleolus* $\mathscr{N}(\mathcal{G})$ of a TU game $\mathcal{G} = \langle N, v \rangle$ is the set

$$\mathscr{N}(\mathcal{G}) = \{x \in X(\langle N, V_v \rangle) \mid \nexists y \in X(\mathcal{G}) \text{ such that } \theta(y) \prec \theta(x)\}.$$ □

For games that do not fit the TU framework, the above definition can still be used provided that a suitable generalization of the concept of excess is conceived. The most influential approach to define excess functions for NTU games was proposed by Kalai (1975), who axiomatized the properties that such functions should satisfy as to retain some of the nice features of the underlying TU specifications. These properties are as follows:

1. Let $x, y \in \mathbb{R}^N$. If $x_i = y_i$ for all players $i \in S$, then $e(S, x, V) = e(S, y, V)$ holds, for each function $V$;

2. Let $x, y \in \mathbb{R}^N$. If $x_i < y_i$ for all players $i \in S$, then $e(S, x, V) > e(S, y, V)$ holds, for each function $V$;

3. Let $x \in \mathbb{R}^S$. If there is no vector $y \in V(S)$ such that, $\forall i \in S$, $y_i > x_i$, then $e(S, x, V) = 0$ holds, for each function $V$;

4. $e(S, x, V)$ is continuous jointly in $x$ and $V$.

As an example, a prototypical excess function discussed by Kalai is the following:

$$e_K(S, x, V) = \sup\left\{ t \in \mathbb{R} \mid \exists y \in V(S) \text{ such that } y_i = x_i + \frac{t}{|S|}, \forall i \in S \right\}. \qquad (2)$$

This function coincides with the canonical excess function $v(S) - x(S)$ whenever it is applied on TU games (Kalai, 1975).

### 2.4 Kernel

The *kernel* is a solution concept originally introduced in the TU framework by Davis and Maschler (1965) to help to understand the properties of the bargaining set.

For a TU game $\langle N, v \rangle$, define the *surplus* $s_{i,j}(x)$ of player $i$ against player $j$ at an imputation $x$ as the value $s_{i,j}(x) = \max_{S|i \in S, j \notin S} e(S, x, V_v) = \max_{S|i \in S, j \notin S} (v(S) - x(S))$.

**Definition 2.5.** The *kernel* $\mathscr{K}(\mathcal{G})$ of a TU game $\mathcal{G} = \langle N, v \rangle$ is the set:

$$\mathscr{K}(\mathcal{G}) = \{x \in X(\langle N, V_v \rangle) \mid s_{i,j}(x) > s_{j,i}(x) \Rightarrow x_j = v(\{j\}), \forall i, j \in N, i \neq j\}. \qquad \square$$

Note that the above definition for TU games is again based on the notion of excess. Intuitively, the surplus of player $i$ against $j$ at $x$ is the highest payoff that player $i$ can gain (or the minimal amount $i$ can lose, if it is a negative value) without the cooperation of $j$, by forming coalitions with other players that are satisfied at $x$; thus, $s_{i,j}(x)$ is the weight of a possible threat of $i$ against $j$. In particular, player $i$ has more "bargaining power" than $j$ at $x$ if $s_{i,j}(x) > s_{j,i}(x)$; however, player $j$ is immune to such threat whenever $x_j = v(\{j\})$, since in this case $j$ can obtain $v(\{j\})$ even by operating alone. We say that player $i$ outweighs player $j$ at $x$ if $s_{i,j}(x) > s_{j,i}(x)$ and $x_j > v(\{j\})$. The kernel is then the set of all imputations where no player outweighs another one.

Generalizing the kernel to NTU games is based on considering generalizations of the excess function, as for the nucleolus. Again, an influential approach, which is recalled next, is due to Kalai. However, it is worthwhile noticing here that other approaches have also been proposed in the literature (see, e.g., Orshan & Zarzuelo, 2000; Peleg & Sudhölter, 2007). Indeed, differently from the solution concepts discussed so far, variations of the kernel (and related concepts, such as the *prekernel*, which is the focus of the extensions cited above) to NTU games are still subject of research and debate (cf. Serrano, 1997).

Let $\mathcal{G} = \langle N, V \rangle$ be an NTU game. We say that a payoff vector $t \in \mathbb{R}^N$ is a *transfer* from player $j$ to player $i$ if $t_j \leq 0$, $t_i \geq 0$, and $t_k = 0$, for each player $k \in N \setminus \{i, j\}$. The transfer $t$ is *justified* at an imputation $x$, if for every real number $\lambda$, $0 < \lambda < 1$, the vector $y = x + \lambda t$ (such that $y_k = x_k + \lambda t_k$, for each $k \in N$) is an individually rational vector in $V(N)$ and $\theta(x + \lambda t) \prec \theta(x)$—of course, an excess function for $\mathcal{G}$ satisfying Kalai's axiomatization must be used in order to define the excess vectors. The *kernel* $\mathcal{K}(\mathcal{G})$ of $\mathcal{G}$ is the set: $\mathcal{K}(\mathcal{G}) = \{x \in X(\mathcal{G}) \mid$ there is no justified transfer from player $j$ to player $i$ at $x$, $\forall i, j \in N, i \neq j\}$.

## 2.5 Shapley Value

The *Shapley value* is a solution concept introduced in the TU framework by Shapley (1953). This concept associates with every TU game $\mathcal{G} = \langle N, v \rangle$ a unique payoff vector $\phi(\mathcal{G}) \in \mathbb{R}^N$, where each component $\phi(\mathcal{G})_i$, which is called the Shapley value of player $i$, indicates the worth to be assigned to player $i$, based upon her ability in cooperation as measured by the expected marginal contribution of player $i$ to forming coalitions (as formalized below).

Let $\pi$ be a permutation on the set $N$ of players. For any player $i$, we denote by $p_\pi^i$ the set of players preceding $i$ in $\pi$. The marginal contribution of player $i$ to the coalition $p_\pi^i$ is $v(p_\pi^i \cup \{i\}) - v(p_\pi^i)$. If permutations are chosen uniformly at random from the set $\Pi$ of all possible permutations, the expected marginal contribution of player $i$ in the game $\mathcal{G}$ is the value $\varphi(\mathcal{G})_i = \frac{1}{|N|!} \sum_{\pi \in \Pi} \left( v(p_\pi^i \cup \{i\}) - v(p_\pi^i) \right)$ or, equivalently:

$$\varphi(\mathcal{G})_i = \sum_{S \subseteq N \setminus \{i\}} \frac{|S|!(|N| - |S| - 1)!}{|N|!} \left( v(S \cup \{i\}) - v(S) \right).$$

The Shapley value is the unique payoff vector satisfying the following properties, which constitute its axiomatic characterization[3]:

*(1) Efficiency:* $\sum_{i \in N} \varphi(\mathcal{G})_i = v(N)$.

*(2) Symmetry:* Two players $i$ and $j$ are *symmetric* if, for each $S \subset N$ with $i, j \notin S$, $v(S \cup \{i\}) = v(S \cup \{j\})$. If players $i$ and players $j$ are symmetric, then $\varphi(\mathcal{G})_i = \varphi(\mathcal{G})_j$.

*(3) Dummy:* A player $i$ is *dummy* if, for each $S \subseteq N \setminus \{i\}$, $v(S \cup \{i\}) - v(S) = v(\{i\})$. If player $i$ is a dummy player, then $\varphi(\mathcal{G})_i = v(\{i\})$.

*(4) Additivity:* Let $\mathcal{G}' = \langle N, w \rangle$ be a TU game, and $\mathcal{G}'' = \langle N, v + w \rangle$ be the TU game such that $(v + w)(S) = v(S) + w(S)$ for each coalition $S \subseteq N$. Then, $\varphi(\mathcal{G}'')_i = \varphi(\mathcal{G})_i + \varphi(\mathcal{G}')_i$.

Note that the Shapley value might not satisfy the individual rationality, and thus it is not necessarily an imputation. Payoff distributions that are efficient, but not necessarily individually rational, are called *pre-imputations* in the literature.

Generalizing the Shapley value to the NTU framework is not straightforward. Different extensions of the Shapley value for NTU games have been proposed. Each of them, when evaluated on the NTU version of a TU game, coincides with the standard Shapley value for TU games. Here we discuss the generalization proposed by Shapley (1969) himself in the formulation reported by McLean (2002), and we refer the interested reader to this latter work for an extended treatment of the subject of values for NTU games, and to the paper by Hart (2004) for a comparison between the most notable three of them.

---

3. The characterization reported here is the one that can be found most often in the literature. However, the original axiomatic formulation of Shapley requires the *carrier axiom* instead of the efficiency and dummy axioms; the two axiomatizations are equivalent (Shapley, 1953; Winter, 2002).

Let $\mathcal{G} = \langle N, V \rangle$ be an NTU game. For a vector $\lambda \in \mathbb{R}^N$ of strictly positive real numbers, let $\mathcal{G}_\lambda$ be the game $\langle N, v_\lambda \rangle$ where

$$v_\lambda(S) = \sup \left\{ \sum_{i \in S} \lambda_i z_i \mid z \in V(S) \right\}.$$

The TU game $\mathcal{G}_\lambda$ is said to be *defined* for $\mathcal{G}$ if $v_\lambda(S)$ is finite for each $S$.

**Definition 2.6.** Let $\mathcal{G} = \langle N, V \rangle$ be an NTU game. A vector $x \in \mathbb{R}^N$ is a *Shapley NTU value payoff* for $\mathcal{G}$ if there exists a vector $\lambda \in \mathbb{R}^N$ of strictly positive real numbers such that: $x \in V(N)$; $\mathcal{G}_\lambda$ is defined for $\mathcal{G}$; and $\lambda_i x_i = \varphi(\mathcal{G}_\lambda)_i$ for each player $i \in N$. The set of all Shapley NTU values for $\mathcal{G}$ is denoted by $\varphi(\mathcal{G})$. □

Shapley NTU values fulfill, with some adaptations, the same axioms characterizing standard TU Shapley values. Actually, these axioms do not suffice to uniquely characterize the NTU counterpart, and other axioms have to be issued in order to define unambiguously the NTU Shapley value. An axiomatization for the NTU case was given by Aumann (1985). The interested reader is referred again to the work by McLean (2002), for more on this issue.

## 2.6 Properties of Solution Concepts for TU Games

We conclude by recalling some well-known properties of the solution concepts discussed above, when they are applied over TU games.

**Proposition 2.7** (see, e.g., Osborne & Rubinstein, 1994)**.** *Let $\mathcal{G} = \langle N, v \rangle$ be a TU game such that $X(\mathcal{G}) \neq \varnothing$. Then:*

*(1) $|\mathcal{N}(\mathcal{G})| = 1$;*

*(2) $\mathcal{N}(\mathcal{G}) \subseteq \mathcal{K}(\mathcal{G})$ (hence, $\mathcal{K}(\mathcal{G}) \neq \varnothing$);*

*(3) $\mathcal{K}(\mathcal{G}) \subseteq \mathcal{B}(\mathcal{G})$ (hence, $\mathcal{B}(\mathcal{G}) \neq \varnothing$);*

*(4) $\mathcal{C}(\mathcal{G}) \subseteq \mathcal{B}(\mathcal{G})$; and,*

*(5) $\mathcal{C}(\mathcal{G}) \neq \varnothing$ implies $\mathcal{N}(\mathcal{G}) \subseteq \mathcal{C}(\mathcal{G})$.*

Note that there is no relationship between the Shapley value and the other solution concepts (just recall that the Shapley value is not necessarily an imputation).

## 3. Constrained Games via Mixed-Integer Linear (In)Equalities

Assume that a TU game $\mathcal{G} = \langle N, v \rangle$ is given and consider the problem of modeling and dealing with constraints to be imposed on feasible worth distributions amongst players in $\mathcal{G}$.

These constraints might be implied by the very nature of the domain at hand (e.g., when the worth is not arbitrarily divisible), or because they reflect some hard preferences expressed by the players or by some regulation authority—recall again Example 1.1. Our approach to encode application-oriented constraints "within" a classical coalitional TU game setting is by defining a set of *mixed-integer linear (in)equalities*, which have to be satisfied by the imputations of the game $\mathcal{G}$. This approach is first formalized below; subsequently, we shall illustrate its modeling capabilities and discuss its relationships with the TU framework.

We start by recalling that a mixed-integer linear (in)equality is a linear (in)equality where some variables might be constrained to take values from the set $\mathbb{Z}$ of integers. For a set LC of mixed-integer linear (in)equalities, we denote by $real(\texttt{LC})$ and $int(\texttt{LC})$ the sets of all the variables in LC defined over $\mathbb{R}$ and $\mathbb{Z}$, respectively. Moreover, we assume that worth distributions can be constrained by defining inequalities via *player* and *auxiliary* variables. A player variable has the form $x_i$, where $i \in N$ is a player in the underlying TU game, and it is meant to encode the worth that has to be assigned to player $i$ in the outcomes of the game. The (possibly empty) set of the auxiliary variables in LC is then the set $real(\texttt{LC}) \cup int(\texttt{LC}) \setminus \{x_i \mid i \in N\}$. Auxiliary variables are sometimes useful for the modeling purposes, as we illustrated in Example 1.1.

Let us now proceed with our formalization. Let $\mathcal{G} = \langle N, v \rangle$ be a TU game, and recall from Section 2 that $\mathcal{G}$ can be viewed as the NTU game $\langle N, V_v \rangle$. Let LC be a set of mixed-integer linear (in)equalities. Define $\Omega(\texttt{LC})$ as the set of all the solutions to LC. Moreover, for a coalition $S \subseteq N$, let $\Omega(\texttt{LC})[S]$ denote the projection of $\Omega(\texttt{LC})$ onto the subspace associated with payoff domains for players in $S$; that is, a vector $y$ with index set $S$ belongs to $\Omega(\texttt{LC})[S]$ if and only if there is a vector $x \in \Omega(\texttt{LC})$ such that $x_i = y_i$ holds, for each $i \in S$.

Intuitively, a constrained game on LC is defined by restricting the consequences of an underlying TU game $\mathcal{G}$ to those belonging to the solution space of the set LC of mixed-integer linear (in)equalities projected onto the subspace associated with player variables—recall that further auxiliary variables may occur in LC. This is formalized below.

**Definition 3.1** (Mixed-Integer Constrained Games). Let $\mathcal{G} = \langle N, v \rangle$ be a TU game and let LC be a set of mixed-integer linear (in)equalities. Then, the (mixed-integer) *constrained game* $\mathcal{G}|_{\texttt{LC}}$ is the NTU game $\langle N, V_{\texttt{LC}} \rangle$ where $V_{\texttt{LC}}(S) = V_v(S) \cap \Omega(\texttt{LC})[S]$. That is,

$$V_{\texttt{LC}}(S) = \left\{ x \in \mathbb{R}^S \Big| \sum_{i \in S} x_i \leq v(S) \text{ and } x \in \Omega(\texttt{LC})[S] \right\}, \ \forall \ S \subseteq N.$$

$\square$

## 3.1 Modeling Capabilities of Constrained Games

Constraining a TU game $\mathcal{G}$ via a set LC of (in)equalities that do not involve integer variables (i.e., $int(\texttt{LC}) = \varnothing$) is an abstraction of those approaches in the literature that consider specific sets of (in)equalities over real variables (such as Aumann & Dreze, 1974; Byford, 2007; Jiang & Baras, 2007). In particular, this capability might be exploited to:

(1) *State hard preferences on the worth distributions.*
    As an example, consider a game $\mathcal{G} = \langle N, v \rangle$ over the set of players $N = \{1, 2, 3, 4\}$, and where $v(N) = 10$. Assume that players 3 and 4 together require to get at least a

half of the worth. Then, this requirement can be modeled as:

$$\begin{cases} x_3 + x_4 \geq 5 \\ x_1, x_2, x_3, x_4 \in \mathbb{R} \end{cases}$$

Moreover, by allowing integer variables, completely novel modeling capabilities emerge in our setting w.r.t. earlier approaches. Indeed, integer variables can be used to isolate non-convex regions, which might be needed to model specific application requirements that are NTU in their very nature, as exemplified below.

*(2) Consider alternative scenarios.*

By allowing integer variables, we may model alternative preferences of the players, i.e., we may enforce disjunctions of preferences. For instance, consider a scenario where *either* players 1 and 2 must get together no more than 3, *or* players 2 and 3 must get together no more than 5. In this case, we have two constraints (i.e., $x_1 + x_2 \leq 3$ and $x_2 + x_3 \leq 5$) and the goal is to define a set of (in)equalities prescribing that at least one of them is satisfied. To this end, an auxiliary integer variable can be used:

$$\begin{cases} x_1 + x_2 \leq 3 + U \times y \\ x_2 + x_3 \leq 5 + U \times (1 - y) \\ 0 \leq y \leq 1 \\ x_1, x_2, x_3 \in \mathbb{R} \\ y \in \mathbb{Z} \end{cases}$$

where the constant value $U$ is an upper bound on the worth of any coalition. Indeed, in the case where $y = 1$, the constraint $x_1 + x_2 \leq 3 + U$ is trivially satisfied (because $U$ is sufficiently large), and thus, we just enforce $x_2 + x_3 \leq 5$. Symmetrically, if $y = 0$, the constraint $x_2 + x_3 \leq 5 + U$ is trivially satisfied, and thus, we just enforce $x_1 + x_2 \leq 3$.

Of course, with simple manipulations, one may easily specify other kinds of alternatives, e.g., the fact that at least (or at most) $k$ given constraints are to be satisfied.

*(3) Restrict worth functions over specific domains.*

When domains are required to be integer intervals, this is rather obvious. For instance, assume that $x_3$ should take values from the domain $\{4, 5, 6, 7, 8, 9, 10\}$. Then, we may simply consider the following constraints:

$$\begin{cases} 4 \leq x_3 \leq 10 \\ x_1, x_2, x_4 \in \mathbb{R} \\ x_3 \in \mathbb{Z} \end{cases}$$

In order to model more general scenarios, we can have (as in point *(2)* above) mixed-integer linear (in)equalities to be defined over auxiliary variables. For instance, assume that player 2 wants either to take the whole worth for herself (even when forming coalitions with other players) or, whenever this is not possible, to get nothing. This can be modeled by a few constraints, over an additional variable $w$—in fact, notice that $v(N)$ is a constant value for a game given at hand:

$$\begin{cases} x_2 = v(N) \times w \\ 0 \le w \le 1 \\ x_1, x_2, x_3, x_4 \in \mathbb{R} \\ w \in \mathbb{Z} \end{cases}$$

Note that Example 1.1 basically presents a more realistic case, where several auxiliary variables are used to restrict money distributions to the available coin set.

Now that the basic modeling capabilities of constrained games have been discussed, in order to illustrate possible applications of the resulting framework, it is convenient to preliminarily observe two of its properties (which are related to the use of integer variables).

First, it is easy to check that, over constrained games, we may deal with imputation sets of arbitrary sizes.[4]

**Proposition 3.2.** *Let* $\mathcal{G} = \langle N, v \rangle$ *be a TU game and let* $\mathcal{X} \subseteq X(\mathcal{G})$ *be an arbitrary finite set of imputations. Then, there is a finite set of constraints* LC *such that* $X(\mathcal{G}|_{\mathtt{LC}}) = \mathcal{X}$.

In addition, integer variables might be used to succinctly specify exponentially many imputations via polynomially many (in)equalities.

**Proposition 3.3.** *There exists a class* $\mathcal{C} = \{\mathcal{G}|_{\mathtt{LC}}^n\}_{n>0}$ *of constrained games such that each game* $\mathcal{G}|_{\mathtt{LC}}^n$ *is over* $n + 1$ *players,* LC *consists of* $2 \times n + 1$ *inequalities, and* $|X(\mathcal{G}|_{\mathtt{LC}})| = 2^n$.

We believe that the setting emerging from the above properties is rather appealing from a knowledge representation perspective. Indeed, one may exploit constrained games to naturally model scenarios where non-transferable conditions emerge, by devising very compact specifications of the desired restrictions on how utilities may be transferred among coalition members. In fact, various circumstances can be envisaged where the usage of constrained games is a natural choice; for instance, whenever the worth to be distributed among the agents comes as a set of indivisible goods, as exemplified below.

**Example 3.4 (Distributing indivisible goods).** A certain region of the country is famous for hosting several producers of two kinds of goods, named $\alpha$ and $\beta$. For each producer $i \in \{1, \ldots, n\}$, let $\alpha_i$ and $\beta_i$ denote the quantity of $\alpha$ and $\beta$ pieces produced by $i$, respectively. By assembling together one piece $\alpha$ and one piece $\beta$, a novel kind of indivisible good can be obtained and, in fact, commercializing the assembled product is a much more advantageous business than selling $\alpha$ and $\beta$ separately. Therefore, an agreement is found amongst producers in the area in order to assemble the pieces of $\alpha$ and $\beta$ that are overall available, provided that the resulting units of the assembled product are (fairly) distributed amongst the involved producers, which would like to independently commercialize them.

This scenario can be easily modeled within our framework as follows. First, we associate with every coalition $S \subseteq \{1, \ldots, n\}$, the number of pieces of the assembled product that $S$ can produce. Thus, we just define:

$$v(S) = \min(\sum_{i \in S} \alpha_i, \sum_{i \in S} \beta_i).$$

---

4. For the sake of exposition, proofs of the propositions stated in this section are reported in the Appendix.

Then, since the assembled product is indivisible, any possible worth distribution is a vector of non-negative integers, which can immediately be modeled via the following set of constraints $\mathtt{LC} = \{x_i \geq 0, x_i \in \mathbb{Z}, \ \forall 1 \leq i \leq n\}$. In particular, $\Omega(\mathtt{LC})$ is not a convex region, so that earlier modeling perspectives on NTU games, such as those present in the handbook edited by Aumann and Hart (2002), do not apply here.                                               ◁

There are cases, however, where the worth might practically be assumed to be divisible, but specific constraints regulate its actual distribution. Notably, even in these cases, integer variables may play a crucial role as illustrated next.

**Example 3.5** (**Service composition**). Assume that a service $T$ can be acquired for 100 Euros—for the sake of simplicity, we assume that money is divisible, for otherwise worth distributions might simply be restricted over a discrete domain as in Example 3.4 or in Example 1.1. Supplying the service $T$ implies executing $m$ tasks, named $t_1, \ldots, t_m$.

Assume also that there is a set $\{1, \ldots, n\}$ of agents, each one capable of carrying out some of those $m$ tasks, and let $s_j^i$ denote the ability of agent $i$ to perform the task $t_j$ ($s_j^i = 1$ means that agent $i$ is able to perform $t_j$, whereas $s_j^i = 0$ means that she is not capable to do so). Thus, a coalition $S \subseteq \{1, \ldots, n\}$ is capable of supporting the service $T$ in the case where $\sum_{i \in S} s_j^i \geq 1$, for each $j \in \{1, \ldots, m\}$.

Assume, moreover, that in order to complete $T$, not only all of its tasks must be completed, but agents contributing to $T$ must be able to exchange some partial results returned by performing required tasks. Establishing a communication infrastructure guaranteeing the needed result-transfers to take place has a specific cost for each coalition $S$, which we denote by $com(S) < 100$. Hence, the amount of money that might finally be distributed amongst players in $S$ is described by the following worth function:

$$v(S) = \begin{cases} 100 - com(S) & \text{if } \sum_{i \in S} s_j^i \geq 1, \ \forall j \in \{1, \ldots, m\} \\ 0 & \text{otherwise.} \end{cases}$$

Note that the above scenario defines a classical TU game $\mathcal{G} = \langle \{1, \ldots, n\}, v \rangle$. However, things may be significantly different if we assume that each agent $i$ has to sustain an internal cost, say $c_j^i$, whenever actually performing the task $t_j$, and that, hence, she may decide not to perform the task at all, if it is not convenient. Indeed, in this case, letting $\gamma_j^i \in \{0, 1\}$ be a variable denoting whether $i$ is actually performing $t_j$, it is natural to state that the total internal cost for agent $i$ (which is given by the expression $\sum_{j=1}^{m} \gamma_j^i \times c_j^i$) should not exceed what the agent gets from the worth distribution. Hence, utilities cannot be freely distributed and, for a proper modeling of this more realistic scenario, the game has to be enriched with the following set of constraints:

$$\mathtt{LC} = \begin{cases} \displaystyle\sum_{i \in N} \gamma_j^i = 1, \ \forall j \in \{1, \ldots, m\} \\[2mm] \displaystyle x_i \geq \sum_{j=1}^{m} \gamma_j^i \times c_j^i, \forall i \in \{1, \ldots, n\} \\[2mm] 0 \leq \gamma_j^i \leq s_j^i, \forall i \in \{1, \ldots, n\}, \forall j \in \{1, \ldots, m\} \\[1mm] x_1, \ldots, x_n \in \mathbb{R} \\[1mm] \gamma_j^i \in \mathbb{Z}, \forall i \in \{1, \ldots, n\}, \forall j \in \{1, \ldots, m\} \end{cases}$$

With respect to this formalization, it is worthwhile noting that if $\Omega(\texttt{LC})$ is empty then the service cannot be provided at all, and indeed $X(\mathcal{G}|_{\texttt{LC}})$ would be empty in its turn. Otherwise, i.e., if $\Omega(\texttt{LC}) \neq \varnothing$, the imputations of $\mathcal{G}|_{\texttt{LC}}$ correspond to those worth distributions associated with some legal staffing for the tasks, rather than to all arbitrary possible worth distributions (as it would be in the plain TU case).

It is worthwhile contrasting the above formulation with an alternative TU one, where the constraints in $\texttt{LC}$ are directly encoded in the definition of worth function, instead of using a separate component thereof, as done in the NTU framework we are proposing here. For instance, one may add to the condition for $v(S) = 100 - com(S)$ the requirement that there exists an element $\hat{x} \in \Omega(\texttt{LC})$ such that for each task $t_j$, there is a player $i \in S$ with $\gamma_j^i \neq 0$ (in $\hat{x}$). This way, we can ensure that the payoff $100 - com(S)$ is assigned to each coalition $S$ which is formed by players that can perform some task conforming with cost constraints, and that can jointly complete $T$—this refined modeling perspective is exactly the one underlying the class of *linear programming games* (see, e.g., Owen, 1975). However, this approach would not prescribe how the payoff $100 - com(S)$ has to be actually distributed amongst the players in $S$. In fact, while focusing on accurately modeling the worth function, it cannot guarantee that the outcome of the game (according to any chosen solution concept) fulfils the desired constraints on the distribution of payoffs for the single players. In other words, using constraints in the definition of the worth function may be useful in certain cases for more careful modeling purposes, but cannot in general replace the use of "external" constraints to actually constrain the allowed worth distributions. ◁

As an important remark, we note here that the "structure" of the above example may well be used as a guideline in the formalization of other application scenarios. Indeed, the basic idea is to use mixed-integer linear constraints to define solutions for combinatorial problems associated with feasible worth distributions (reflecting, e.g., the costs of such solutions). Thus, while we have contextualized the approach to the case of a staffing problem, very similar encodings can be used to define constrained games suited to deal with scheduling and planning problems, just to cite a few.

## 3.2 A Closer Look at Constrained Games

We can now resort to the presentation of the framework of constrained games by analyzing the structure of their consequences, and the role played by the individual rationality requirement in this context. This analysis will provide important bases for our subsequent treatment of the analytical and the computational properties of NTU solution concepts as applied on constrained games.

### 3.2.1 CONSEQUENCES IN CONSTRAINED GAMES

In order to understand the nature of constrained games, it is convenient to take a closer look at the structure of the function $V_{\texttt{LC}}$. In fact, while in principle an NTU game $\langle N, V \rangle$ does not impose any requirement on the function associating a set of consequence $V(S)$ with each coalition $S \subseteq N$, those NTU games that attracted much attention in the literature do not actually allow arbitrary consequences to be specified (Aumann & Hart, 2002). Indeed, the sets of consequences are usually required to satisfy some additional conditions, which are (de facto) conceived as to guarantee that nice properties hold over the solution concepts of interest. Some of the assumptions that have been most often considered in the literature (not necessarily required to simultaneously hold) are recalled next. For each $S \subseteq N$, $V(S)$ might be required to be:

(1) *(Upper) Bounded:* there is a real number $a \in \mathbb{R}$ such that for each $x \in V(S)$, and for each $i \in S$, $x_i \leq a$ holds;

(2) *Closed:* $V(S)$ contains its own boundaries;

(3) *Compact:* $V(S)$ is both closed and bounded;

(4) *Comprehensive:* if $x \in V(S)$, $y \in \mathbb{R}^N$ and $(\forall i \in S)(y_i \leq x_i)$, then $y \in V(S)$;

(5) *Convex:* for each pair $x, y \in V(S)$, and for each real number $t$, $0 < t < 1$, the vector $(1-t)x + ty$ belongs to $V(S)$.

(6) *Non-empty:* $|V(S)| > 0$.

In the case of constrained games, we do not explicitly ask for the above requirements to be satisfied, thereby giving rise to a setting with very powerful modeling capabilities (as we discussed in Section 3.1). The differences between constrained games and classical NTU games are illustrated next.

Consider the function $V_{\mathtt{LC}}$ associated with a constrained game $\mathcal{G}|_{\mathtt{LC}}$. The first difference concerns property *(1)*, because $V_{\mathtt{LC}}(S)$ is not required to be bounded (as for the TU case, where individual payoffs for players in $S$ are not bounded in general, since the only requirement is that their sum does not exceed the worth $v(S)$ associated with $S$—see Equation (1) in Section 2). Actually, this is not a substantial difference given that, in any possible worth distribution for $\mathcal{G}|_{\mathtt{LC}}$ corresponding to any solution concept illustrated in Section 2, the payoff of each player is bounded.

Similarly, it is easy to see that there might be cases where property *(2)* does not hold in the context of constrained games. Indeed, $V_{\mathtt{LC}}(S)$ might be not closed whenever $\mathtt{LC}$ contains some strict inequality that excludes the boundary of $V_{\mathtt{LC}}(S)$. However, it is known that such cases are undesirable for some solution concepts, and hence we shall consider constraints based on non-strict inequalities only. Property *(3)* is the combination of the first two properties, and thus the above lines of reasoning still apply.

The differences in the remaining three properties, instead, do characterize the framework of constrained games and are at the basis of its modeling capabilities. In fact, the ability to handle sets of consequences that are not comprehensive and convex, and that are possibly empty for some coalition, is an important peculiarity of constrained games from a knowledge representation perspective. Indeed, we may lose comprehensiveness each time a constraint on the payoff distribution is given which states that some players are required to get at least a certain worth; we may lose convexity (as well as comprehensiveness) each time integrality constraints are involved. Moreover, we may deal with an empty set of consequences for some coalition $S$ whenever there is no feasible way to distribute the worth associated with $S$ according to the constraints that players in $S$ must satisfy.

**Example 3.6.** Consider the TU game $\mathcal{G} = \langle N, v \rangle$ such that $N = \{1, 2, 3\}$, $v(\{1, 2, 3\}) = 3$, and $v(S) = 0$ for each $S \subset \{1, 2, 3\}$. Consider a scenario where the worth in $\mathcal{G}$ is restricted to be an integer value (i.e., payoff distributions are taken from $\mathbb{Z}^{\{1,2,3\}}$), and where players 1 and 2 require to get at least 2. These constraints can be modeled as follows:

$$\mathtt{LC} = \left\{ \begin{array}{l} x_1 + x_2 \geq 2 \\ x_1, x_2, x_3 \in \mathbb{Z} \end{array} \right.$$

Note that $\Omega(\mathtt{LC})[\{1,2,3\}] = \{x \in \mathbb{Z}^{\{1,2,3\}} \mid x_1 + x_2 \geq 2\}$, $\Omega(\mathtt{LC})[\{1,2\}] = \{x \in \mathbb{Z}^{\{1,2\}} \mid x_1 + x_2 \geq 2\}$, $\Omega(\mathtt{LC})[\{1,3\}] = \mathbb{Z}^{\{1,3\}}$, $\Omega(\mathtt{LC})[\{2,3\}] = \mathbb{Z}^{\{2,3\}}$, $\Omega(\mathtt{LC})[\{1\}] = \mathbb{Z}^{\{1\}}$, $\Omega(\mathtt{LC})[\{2\}] = \mathbb{Z}^{\{2\}}$, and $\Omega(\mathtt{LC})[\{3\}] = \mathbb{Z}^{\{3\}}$. Then, the constrained game $\mathcal{G}|_{\mathtt{LC}} = \langle N, V_{\mathtt{LC}} \rangle$ is such that:

$$
\begin{aligned}
V_{\mathtt{LC}}(\{1,2,3\}) &= \{x \in \mathbb{R}^{\{1,2,3\}} \mid x_1 + x_2 + x_3 \leq 3\} \cap \Omega(\mathtt{LC})[\{1,2,3\}] = \\
&= \{x \in \mathbb{Z}^{\{1,2,3\}} \mid x_1 + x_2 + x_3 \leq 3, x_1 + x_2 \geq 2\}; \\
V_{\mathtt{LC}}(\{1,2\}) &= \{x \in \mathbb{R}^{\{1,2\}} \mid x_1 + x_2 \leq 0\} \cap \Omega(\mathtt{LC})[\{1,2\}] = \varnothing; \\
V_{\mathtt{LC}}(\{1,3\}) &= \{x \in \mathbb{R}^{\{1,3\}} \mid x_1 + x_3 \leq 0\} \cap \Omega(\mathtt{LC})[\{1,3\}] = \{x \in \mathbb{Z}^{\{1,3\}} \mid x_1 + x_3 \leq 0\}; \\
V_{\mathtt{LC}}(\{2,3\}) &= \{x \in \mathbb{R}^{\{2,3\}} \mid x_2 + x_3 \leq 0\} \cap \Omega(\mathtt{LC})[\{2,3\}] = \{x \in \mathbb{Z}^{\{2,3\}} \mid x_2 + x_3 \leq 0\}; \\
V_{\mathtt{LC}}(\{i\}) &= \{x \in \mathbb{R}^{\{i\}} \mid x_i \leq 0\} \cap \Omega(\mathtt{LC})[\{i\}] = \{x \in \mathbb{Z}^{\{i\}} \mid x_i \leq 0\}; \ (i \in \{1,2,3\})
\end{aligned}
$$

Despite the very simple constraints considered for $\mathcal{G}$, it is immediate to check that (e.g.) $V_{\mathtt{LC}}(N)$ is not comprehensive and not convex, and that $V_{\mathtt{LC}}(\{1,2\})$ is empty. Indeed, the integrality constraints immediately lead to loose comprehensiveness and convexity. Moreover, the fact that players 1 and 2 require to get at least 2 implies that the coalition $\{1,2\}$ will never form to deviate from the grand-coalition, given that these two players cannot guarantee themselves any worth when acting without player 3 (indeed, $v(\{1,2\}) = 0$). ◁

### 3.2.2 INDIVIDUAL RATIONALITY IN CONSTRAINED GAMES

A further important issue to be pointed out for constrained games is related to the individual rationality requirement over the set of imputations. Let $\mathcal{G} = \langle N, V \rangle$ be an NTU game, and recall from Section 2 that any imputation $x \in X(\mathcal{G})$ must be such that for each player $i \in N$, $x_i \geq \max\{y_i \mid y_i \in V(\{i\})\}$.

Consider a constrained game $\mathcal{G}|_{\mathtt{LC}}$ where $\mathcal{G} = \langle N, V_v \rangle$ is the underlying TU game. By Definition 3.1, the set $V_{\mathtt{LC}}(\{i\})$ coincides with $V_v(\{i\}) \cap \Omega(\mathtt{LC})[\{i\}]$. Then, because of the individual rationality requirement, for each player $i$, $x_i \geq \max\{y_i \mid y_i \in V_v(\{i\}) \cap \Omega(\mathtt{LC})[\{i\}]\}$.

Note that a special case occurs when $V_{\mathtt{LC}}(\{i\}) = V_v(\{i\}) \cap \Omega(\mathtt{LC})[\{i\}] = \varnothing$. In this case, indeed, $\max\{y_i \mid y_i \in V_{\mathtt{LC}}(\{i\})\}$ is not defined (as a real value). An approach might be therefore to observe that the game is "over-constrained", and then stop the analysis. This approach is, in fact, consistent with several NTU formalizations requiring that the set of consequences is non-empty, for each possible coalition (see Section 3.2.1).

However, in Example 3.6 we have already pointed out that empty sets of consequences may naturally emerge from constrained games, because $V_{\mathtt{LC}}(S) = \varnothing$ reflects the fact that coalition $S$ cannot form to deviate from the grand-coalition, for there is no worth distribution that can be in principle granted to its members alone (according to the underlying TU game), and that satisfies the constraints at hand. Consequently, a finer-grained perspective should be considered to deal with the individual rationality requirement, in this special case where $V_{\mathtt{LC}}(\{i\}) = \varnothing$, for some player $i \in N$.

The basic observation is that $V_{\mathtt{LC}}(\{i\}) = \varnothing$ necessarily implies that $x_i > v(\{i\})$ holds, for each imputation $x \in V_{\mathtt{LC}}(N)$. Thus, in these extreme scenarios, the individual rationality constraint is conceptually satisfied (though formally undefined) for each possible imputation $x$, since constraints in LC require $x_i$ to be larger than $v(\{i\})$. Technically, we stress here that the same conclusion is implied by defining $\max\{\varnothing\} = -\infty$, which is the standard extension of "max" over empty sets.

In the light of the above perspective, we can show that individual rationality is "preserved" when constraints are added to a given TU game.

**Proposition 3.7.** *Let* $\mathcal{G} = \langle N, v \rangle$ *be a TU game and let* $x$ *be a payoff vector that is individually rational w.r.t.* $\mathcal{G}$ *(i.e.,* $x_i \geq v(\{i\})$, *for each player* $i \in N$*). Then, for each set* LC *of constraints,* $x$ *is individually rational w.r.t. the constrained game* $\mathcal{G}|_{LC}$.

## 4. Solution Concepts for Constrained Games

Constrained coalitional games are special cases of NTU games, and therefore they inherit from them the various solution concepts we have discussed in Section 2. Thus, for a constrained (and, as such, NTU) game $\mathcal{G}|_{LC}$, it is of interest to compute the core, the bargaining set, the nucleolus, the kernel, and the Shapley value. In this section, we will study the properties of these concepts, by highlighting similarities and differences featured by constrained games as opposed to TU games.

In a nutshell, we will show that the properties of TU games stated in Proposition 2.7 still hold over constrained games for all solution concepts, but for the bargaining set that might be empty in some games. Moreover, the portion of the core of a TU game $\mathcal{G}$ that satisfies all the constraints is "preserved", in the sense that it is a subset of the core of the constrained game $\mathcal{G}|_{LC}$ built on top of $\mathcal{G}$. On the other hand, for the other solutions concepts, the preservation property holds in special cases only.

Before illustrating our results, it is useful to state a property relating the imputation set of a constrained game with the imputation set of the underlying TU game.

**Proposition 4.1.** *Let* $\mathcal{G}|_{LC}$ *be a constrained game where* $\mathcal{G} = \langle N, v \rangle$. *Then,* $X(\mathcal{G}) \cap \Omega(LC)[N] \subseteq X(\mathcal{G}|_{LC})$.

*Proof.* Let $x$ be a payoff vector in $X(\mathcal{G}) \cap \Omega(LC)[N]$. Since $x \in X(\mathcal{G})$ then $x \in V_v(N)$. We are also assuming that $x \in \Omega(LC)[N]$ and hence we have that $x \in V_{LC}(N)$—recall, by Definition 3.1, that $V_{LC}(N) = V_v(N) \cap \Omega(LC)[N]$. Being $x \in X(\mathcal{G})$, $x$ is also efficient w.r.t. $\mathcal{G}$, meaning that for each $y \in V_v(N)$, there is a player $i \in N$ with $x_i \geq y_i$. Moreover, since $V_{LC}(N) \subseteq V_v(N)$, $x$ is also efficient w.r.t. $\mathcal{G}|_{LC}$. Finally, by Proposition 3.7, we know that $x$ is individually rational w.r.t. $\mathcal{G}|_{LC}$. Thus, $x \in X(\mathcal{G}|_{LC})$. ☐

Based on the above result, to show that an imputation $x \in X(\mathcal{G})$ also belongs to $X(\mathcal{G}|_{LC})$, in the following we shall just show that $x$ satisfies the constraints in LC, thereby avoiding to explicitly reason on the efficiency and the individual rationality of $x$.

### 4.1 Relationships Among Solution Concepts

We start our analysis by investigating whether the properties of the basic solution concepts (as they appear from Proposition 2.7) are preserved in the setting of constrained games.

#### 4.1.1 Counterparts of Proposition 2.7.(1) and Proposition 2.7.(2)

Let us begin by focusing on the first two properties in Proposition 2.7, which pertain to the nucleolus. In the TU framework, the nucleolus always consists of exactly one imputation. In the constrained framework, the properties of this solution concept are intimately related to the closeness of $\Omega(\text{LC})$ (and, in turn, of $X(\mathcal{G}|_{\text{LC}})$), i.e., to whether $\Omega(\text{LC})$ contains its own boundaries. Recall from Section 3 that $\Omega(\text{LC})$ might not be closed only due to the occurrence of some strict inequalities in $\text{LC}$.

**Proposition 4.2.** *There exists a constrained game $\bar{\mathcal{G}}|_{\text{LC}}$ (with $int(\text{LC}) = \varnothing$) such that $X(\bar{\mathcal{G}}|_{\text{LC}}) \neq \varnothing$ and where $\mathcal{N}(\bar{\mathcal{G}}|_{\text{LC}}) = \varnothing$ (for the excess function $e_K$ in Equation (2) on page 640).*

*Proof.* Consider the game $\bar{\mathcal{G}}$ over players $\{1, 2\}$, where $v(\{1, 2\}) = 1$ and $v(\{1\}) = v(\{2\}) = 0$. Given the constraint $\text{LC} = \{x_1 < \frac{1}{2}, x_1 \in \mathbb{R}\}$, one may note that $X(\bar{\mathcal{G}}|_{\text{LC}}) \neq \varnothing$. Indeed, observe that $\max\{x_1 \mid x_1 \in V_{\text{LC}}(\{1\})\} = \max\{x_2 \mid x_2 \in V_{\text{LC}}(\{2\})\} = 0$, since $V_{\text{LC}}(\{i\}) = V_v(\{i\}) \cap \Omega(\text{LC})[\{i\}] = \{x_i \in \mathbb{R}^{\{i\}} \mid x_i \leq 0\} \cap \Omega(\text{LC})[\{i\}] = \{x_i \in \mathbb{R}^{\{i\}} \mid x_i \leq 0\}$, for each $i \in \{1, 2\}$. Thus, a payoff vector $x \in X(\bar{\mathcal{G}}|_{\text{LC}})$ is just required to satisfy $x_1 \geq 0$ and $x_2 \geq 0$, in order to be individually rational. In particular, we claim that $X(\bar{\mathcal{G}}|_{\text{LC}}) = \{x \in \mathbb{R}^{\{1, 2\}} \mid x_1 + x_2 = 1, x_1 < \frac{1}{2}, x_1 \geq 0, x_2 \geq 0\}$. Indeed, any vector $x \in V_{\text{LC}}(N) = \{x \in \mathbb{R}^N \mid x_1 + x_2 \leq 1, x_1 < \frac{1}{2}\}$ such that $x_1 < \frac{1}{2}$ and $x_2 < \frac{1}{2}$ is not efficient, given that there is a vector $y \in V_{\text{LC}}(N)$ such that $x_1 < y_1 < \frac{1}{2}$ and $x_2 < y_2 < \frac{1}{2}$, with $y_1 + y_2 < 1$. Moreover, any vector $x \in V_{\text{LC}}(N)$ such that $x_1 < \frac{1}{2}$, $x_2 > \frac{1}{2}$, and $x_1 + x_2 < 1$ is also not efficient, given the existence of a vector $y \in V_{\text{LC}}(N)$ such that $x_1 < y_1 = x_1 + \frac{1-x_2-x_1}{2} < \frac{1}{2}$ and $x_2 < y_2 = x_2 + \frac{1-x_2-x_1}{2}$, with (therefore) $y_1 + y_2 = 1$.

Consider now an imputation $\hat{x}$ with $\hat{x}_1 + \hat{x}_2 = 1$ and $\hat{x}_1 < \frac{1}{2}$ (and hence $\hat{x}_2 > \frac{1}{2}$). Then, we can always build an imputation $\hat{y} \neq \hat{x}$ such that $\hat{y}_1 = \hat{x}_1 + (\frac{1}{2} - \hat{x}_1)/2 > \hat{x}_1$ and $\hat{y}_2 = 1 - \hat{y}_1 < \hat{x}_2$; just notice here that $\hat{y}_1 < \frac{1}{2}$ holds. For this new imputation, it is the case that $\theta(\hat{y}) \prec \theta(\hat{x})$ holds w.r.t. the excess function $e_K$ in Equation (2). Indeed, by recalling that $V_{\text{LC}}(\{i\}) = \{x_i \in \mathbb{R}^{\{i\}} \mid x_i \leq 0\}$ and that $\hat{x}_i \geq 0$ and $\hat{y}_i \geq 0$ hold, for each $i \in \{1, 2\}$, we have $\theta(\hat{x}) = (0, -\hat{x}_1, -\hat{x}_2)$ and $\theta(\hat{y}) = (0, -\hat{y}_1, -\hat{y}_2)$. Thus, there is no imputation $\hat{x}$ that belongs to $\mathcal{N}(\bar{\mathcal{G}}|_{\text{LC}})$, i.e., $\mathcal{N}(\bar{\mathcal{G}}|_{\text{LC}}) = \varnothing$. Note that $\Omega(\text{LC})$ is not closed. $\qquad\square$

In practical applications of linear programming, one may deal with non-strict inequalities only (see, e.g., Papadimitriou & Steiglitz, 1998); in these cases (i.e., whenever $X(\mathcal{G}|_{\text{LC}})$ is closed and hence compact, since it is always bounded), the nucleolus is not empty. This property was observed to hold by Kalai (1975) along with the relationship of the nucleolus with the kernel. These properties can be restated in our context as follows.

**Proposition 4.3** (cf. Kalai, 1975). *Let $\mathcal{G}|_{\text{LC}}$ be a constrained game with $X(\mathcal{G}|_{\text{LC}}) \neq \varnothing$. Then,*

(1) *if $X(\mathcal{G}|_{\text{LC}})$ is compact, then $|\mathcal{N}(\mathcal{G}|_{\text{LC}})| \geq 1$;*

(2) *$\mathcal{N}(\mathcal{G}|_{\text{LC}}) \subseteq \mathcal{K}(\mathcal{G}|_{\text{LC}})$ (hence, $\mathcal{K}(\mathcal{G}|_{\text{LC}}) \neq \varnothing$ whenever $X(\mathcal{G}|_{\text{LC}})$ is compact).*

In the following, examples and counterexamples will be built avoiding the use of strict inequalities.

### 4.1.2 COUNTERPARTS OF PROPOSITIONS 2.7.*(3)*, *(4)*, AND *(5)*

Let us now move to analyze the counterpart of Proposition 2.7.*(3)*. To this end, we first notice that, unlike in the TU case, the bargaining set can sometimes be empty.

**Proposition 4.4.** *There exists a constrained game* $\bar{\mathcal{G}}|_{\mathsf{LC}}$ *(with* $\mathrm{int}(\mathsf{LC}) = \varnothing$*) such that* $X(\bar{\mathcal{G}}|_{\mathsf{LC}}) \neq \varnothing$ *and* $\mathscr{B}(\bar{\mathcal{G}}|_{\mathsf{LC}}) = \varnothing$.

*Proof.* Consider the TU game $\bar{\mathcal{G}}$ over players $\{1, 2, 3, 4\}$, whose worths are as follows: $v(\{1, 2, 3, 4\}) = 3$, $v(\{1, 2\}) = 2$, $v(\{2, 3, 4\}) = 3$, $v(\{1, 3, 4\}) = 3$, $v(\{2\}) = 1$, and $v(S) = 0$ for any other coalition $S \subset \{1, 2, 3, 4\}$. Consider moreover the following set of constraints:

$$\mathsf{LC} = \begin{cases} x_1 + x_2 + x_3 + x_4 = 3 \\ x_2 + x_3 + x_4 = 3 \\ x_1 + x_3 = 1 \\ x_1 + x_4 = 1 \\ x_1, x_2, x_3, x_4 \in \mathbb{R} \end{cases}$$

Let $\hat{x}$ be the payoff vector with $\hat{x}_1 = 0$ and $\hat{x}_2 = \hat{x}_3 = \hat{x}_4 = 1$. Observe that $\hat{x}$ satisfies the constraints in $\mathsf{LC}$. Moreover, $\hat{x}$ is an imputation of $X(\bar{\mathcal{G}})$; thus, by Proposition 4.1, $\hat{x}$ belongs to $X(\bar{\mathcal{G}}|_{\mathsf{LC}})$. In fact, $\hat{x}$ is the only vector in $\Omega(\mathsf{LC})$ and, therefore, it is the only imputation in $X(\bar{\mathcal{G}}|_{\mathsf{LC}})$. Thus, to prove that $\mathscr{B}(\bar{\mathcal{G}}|_{\mathsf{LC}}) = \varnothing$, we have just to show that $\hat{x}$ is not contained in $\mathscr{B}(\bar{\mathcal{G}}|_{\mathsf{LC}})$. To this end, consider the objection $(y, \{1, 2\})$ of player 1 against player 3, where $y_1 = \frac{1}{3}$ and $y_2 = \frac{5}{3}$. Note that $y_1 + y_2 = v(\{1, 2\}) = 2$, $y_2 > 1$ and $y_1 > 0$, and thus $y$ is $\{1, 2\}$-feasible. Moreover, recall that $v(\{3\}) = v(\{3, 4\}) = v(\{2, 3\}) = 0$. It follows that counterobjections of 3 against 1 to $(y, \{1, 2\})$ may be only constructed over the coalition $\{2, 3, 4\}$. Assume that $(z, \{2, 3, 4\})$ is one such a counterobjection. For player 2, which belongs to the intersection of the two coalitions $\{1, 2\}$ and $\{2, 3, 4\}$, $z_2 \geq y_2 > 1$ holds. Because of the constraint $z_2 + z_3 + z_4 = 3$, this entails $z_3 + z_4 < 2$. However, this is impossible since we should have also $z_3 \geq \hat{x}_3 = 1$ and $z_4 \geq \hat{x}_4 = 1$. Thus, there are no possible counterobjections to the objection $(y, \{1, 2\})$ to $\hat{x}$. It follows that $\hat{x}$ does not belong to $\mathscr{B}(\bar{\mathcal{G}}|_{\mathsf{LC}})$ and, hence, $\mathscr{B}(\bar{\mathcal{G}}|_{\mathsf{LC}}) = \varnothing$, even though $X(\bar{\mathcal{G}}|_{\mathsf{LC}}) \neq \varnothing$. $\qquad\blacksquare$

As a consequence, we derive that the counterpart of Proposition 2.7.*(3)* does not hold over constrained games. Indeed, we may just consider the game $\bar{\mathcal{G}}|_{\mathsf{LC}}$ defined in the proof of Proposition 4.4 and observe that, since $X(\bar{\mathcal{G}}|_{\mathsf{LC}}) \neq \varnothing$, by Proposition 4.3, we know that $\mathscr{K}(\bar{\mathcal{G}}|_{\mathsf{LC}}) \neq \varnothing$.

**Corollary 4.5.** *There is a constrained game* $\bar{\mathcal{G}}|_{\mathsf{LC}}$ *(with* $\mathrm{int}(\mathsf{LC}) = \varnothing$*) such that* $\mathscr{B}(\bar{\mathcal{G}}|_{\mathsf{LC}}) = \varnothing$ *while* $\mathscr{K}(\bar{\mathcal{G}}|_{\mathsf{LC}}) \neq \varnothing$ *(and thus* $\mathscr{K}(\bar{\mathcal{G}}|_{\mathsf{LC}}) \not\subseteq \mathscr{B}(\bar{\mathcal{G}}|_{\mathsf{LC}})$*).*

Below we complete the picture pertaining to the bargaining set, by showing that the core is always included in it. This provides the counterpart of Proposition 2.7.*(4)*.

**Proposition 4.6.** *Let* $\mathcal{G}|_{\mathsf{LC}}$ *be a constrained game. Then,* $\mathscr{C}(\mathcal{G}|_{\mathsf{LC}}) \subseteq \mathscr{B}(\mathcal{G}|_{\mathsf{LC}})$.

*Proof.* Consider an imputation $x \in \mathscr{C}(\mathcal{G}|_{\mathsf{LC}})$ and assume by contradiction that $x \notin \mathscr{B}(\mathcal{G}|_{\mathsf{LC}})$. By this, there must exist an objection $(y, S)$ to $x$. Therefore, it must be the case that $y$ is an $S$-feasible payoff vector in $\mathcal{G}|_{\mathsf{LC}}$ and $y_k > x_k$, for each $k \in S$. This implies that $x \notin \mathscr{C}(\mathcal{G}|_{\mathsf{LC}})$: a contradiction. Thus, $x \in \mathscr{B}(\mathcal{G}|_{\mathsf{LC}})$. $\qquad\square$

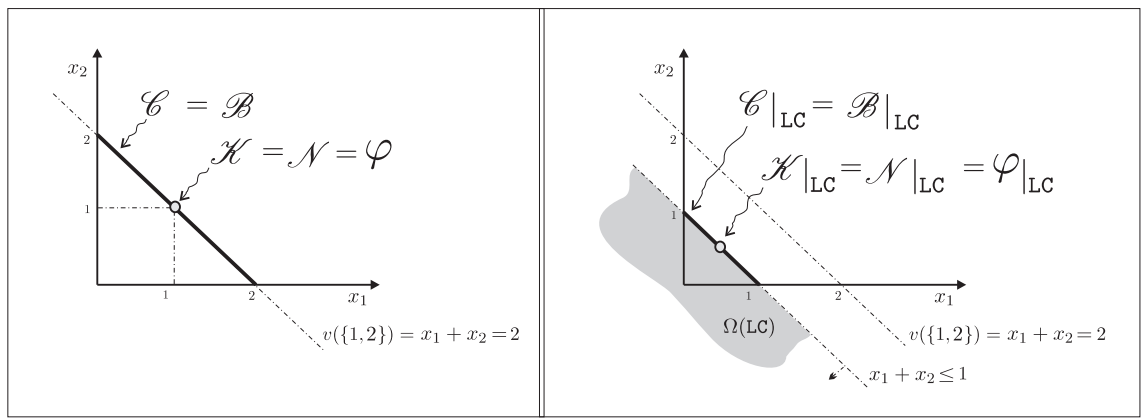

Figure 1: Illustration of the Solution Concepts in Example 4.8.

We finally stress that the counterpart of Proposition 2.7.(5) is already known from the work by Kalai (1975), and can be restated in the settings of constrained games as follows.

**Proposition 4.7** (cf. Kalai, 1975). *Let $\mathcal{G}|_{\mathsf{LC}}$ be a constrained game. Then, $\mathscr{C}(\mathcal{G}|_{\mathsf{LC}}) \neq \varnothing$ implies $\mathscr{N}(\mathcal{G}|_{\mathsf{LC}}) \subseteq \mathscr{C}(\mathcal{G}|_{\mathsf{LC}})$.*

### 4.2 Preservation of Solution Concepts

We continue our investigation by turning to the problem of assessing whether an outcome that is stable (under some solution concept) in a TU game remains stable when constraints are issued. A crucial issue here is to what extent the imputation set is affected by the constraints imposed over the game. This issue is illustrated next.

**Example 4.8.** Consider the TU game $\mathcal{G} = \langle N, v \rangle$ where $N = \{1, 2\}$, $v(\{1, 2\}) = 2$, and $v(\{1\}) = v(\{2\}) = 0$. It is immediate to check that $X(\mathcal{G}) = \{x \in \mathbb{R}^{\{1,2\}} \mid x_1 + x_2 = 2 \wedge x_1 \geq 0 \wedge x_2 \geq 0\}$. The solution concepts for $\mathcal{G}$ are as follows (see Figure 1 for an illustration):

**Core.** For any imputation $x \in X(\mathcal{G})$ and for each coalition $S \subseteq \{1, 2\}$, it is the case that $x(S) \geq v(S)$. Thus, $\mathscr{C}(\mathcal{G}) = X(\mathcal{G})$.

**Bargaining Set.** Since $\mathscr{C}(\mathcal{G}) \subseteq \mathscr{B}(\mathcal{G})$ (recall Proposition 2.7) and since $\mathscr{B}(\mathcal{G}) \subseteq X(\mathcal{G})$, we immediately have that $\mathscr{B}(\mathcal{G}) = \mathscr{C}(\mathcal{G}) = X(\mathcal{G})$.

**Nucleolus.** Let $x \in X(\mathcal{G})$ be an imputation. Considering the standard excess function for TU games, we have that either $\theta(x) = (0, -x_2, -x_1)$ or $\theta(x) = (0, -x_1, -x_2)$, depending on whether $x_1 \geq x_2$ or $x_2 > x_1$. Indeed, just recall that $v(\{1, 2\}) = x(\{1, 2\}) = 2$ and $v(\{1\}) = v(\{2\}) = 0$. Thus, the lexicographically minimum excess vector is obtained at the imputation $\hat{x}$ such that $\hat{x}_1 = \hat{x}_2 = 1$, i.e., $\mathscr{N}(\mathcal{G}) = \{\hat{x}\}$.

**Kernel.** Since $\mathscr{N}(\mathcal{G}) \subseteq \mathscr{K}(\mathcal{G})$ (recall Proposition 2.7), we have that $\hat{x} \in \mathscr{K}(\mathcal{G})$. Consider now an imputation $x \in X(\mathcal{G})$ such that $x \neq \hat{x}$. Assume that $x_1 > 1$ (the same line of reasoning applies to the case where $x_2 > 1$), and thus $x_2 < 1$. For the standard excess function on TU games, $s_{1,2}(x) = -x_1$ and $s_{2,1}(x) = -x_2$ are the surpluses at $x$, because $v(\{1\}) = v(\{2\}) = 0$. Then, $s_{2,1} > s_{1,2}$ holds, and in order to have $x \in \mathscr{K}(\mathcal{G})$, it should be $x_1 = v(\{1\}) = 0$, which is not the case, because $x_1 > 1$. It follows that $\mathscr{K}(\mathcal{G}) = \{\hat{x}\}$.

**Shapley Value.** Note that the two players of $\mathcal{G}$ are symmetric and hence their Shapley value must be the same. Thus, $\varphi(\mathcal{G}) = (1, 1)$.

Now, consider the constraints $\mathtt{LC} = \{x_1 + x_2 \leq 1, x_1, x_2 \in \mathbb{R}\}$. Then, it is easily checked that $X(\mathcal{G}|_{\mathtt{LC}}) = \{x \in \mathbb{R}^{\{1,2\}} \mid x_1 + x_2 = 1 \wedge x_1 \geq 0 \wedge x_2 \geq 0\}$, and hence $X(\mathcal{G}) \cap X(\mathcal{G}|_{\mathtt{LC}}) = \varnothing$. By applying the same line of reasoning as above (and by considering Kalai's excess function in Equation (2) for the nucleolus and the kernel), we derive that $\mathscr{C}(\mathcal{G}|_{\mathtt{LC}}) = \mathscr{B}(\mathcal{G}|_{\mathtt{LC}}) = X(\mathcal{G}|_{\mathtt{LC}})$ and $\mathscr{N}(\mathcal{G}|_{\mathtt{LC}}) = \mathscr{K}(\mathcal{G}|_{\mathtt{LC}}) = \{\hat{y}\}$, where $\hat{y}_1 = \hat{y}_2 = \frac{1}{2}$ (see, again, Figure 1).

Moreover, as for the Shapley NTU values of $\mathcal{G}|_{\mathtt{LC}}$, note that only vectors $\lambda = (\lambda_1, \lambda_2)$ with $\lambda_1 = \lambda_2 = \alpha > 0$ are such that $\mathcal{G}|_{\mathtt{LC}_\lambda}$ is defined for $\mathcal{G}|_{\mathtt{LC}}$. In fact, for every vector $\lambda = (\lambda_1, \lambda_2)$ with $\lambda_1 \neq \lambda_2$ the value $v|_{\mathtt{LC}_\lambda}(N) = \sup \left\{ \sum_{i \in N} \lambda_i z_i \mid z \in V(N) \right\}$ is infinite. Indeed, in pre-imputations players are not necessarily individually rational, and hence in this game one player may get an unbounded negative value, as long as the other one gets an unbounded positive value such that their sum is 2. Now, for every $\lambda = (\alpha, \alpha)$, the worth function of the TU game $\mathcal{G}|_{\mathtt{LC}_\lambda}$ is $v|_{\mathtt{LC}_\lambda}(\{1\}) = v|_{\mathtt{LC}_\lambda}(\{2\}) = 0$, and $v|_{\mathtt{LC}_\lambda}(N) = \alpha$. Since the two players are symmetric, the Shapley values of this family of games are of the form $\varphi(\mathcal{G}|_{\mathtt{LC}_\lambda}) = (\frac{\alpha}{2}, \frac{\alpha}{2})$. Therefore, no consequence $x \in V|_{\mathtt{LC}}(N)$ different from $(\frac{1}{2}, \frac{1}{2})$ admits a vector $\lambda = (\lambda_1, \lambda_2)$, with $\lambda_1 = \lambda_2$, such that $\lambda_i x_i = \varphi(\mathcal{G}|_{\mathtt{LC}_\lambda})_i$ for both players 1 and 2. We conclude that the singleton $\{\hat{y}\}$ is also the set of all Shapley NTU values of $\mathcal{G}|_{\mathtt{LC}}$.

Thus, all the solutions concepts for the constrained game $\mathcal{G}|_{\mathtt{LC}}$ are completely unrelated to those of $\mathcal{G}$. ◁

Note that, in the above example, the fact that no solution concept is preserved is not by chance. Indeed, recall that core, bargaining set, nucleolus, and kernel are defined as refinements of the set of all possible imputations. Therefore, in the extreme scenario where $X(\mathcal{G}) \cap X(\mathcal{G}|_{\mathtt{LC}}) = \varnothing$ holds (for a constrained game $\mathcal{G}|_{\mathtt{LC}}$ built on top of the TU game $\mathcal{G}$), none of these solution concepts can be preserved.

**Fact 4.9.** *Let $\mathcal{G}|_{\mathtt{LC}}$ be a constrained game such that $X(\mathcal{G}) \cap X(\mathcal{G}|_{\mathtt{LC}}) = \varnothing$. Then,*

*(1) $\mathscr{C}(\mathcal{G}) \cap \mathscr{C}(\mathcal{G}|_{\mathtt{LC}}) = \varnothing$;*

*(2) $\mathscr{B}(\mathcal{G}) \cap \mathscr{B}(\mathcal{G}|_{\mathtt{LC}}) = \varnothing$;*

*(3) $\mathscr{N}(\mathcal{G}) \cap \mathscr{N}(\mathcal{G}|_{\mathtt{LC}}) = \varnothing$; and,*

*(4) $\mathscr{K}(\mathcal{G}) \cap \mathscr{K}(\mathcal{G}|_{\mathtt{LC}}) = \varnothing$.*

Moreover, recall that Shapely NTU values are refinements of the set of all possible payoff distributions associated with the grand-coalition (and, in particular, of the pre-imputations for TU games). Thus, if $\varphi(\mathcal{G}) \notin V_{\mathtt{LC}}(N)$ (as in Example 4.8), this solution concept cannot be preserved. In general, since $\varphi(\mathcal{G})$ is a pre-imputation, the following holds.

**Fact 4.10.** *Let $\mathcal{G}|_{\mathtt{LC}}$ be a constrained game, with $\mathcal{G} = \langle N, v \rangle$. Assume that $V_{\mathtt{LC}}(N) \cap \{x \in V_v(N) \mid x \text{ is efficient}\} = \varnothing$. Then, $\varphi(\mathcal{G}) \notin \varphi(\mathcal{G}|_{\mathtt{LC}})$.*

In the light of the above observations, it is however of interest to analyze whether "preservation" properties hold with respect to (pre)imputations of both $\mathcal{G}$ and $\mathcal{G}|_{\mathtt{LC}}$.

As an example, it is of interest to establish the relationship between payoff vectors in $\mathscr{C}(\mathcal{G}) \cap X(\mathcal{G}|_{\mathtt{LC}})$ (i.e., vectors that are in the core of the TU game—and thus are imputations for this game—and that are also imputations for the constrained game) and payoff vectors in $\mathscr{C}(\mathcal{G}|_{\mathtt{LC}}) \cap X(\mathcal{G})$ (i.e., vectors that are in the core of the constrained game—and thus are imputations for this game—and that are also imputations for the TU game). Exploring these relationships, for each solution concept, is addressed in the rest of this section.

### 4.2.1 Core

Our first result concerns the core, and shows that imputations that are in the core of the TU game and satisfy the constraints are also in the core of the resulting constrained game.

**Proposition 4.11.** *Let $\mathcal{G}|_{\mathtt{LC}}$ be a constrained game. Then, $\mathscr{C}(\mathcal{G}) \cap X(\mathcal{G}|_{\mathtt{LC}}) \subseteq \mathscr{C}(\mathcal{G}|_{\mathtt{LC}}) \cap X(\mathcal{G})$.*

*Proof.* Let $\mathcal{G} = \langle N, v \rangle$ be a TU game, and recall from Section 2 that $\mathcal{G}$ can be equivalently viewed as the NTU game $\langle N, V_v \rangle$. Assume that $x$ is a payoff vector in $\mathscr{C}(\mathcal{G})$, and hence in $X(\mathcal{G})$. Then, there is no coalition $S \subseteq N$ and no vector $y \in V_v(S)$ such that $y_i > x_i, \forall i \in S$. By Definition 3.1, for the NTU game $\mathcal{G}|_{\mathtt{LC}} = \langle N, V_{\mathtt{LC}} \rangle$, it is the case that $V_{\mathtt{LC}}(S) \subseteq V_v(S)$, for each $S \subseteq N$. Therefore, there is no coalition $S \subseteq N$ and no vector $y \in V_{\mathtt{LC}}(S)$ such that $y_i > x_i, \forall i \in S$. That is, if $x \in X(\mathcal{G}|_{\mathtt{LC}})$, then $x$ belongs to $\mathscr{C}(\mathcal{G}|_{\mathtt{LC}})$. □

However, the above inclusion can be strict in some cases, even if no imputation is affected by the constraints, i.e., even if $X(\mathcal{G}) = X(\mathcal{G}|_{\mathtt{LC}})$.

**Proposition 4.12.** *There exists a constrained game $\bar{\mathcal{G}}|_{\mathtt{LC}}$ (with $int(\mathtt{LC}) = \varnothing$) such that $X(\bar{\mathcal{G}}) = X(\bar{\mathcal{G}}|_{\mathtt{LC}})$, $\mathscr{C}(\bar{\mathcal{G}}) = \varnothing$ and $\mathscr{C}(\bar{\mathcal{G}}|_{\mathtt{LC}}) \neq \varnothing$. Thus, $\mathscr{C}(\bar{\mathcal{G}}) \cap X(\bar{\mathcal{G}}|_{\mathtt{LC}}) \subset \mathscr{C}(\bar{\mathcal{G}}|_{\mathtt{LC}}) \cap X(\bar{\mathcal{G}})$.*

*Proof.* Consider the TU game $\bar{\mathcal{G}} = \langle N, v \rangle$ such that $N = \{1, 2, 3\}$, $v(\{1\}) = 1$, $v(\{2\}) = 1$, $v(\{3\}) = 2$, $v(\{1, 2\}) = 3$, $v(\{1, 3\}) = 0$, $v(\{2, 3\}) = 0$, and $v(\{1, 2, 3\}) = 4$. Notice that $X(\bar{\mathcal{G}}) = \{\hat{x}\}$ with $\hat{x}_1 = 1$, $\hat{x}_2 = 1$, and $\hat{x}_3 = 2$; and that $\mathscr{C}(\bar{\mathcal{G}}) = \varnothing$. In particular, for the latter equality, consider the pair $(y, \{1, 2\})$ such that $y_1 = y_2 = \frac{3}{2}$. Since $y(\{1, 2\}) = v(\{1, 2\})$, $y_1 > \hat{x}_1$ and $y_2 > \hat{x}_2$, we have that $(y, \{1, 2\})$ is an objection to $\hat{x}$, which therefore does not belong to $\mathscr{C}(\bar{\mathcal{G}})$.

Consider now the following set of constraints:

$$\mathtt{LC} = \left\{ \begin{array}{l} x_1 + x_2 \leq 2 \\ x_1, x_2 \in \mathbb{R} \end{array} \right.$$

It is easily seen that $\hat{x}$ satisfies $\mathtt{LC}$. Thus, $x \in X(\bar{\mathcal{G}}|_{\mathtt{LC}})$ holds by Proposition 4.1. Moreover, since $\Omega(\mathtt{LC})[\{i\}] = \mathbb{R}^{\{i\}}$ holds for each player $i \in N$, because there is no constraint on worths of singleton coalitions, the individual rationality constraint on $\bar{\mathcal{G}}|_{\mathtt{LC}}$ prescribes that for each $x \in X(\bar{\mathcal{G}}|_{\mathtt{LC}})$: $x_1 \geq v(\{1\}) = 1$, $x_2 \geq v(\{2\}) = 1$, and $x_3 \geq v(\{3\}) = 2$. Since $v(\{1, 2, 3\}) = 4$, $\hat{x}$ is in fact the only imputation in $X(\bar{\mathcal{G}}|_{\mathtt{LC}})$. Thus, $X(\bar{\mathcal{G}}) = X(\bar{\mathcal{G}}|_{\mathtt{LC}})$.

To conclude the proof, let us now observe that, in the constrained game $\bar{\mathcal{G}}|_{\mathtt{LC}}$, there is no $\{1, 2\}$-feasible vector $z$ with $z_1 > \hat{x}_1$ and $z_2 > \hat{x}_2$; indeed, just observe that $z_1 + z_2 \leq 2$ holds because of the constraints, while $\hat{x}_1 + \hat{x}_2 = 2$. That is, there is no objection to $\hat{x}$, which is therefore an imputation in $\mathscr{C}(\bar{\mathcal{G}}|_{\mathtt{LC}})$. □

### 4.2.2 BARGAINING SET

As far as the bargaining set is concerned, we can show that there are constrained games whose bargaining set is completely unrelated with that of the underlying TU games. This is because objections and counterobjections are not necessarily restricted to the set of the possible imputations. Thus, constraints may radically alter the feasibility properties of certain payoff vectors, yet without affecting the imputation set. This is shown in the following two propositions.

**Proposition 4.13.** *There exists a constrained game $\bar{\mathcal{G}}|_{\mathsf{LC}}$ (with $int(\mathsf{LC}) = \varnothing$) such that $X(\bar{\mathcal{G}}) = X(\bar{\mathcal{G}}|_{\mathsf{LC}})$, and $\mathscr{B}(\bar{\mathcal{G}}) \cap X(\bar{\mathcal{G}}|_{\mathsf{LC}}) \nsubseteq \mathscr{B}(\bar{\mathcal{G}}|_{\mathsf{LC}}) \cap X(\bar{\mathcal{G}})$.*

*Proof.* Consider the TU game $\bar{\mathcal{G}} = \langle N, v \rangle$ such that $N = \{1, 2, 3, 4, 5\}$, $v(\{1, 2, 3, 4, 5\}) = 8$, $v(\{2, 3, 4\}) = 8$, $v(\{1, 3, 4\}) = 7$, $v(\{1, 2\}) = 2$, $v(\{5\}) = 1$, and $v(S) = 0$ for each other coalition $S \subset N$. Consider the imputation $\hat{x}$ such that $\hat{x}_1 = 0$, $\hat{x}_2 = 1$, $\hat{x}_3 = 3$, $\hat{x}_4 = 3$, and $\hat{x}_5 = 1$. We claim that $\hat{x} \in \mathscr{B}(\bar{\mathcal{G}})$. Indeed, let $(y, S)$ be an objection to $\hat{x}$. This objection can be carried out through three different coalitions, each of them having a counterobjection:

- First, we can have $S = \{2, 3, 4\}$, $y_2 > \hat{x}_2 = 1$, $y_3 > \hat{x}_3 = 3$, $y_4 > \hat{x}_4 = 3$, and $y_2 + y_3 + y_4 \leq v(\{2, 3, 4\}) = 8$. In this case $(y, S)$ is an objection against player 1 or against player 5. In the former case, $(z, \{1\})$ with $z_1 = \hat{x}_1 = 0$ is a trivial counterobjection to $(y, S)$; in the latter case, $(z, \{5\})$ with $z_5 = \hat{x}_5 = 1$ is a counterobjection to $(y, S)$.

- Second, we can have $S = \{1, 3, 4\}$, $y_1 > \hat{x}_1 = 0$, $y_3 > \hat{x}_3 = 3$, $y_4 > \hat{x}_4 = 3$, and $y_1 + y_3 + y_4 \leq v(\{1, 3, 4\}) = 7$. In this case, $(y, S)$ is an objection of some player in $S$ against player 2 or against player 5. As we observed above, $(z, \{5\})$ with $z_5 = \hat{x}_5 = 1$ is a trivial counterobjection to any objection against 5. Thus, let us assume that $(y, S)$ is an objection against player 2. If $(y, S)$ is an objection of player 3 or 4, we may just consider the pair $(z, \{1, 2\})$ with $z_1 = y_1$ and $z_2 = \hat{x}_2$. Indeed, note that $y_1 < 1$ holds and, thus, $z_1 + z_2 < 1 + \hat{x}_2 = 2 = v(\{1, 2\})$. Therefore, $z$ is $\{1, 2\}$-feasible, and $(z, \{1, 2\})$ is a counterobjection to $(y, S)$. On the other hand, if $(y, S)$ is an objection of player 1 against player 2 to $\hat{x}$, we may consider the pair $(w, \{2, 3, 4\})$ such that $w_2 = \hat{x}_2$, $w_3 = y_3$, and $w_4 = y_4$. Note that $y_3 + y_4 < 7$ and, thus, $w_2 + w_3 + w_4 < \hat{x}_2 + 7 = 1 + 7 = v(\{2, 3, 4\})$. Then, $w$ is $\{2, 3, 4\}$-feasible, and $(w, \{2, 3, 4\})$ is a counterobjection to $(y, S)$.

- Finally, we can have $S = \{1, 2\}$, $y_1 > \hat{x}_1 = 0$, $y_2 > \hat{x}_2 = 1$, and $y_1 + y_2 \leq v(\{1, 2\}) = 2$. In this case, $(y, S)$ is an objection of some player in $S$ against 3, 4, or 5. Let us consider the first two cases, since $(z, \{5\})$ with $z_5 = \hat{x}_5 = 1$ is a trivial counterobjection to objections against 5. If $(y, S)$ is an objection of player 1 (against player 3 or 4), we may consider the pair $(z, \{2, 3, 4\})$ such that $z_2 = y_2$, $z_3 = \hat{x}_3$, and $z_4 = \hat{x}_4$. Note that $y_2 < 2$ and, thus, $z_2 + z_3 + z_4 < 2 + \hat{x}_3 + \hat{x}_4 = 2 + 6 = v(\{2, 3, 4\}) = 8$. Hence, $(z, \{2, 3, 4\})$ is a counterobjection to $(y, S)$. If $(y, S)$ is an objection of player 2 (against player 3 or 4), we may consider the pair $(w, \{1, 3, 4\})$ such that $w_1 = y_1$, $w_3 = \hat{x}_3$, and $w_4 = \hat{x}_4$. Note that $y_1 < 1$ and, thus, $w_1 + w_3 + w_4 < 1 + \hat{x}_3 + \hat{x}_4 = 1 + 6 = v(\{1, 3, 4\}) = 7$. Hence, $(w, \{1, 3, 4\})$ is a counterobjection to $(y, S)$.

Consider now the following set of constraints:

$$\mathtt{LC} = \left\{ \begin{array}{l} x_2 + x_3 + x_4 \leq 7 \\ x_2, x_3, x_4 \in \mathbb{R} \end{array} \right.$$

It is immediate to check that $X(\bar{\mathcal{G}}) = X(\bar{\mathcal{G}}|_{\mathtt{LC}})$; indeed, the individual rationality over player 5 forces $\hat{x}_5 \geq 1$; given that $v(\{1,2,3,4,5\}) = 8$, the above constraint is therefore logically implied for all individually rational vectors in $V_{\mathtt{LC}}(N)$. However, $\mathtt{LC}$ plays a crucial role concerning the formation of the coalition $\{2,3,4\}$. Indeed, consider the objection $(y, \{1,2\})$ of player 1 against player 3 to $\hat{x}$, where $y_1 = \frac{1}{2}$ and $y_2 = \frac{3}{2}$. Any counterobjection $(z, T)$ to $(y, \{1,2\})$ must be such that $T = \{2,3,4\}$. Thus, $z_2 \geq y_2 = \frac{3}{2}$, $z_3 \geq \hat{x}_3 = 3$, and $z_4 \geq \hat{x}_4 = 3$ must hold. It follows that $z_2 + z_3 + z_4 > 7$ and, hence, $z \notin V_{\mathtt{LC}}(T)$. Since $(y, S)$ is a justified objection, $\hat{x} \notin \mathscr{B}(\bar{\mathcal{G}}|_{\mathtt{LC}})$. $\qquad \square$

**Proposition 4.14.** *There exists a constrained game $\bar{\mathcal{G}}|_{\mathtt{LC}}$ (with $\mathrm{int}(\mathtt{LC}) = \varnothing$) such that $X(\bar{\mathcal{G}}) = X(\bar{\mathcal{G}}|_{\mathtt{LC}})$, and $\mathscr{B}(\bar{\mathcal{G}}|_{\mathtt{LC}}) \cap X(\bar{\mathcal{G}}) \not\subseteq \mathscr{B}(\bar{\mathcal{G}}) \cap X(\bar{\mathcal{G}}|_{\mathtt{LC}})$.*

*Proof.* Consider the TU game $\bar{\mathcal{G}} = \langle N, v \rangle$ such that $N = \{1,2,3\}$, $v(\{1\}) = v(\{2\}) = 1$, $v(\{3\}) = 0$, $v(\{1,3\}) = v(\{2,3\}) = 4$, $v(\{1,2\}) = 5$, and $v(\{1,2,3\}) = 3$. Consider the imputation $\hat{x}$ such that $\hat{x}_1 = \hat{x}_2 = \hat{x}_3 = 1$. We observe that $\hat{x} \notin \mathscr{B}(\bar{\mathcal{G}})$. Indeed, consider an objection $(y, \{1,2\})$ of player 1 against player 3 such that $y_1 = 1 + \frac{1}{2}$ and $y_2 = 3 + \frac{1}{2}$ (observe that $y(\{1,2\}) = v(\{1,2\})$). Player 3 cannot counterobject either as a singleton, since $v(\{3\}) < \hat{x}_3$, or through coalition $\{2,3\}$, since for each vector $z \in \mathbb{R}^{\{2,3\}}$ such that $z_2 \geq y_2 > 3$ and $z_3 \geq \hat{x}_1 = 1$, we have $z(\{2,3\}) > 4 = v(\{2,3\})$. It follows that $\hat{x} \notin \mathscr{B}(\bar{\mathcal{G}})$.

Consider now the following set of constraints:

$$\mathtt{LC} = \left\{ \begin{array}{l} x_1 + x_2 \leq 4 \\ x_1, x_2 \in \mathbb{R} \end{array} \right.$$

It is immediate to check that $X(\bar{\mathcal{G}}) = X(\bar{\mathcal{G}}|_{\mathtt{LC}})$. Moreover, let us notice that no player has a justified objection against player 1 or 2 to $\hat{x}$, since they can counterobject as singletons; indeed, just observe that $v(\{i\}) = 1 = \hat{x}_i$, for $i \in \{1,2\}$. Consider, then, an objection $(y, \{1,2\})$ of player 1 against player 3, such that $y_1 \geq \hat{x}_1 = 1$ and $y_2 \geq \hat{x}_2 = 1$. Since, $y$ must belong to $V_{\mathtt{LC}}(\{1,2\})$, we have that $y_2 \leq 3$ holds. Thus, the pair $(z, \{2,3\})$ with $z_2 = 3 \geq y_2$ and $z_3 = \hat{x}_3 = 1$ is a counterobjection to $(y, \{1,2\})$, because $z(\{2,3\}) = 4 = v(\{2,3\})$. By the symmetry in the game definition, the same line of reasoning as above applies to show that also player 2 has no justified objections against player 3. Therefore, $\hat{x} \in \mathscr{B}(\bar{\mathcal{G}}|_{\mathtt{LC}})$. $\qquad \square$

### 4.2.3 Nucleolus and Kernel

Let us move to analyze the nucleolus and the kernel. As in the case of the bargaining set, no preservation property holds, as demonstrated next.

**Proposition 4.15.** *There exists a constrained game $\bar{\mathcal{G}}|_{\mathtt{LC}}$ (with $\mathrm{int}(\mathtt{LC}) = \varnothing$) such that $X(\bar{\mathcal{G}}) = X(\bar{\mathcal{G}}|_{\mathtt{LC}}) \neq \varnothing$, $\mathscr{K}(\bar{\mathcal{G}}) \cap \mathscr{K}(\bar{\mathcal{G}}|_{\mathtt{LC}}) = \varnothing$, and $\mathscr{N}(\bar{\mathcal{G}}) \cap \mathscr{N}(\bar{\mathcal{G}}|_{\mathtt{LC}}) = \varnothing$ (for Kalai's excess function in Equation* (2) *on page 640).*

*Proof.* Consider the TU game $\bar{\mathcal{G}} = \langle N, v \rangle$ such that $N = \{1, 2, 3\}$, $v(\{1, 2, 3\}) = 3$, $v(\{1, 2\}) = 5$, $v(\{1, 3\}) = 4$, $v(\{2, 3\}) = 3$, and $v(S) = 0$, for each other coalition $S \subset N$.

Consider an imputation $x$ that belongs to $\mathscr{K}(\bar{\mathcal{G}})$, and consider the expressions: $s_{1,3}(x) - s_{3,1}(x) = (5 - x_1 - x_2) - (3 - x_2 - x_3) = 2 - x_1 + x_3$ and $s_{1,2}(x) - s_{2,1}(x) = (4 - x_1 - x_3) - (3 - x_2 - x_3) = 1 - x_1 + x_2$. By Definition 2.5, we then get that $2 - x_1 + x_3 > 0$ implies $x_3 = 0$, that $2 - x_1 + x_3 < 0$ implies $x_1 = 0$, that $1 - x_1 + x_2 > 0$ implies $x_2 = 0$, and that $1 - x_1 + x_2 < 0$ implies $x_1 = 0$. By simple algebraic calculations, the above relationships together with the individual rationality of $x$ (i.e., $x_1 \geq 0$, $x_2 \geq 0$, and $x_3 \geq 0$) entail that $x_1 - x_2 = 1$ and $x_1 - x_3 = 2$ must hold. In turn, since $x_1 + x_2 + x_3 = 3$, the latter two equations uniquely determine the value of $x$. In particular, $\mathscr{K}(\bar{\mathcal{G}})$ is the singleton $\{x\}$ such that $x_1 = 2$, $x_2 = 1$ and $x_3 = 0$. Moreover, since $\mathscr{N}(\bar{\mathcal{G}}) \subseteq \mathscr{K}(\bar{\mathcal{G}})$ and $|\mathscr{N}(\bar{\mathcal{G}})| = 1$ (see Proposition 2.7), we have that $\mathscr{N}(\bar{\mathcal{G}}) = \mathscr{K}(\bar{\mathcal{G}})$.

Consider now the following set of constraints:

$$\mathtt{LC} = \begin{cases} x_1 + x_2 \leq 3 \\ x_1 + x_3 \leq 3 \\ x_2 + x_3 \leq 3 \\ x_1, x_2, x_3 \in \mathbb{R} \end{cases}$$

Notice that the above constraints do not modify the imputation set, that is, $X(\bar{\mathcal{G}}) = X(\bar{\mathcal{G}}|_{\mathtt{LC}}) \neq \varnothing$. Moreover, observe that $V|_{\mathtt{LC}} = V_{v'}$, where $v'$ is the worth function of the game $\bar{\mathcal{G}}' = \langle N, v' \rangle$ in which $v'(\{1, 2, 3\}) = v'(\{1, 2\}) = v'(\{1, 3\}) = v'(\{2, 3\}) = 3$, and $v'(S) = 0$ for all other coalitions $S \subset N$. By this, the excess function $e_K$ reported in Equation (2) coincides with the canonical TU excess, and the definitions of kernel NTU and nucleolus NTU coincide with those for TU games (cf. Kalai, 1975). So the kernel and nucleolus of $\bar{\mathcal{G}}|_{\mathtt{LC}}$ are those of $\bar{\mathcal{G}}'$. Finally, it is easily checked that $\mathscr{K}(\bar{\mathcal{G}}')$ is the singleton $\{x'\}$ such that $x_1' = 1$, $x_2' = 1$ and $x_3' = 1$ (and, thus, $\mathscr{K}(\bar{\mathcal{G}}') = \mathscr{N}(\bar{\mathcal{G}}')$). It follows that $\mathscr{K}(\bar{\mathcal{G}}) \cap \mathscr{K}(\bar{\mathcal{G}}|_{\mathtt{LC}}) = \varnothing$ and $\mathscr{N}(\bar{\mathcal{G}}) \cap \mathscr{N}(\bar{\mathcal{G}}|_{\mathtt{LC}}) = \varnothing$. $\qquad \blacksquare$

### 4.2.4 Shapley Value

Let us conclude our analysis with the Shapley value. Below, we show that this solution concept is preserved whenever the set of all pre-imputations is not modified by the constraints.

**Proposition 4.16.** *Let $\mathcal{G}|_{\mathtt{LC}}$ be a constrained game. If the sets of pre-imputations of $\mathcal{G}$ and $\mathcal{G}|_{\mathtt{LC}}$ coincide, then $\varphi(\mathcal{G}|_{\mathtt{LC}}) = \{\varphi(\mathcal{G})\}$.*

*Proof.* Let $pX(\mathcal{G}) = \{x \in \mathbb{R}^N \mid x(N) = v(N)\}$ be the set of all pre-imputations of the TU $\mathcal{G} = \langle N, v \rangle$; in fact, recall that $pX(\mathcal{G})$ contains all the efficient payoff vectors in $V_v(N) = \{x \in \mathbb{R}^N \mid x(N) \leq v(N)\}$. Let $pX(\mathcal{G}|_{\mathtt{LC}}) = \{x \in \mathbb{R}^n \mid x(N) \leq v(N) \wedge x \in \Omega(\mathtt{LC})[N] \wedge x \text{ is efficient}\}$ be the set of all pre-imputations of $\mathcal{G}|_{\mathtt{LC}}$. Since $pX(\mathcal{G}) = pX(\mathcal{G}|_{\mathtt{LC}})$, it must be the case that $\Omega(\mathtt{LC})[N] \supseteq V_v(N) = \{x \in \mathbb{R}^n \mid x(N) \leq v(N)\}$. Therefore, $V_{\mathtt{LC}}(N) = V_v(N) \cap \Omega(\mathtt{LC})[N] = V_v(N)$. For each coalition $S$ and vector $y \in \mathbb{R}^S$, consider now the vector $x \in \mathbb{R}^N$ with $x_i = y_i$, for each $i \in S$, and $x_i = (v(N) - y(S))/|N \setminus S|$, for each $i \in (N \setminus S)$. Note that $x(N) = v(N)$ and, hence, $x \in V_v(N)$ and $x \in \Omega(\mathtt{LC})[N]$. Therefore, $y$ belongs to $\Omega(\mathtt{LC})[S]$. Thus, $\Omega(\mathtt{LC})[S] = \mathbb{R}^S$ and, hence, $V_{\mathtt{LC}}(S) = V_v(S) \cap \Omega(\mathtt{LC})[S] = V_v(S)$, for each $S \subset N$. Finally, since $V_{\mathtt{LC}}(N) = V_v(N)$ also holds (and thus $V_{\mathtt{LC}}(S) = V_v(S)$, for each $S \subseteq N$), it follows that $\varphi(\mathcal{G}|_{\mathtt{LC}}) = \{\varphi(\mathcal{G})\}$ (see, e.g., McLean, 2002). $\qquad \blacksquare$

However, the Shapley value $\varphi(\mathcal{G})$ is not preserved in general, even if $\varphi(\mathcal{G})$ is an imputation, and if imputation sets are not affected by the constraints.

**Proposition 4.17.** *There exists a constrained game $\bar{\mathcal{G}}|_{\mathsf{LC}}$ (with $int(\mathsf{LC}) = \varnothing$) such that $\varphi(\bar{\mathcal{G}}) \in X(\bar{\mathcal{G}})$, $X(\bar{\mathcal{G}}) = X(\bar{\mathcal{G}}|_{\mathsf{LC}})$, and $\varphi(\bar{\mathcal{G}}) \notin \varphi(\bar{\mathcal{G}}|_{\mathsf{LC}})$.*

*Proof.* Consider the TU game $\bar{\mathcal{G}} = \langle N, v \rangle$ such that $N = \{1, 2, 3\}$, $v(\{1, 2, 3\}) = 3$, $v(\{1, 2\}) = 4$, $v(\{1, 3\}) = 3$, $v(\{2, 3\}) = 3$, and $v(S) = 0$, for each other coalition $S \subset N$. By simple calculations, one may compute the Shapley value $\varphi(\bar{\mathcal{G}})$, and notice that $\varphi(\bar{\mathcal{G}})_1 = 7/6$, $\varphi(\bar{\mathcal{G}})_2 = 7/6$, and $\varphi(\bar{\mathcal{G}})_3 = 4/6$. Thus, $\varphi(\bar{\mathcal{G}}) \in X(\bar{\mathcal{G}})$.

Consider now the following set of constraints:

$$\mathsf{LC} = \left\{ \begin{array}{l} x_1 + x_2 \leq 3 \\ x_1, x_2 \in \mathbb{R} \end{array} \right.$$

and notice that they do not modify the imputation set, that is, $X(\bar{\mathcal{G}}) = X(\bar{\mathcal{G}}|_{\mathsf{LC}}) \neq \varnothing$. Indeed, the inequality $x_1 + x_2 \leq 3$ is logically implied by the worth of the grand-coalition (which forces $x_1 + x_2 + x_3 \leq 3$) and by the individual rationality of all players (i.e., $x_1 \geq 0$, $x_2 \geq 0$, and $x_3 \geq 0$). For the sake of completeness, note that the constrained game $\bar{\mathcal{G}}$ does not fit the hypothesis of Proposition 4.16 since $pX(\bar{\mathcal{G}}) \neq pX(\bar{\mathcal{G}}|_{\mathsf{LC}})$ (indeed, any payoff vector $x$ with $x_1 + x_2 > 3$ and $x(N) \leq v(N) = 3$ is a pre-imputation for $\bar{\mathcal{G}}$ but not for $\bar{\mathcal{G}}|_{\mathsf{LC}}$, because $x_1 + x_2 \leq 3$ is not satisfied).

Moreover, observe that $V|_{\mathsf{LC}} = V_{v'}$, where $v'$ is the worth function of the game $\bar{\mathcal{G}}' = \langle N, v' \rangle$ in which $v'(\{1, 2, 3\}) = v'(\{1, 2\}) = v'(\{1, 3\}) = v'(\{2, 3\}) = 3$, and $v'(S) = 0$ for all other coalitions $S \subset N$. This suffices to conclude that $\varphi(\bar{\mathcal{G}}|_{\mathsf{LC}}) = \{\varphi(\bar{\mathcal{G}}')\}$ (see, e.g., McLean, 2002). However, it is easily checked that $\varphi(\bar{\mathcal{G}}')$ is such that $\varphi(\bar{\mathcal{G}}')_1 = \varphi(\bar{\mathcal{G}}')_2 = \varphi(\bar{\mathcal{G}}')_3 = 1$. Thus, $\varphi(\bar{\mathcal{G}}) \notin \varphi(\bar{\mathcal{G}}|_{\mathsf{LC}})$. □

## 5. Complexity Analysis

In this section, we shall look at the core and the bargaining set for (constrained) coalitional games from a computational viewpoint. In particular, our aim is to shed light on the impact of issuing constraints w.r.t. the intrinsic complexity of these notions, and to assess whether any price has to be paid for the increased expressiveness of constrained games—for the sake of completeness, background notions on complexity theory are reported in the Appendix.

We argue that it is in fact sensible to analyze these computational properties, as this corresponds to analyzing the feasibility of using such concepts under the thesis of *bounded rationality*, that is, that decisions taken by realistic agents cannot involve unbounded resources to support reasoning (Simon, 1972). Moreover, it is worthwhile noting that studying such matters might hopefully guide the design of effective computation algorithms.

We leave as future work a complexity analysis of the other solution concepts, where it would be interesting to consider various kinds of Kalai's excess functions with different computational properties.

## 5.1 Setup and Problems Analyzed

In the analysis that follows, we assume games to be provided in *characteristic function form*, i.e., we deal with scenarios where coalition worths are returned by some given function (von

Neumann & Morgenstern, 1944). For instance, the games discussed in Example 3.4 and in Example 3.5 are in characteristic function form. Moreover, by following the general framework proposed by Bilbao (2000), we assume that the input for any reasoning problem consists of a constrained game $\mathcal{G}|_{\mathtt{LC}}$ where the worth function $v$ is given as an oracle. In particular, we shall consider two types of oracles:

(1) *Oracles computable in polynomial time in the size* $||\mathcal{G}|_{\mathtt{LC}}||$ *of the game representation.* [5] For instance, the game in Example 3.4 fits this framework, as well as the game in Example 3.5, provided that the cost function $com(S)$ (of establishing a communication infrastructure over the agents in $S$) comes as an oracle computable in polynomial time.

(2) *Oracles computable in non-deterministic polynomial time in the size* $||\mathcal{G}|_{\mathtt{LC}}||$ *of the game representation.* For instance, the game in Example 3.5 may fit this setting in the cautious perspective where we require that, for any coalition $S$, the value $v(S) = 100 - com(S)$ can actually be obtained in some imputation. That is, if we add the condition that there is an element $\hat{x} \in \Omega(\mathtt{LC})$ such that for each task $t_j$, there is a player $i \in S$ with $\gamma_j^i \neq 0$ (in $\hat{x}$), i.e., that the task $T$ can actually be performed by coalition $S$ while conforming with all the costs constraints. Here, we also require of course that $com(S)$ is computable in non-deterministic polynomial time. Note that such powerful worth-functions can be used to encode **NP**-complete problems reflecting results of complex algorithmic procedures, such as those arising in allocation, scheduling and routing scenarios, to name a few.

Let us remark that the framework where the worth function is an oracle computable in polynomial time encompasses all those settings where games are (implicitly) described over some kind of compact "structures", and where simple calculations on such encodings are to be performed to compute the worth of any given coalition—noticeable and very influential settings of this type are the *graph* and *hypergraph* games (Deng & Papadimitriou, 1994), the *marginal contribution nets* (Ieong & Shoham, 2005), the games in *multi-issue domains* (Conitzer & Sandholm, 2004), and the *weighted voting games* (Elkind & Pasechnik, 2009; Elkind, Goldberg, Goldberg, & Wooldridge, 2009). Therefore, our membership results will immediately carry over to the various classes of games cited above, whereas hardness results are specific to the oracle setting, and do not hold in general for these (sub)settings.

Within the setting discussed above, we shall next focus on checking whether a given imputation satisfies the conditions needed to be in the core or in the bargaining set. Thus, given a constrained game $\mathcal{G}|_{\mathtt{LC}}$ and a vector $\hat{x}$, the following problems will be considered:

- CORE-CHECK: Is $\hat{x}$ in $\mathscr{C}(\mathcal{G}|_{\mathtt{LC}})$?

- BARGAININGSET-CHECK: Is $\hat{x}$ in $\mathscr{B}(\mathcal{G}|_{\mathtt{LC}})$?

In addition, recall from Section 4 that the core and the bargaining set might be empty for constrained games. Thus, it is sensible as well to study the following problems:

---

5. As usual, it is implicitly assumed that the game representation includes the list of players, so that, for every coalition $S$, $||S|| \leq ||\mathcal{G}|_{\mathtt{LC}}||$. Otherwise, one should more formally say, e.g., that oracles are computable in polynomial time in the combined size of $\mathcal{G}|_{\mathtt{LC}}$ and $S$.

| Problem | Constrained | Constrained $(int(\mathtt{LC}) \subseteq \{x_i \mid i \in N\})$ | | Constrained $(int(\mathtt{LC}) = \varnothing)$ | TU |
|---------|-------------|-------------|--|-------------|----|
| Core-Check | $\mathbf{D}^P$-complete | co-$\mathbf{NP}$-complete | | co-$\mathbf{NP}$-complete | co-$\mathbf{NP}$-complete |
| BargainingSet-Check | $\mathbf{\Pi}_2^P$-complete | $\mathbf{\Pi}_2^P$-complete | | $\mathbf{\Pi}_2^P$-complete | $\mathbf{\Pi}_2^P$-complete |
| Core-NonEmptiness | $\mathbf{\Sigma}_2^P$-complete | $\mathbf{\Sigma}_2^P$-complete | | $\mathbf{\Sigma}_2^P$-complete | co-$\mathbf{NP}$-complete |
| BargainingSet-NonEmptiness | $\mathbf{\Sigma}_3^P$-complete | $\mathbf{\Sigma}_3^P$-complete | | $\mathbf{\Sigma}_3^P$-complete | trivial |

Figure 2: Complexity Results for Constrained Games. Hardness results hold even on cohesive games with worth functions given as polynomial-time oracles. Membership results hold on non-deterministic polynomial-time worth-function oracles, without any assumption on the representation of real numbers.

- Core-NonEmptiness: Is $\mathscr{C}(\mathcal{G}|_{\mathtt{LC}}) \neq \varnothing$?

- BargainingSet-NonEmptiness: Is $\mathscr{B}(\mathcal{G}|_{\mathtt{LC}}) \neq \varnothing$?

**Overview of the Results.** A summary of our results is reported in Figure 2. Note there that four settings emerge from our analysis: TU games, constrained games without integer variables (i.e., $int(\mathtt{LC}) = \varnothing$), constrained games without auxiliary integer variables (i.e., $int(\mathtt{LC}) \subseteq \{x_i \mid i \in N\}$), and arbitrary constrained games. In fact, we stress that hardness results are established without the use of auxiliary real variables, while membership results (for constrained games) hold even if variables of this kind actually occur. Thus, auxiliary real variables play no computational role in the setting of constrained games.

Concerning the checking problems, Figure 2 evidences that Core-Check is co-$\mathbf{NP}$-hard for TU games, and in co-$\mathbf{NP}$ for constrained games where auxiliary integer variables are not allowed—as said above, there is no bound in the membership result on the number of considered auxiliary real variables. By allowing the use of auxiliary integer variables, Core-Check becomes $\mathbf{D}^P$-hard, and in fact complete for this class. Thus, auxiliary integer variables cause a slight increase of complexity for this solution concept. On the other hand, it emerged that the occurrence of real variables—either player variables or auxiliary ones—and of integer player variables is completely immaterial from a computational perspective.

As far as the BargainingSet-Check is concerned, we can deliver the good news that adding constraints does not alter the complexity w.r.t. the TU case. Indeed, this problem is $\mathbf{\Pi}_2^P$-hard for TU games, and it is in $\mathbf{\Pi}_2^P$ whichever constraints are considered.

Concerning the non-emptiness problems, we show instead that constraints may radically alter the computational properties. Indeed, Core-NonEmptiness raises one level up in the polynomial hierarchy, from co-$\mathbf{NP}$ in absence of constraints (Malizia, Palopoli, & Scarcello, 2007) to $\mathbf{\Sigma}_2^P$, while BargainingSet-NonEmptiness is trivial on TU games (since this concept is always non-empty there), but it becomes $\mathbf{\Sigma}_3^P$-complete on constrained games. Interestingly, in both cases, auxiliary and integer variables do not play any role. Indeed, hardness results are established for the basic case where $int(\mathtt{LC}) = \varnothing$ (and without auxiliary variables), while membership results hold for arbitrary constraints.

In the following, all hardness results will be shown to hold in the simplest case of (deterministic) polynomial-time worth-function oracles. Moreover, membership results will not assume any a-priori bound on the representation size of real numbers. To this end, some non-trivial technical matters will be faced next, to show that algorithms can safely work with as few as polynomially many bits, for any solution concept considered in this paper.

## 5.2 Hardness Results (on Cohesive Games with Polynomial-Time Oracles)

In this section, we shall establish our hardness results. In particular, in order to highlight the intrinsic difficulty associated with the solution concepts, constructions are reported over kinds of worth functions that are "simple" not only from a computational viewpoint, in that they are given via oracles computable in polynomial time, but also from an algebraic viewpoint, as they induce *cohesive* games.

We recall here that a (TU) game is cohesive if its worth function $v$ is such that, for each partition $\mathcal{S}$ of the players in $N$, $v(N) \geq \sum_{S \in \mathcal{S}} v(S)$ holds (Osborne & Rubinstein, 1994)—a condition often imposed in order to guarantee that the grand-coalition actually forms. Note that earlier proofs of complexity results on compactly specified games (see, e.g., Deng & Papadimitriou, 1994; Greco, Malizia, Palopoli, & Scarcello, 2009b; Ieong & Shoham, 2005) do generally exploit constructions over games that are not cohesive and, hence, they do not entail the hardness results stated in this paper. In fact, our results interestingly show that cohesivity does not simplify reasoning with solution concepts for coalitional games.

In order to establish hardness results, we exploit a number of reductions that refer to Boolean formulae. Let $\varphi$ be a Boolean formula, and let $vars(\varphi) = \{W_1, \ldots, W_n\}$ be the set of Boolean variables occurring in $\varphi$. Recall that a literal is either a Boolean variable $W_i$ or its negation $\overline{W_i}$. The former is called a positive literal, while the latter is called a negative literal. We denote by $\overline{vars}(\varphi) = \{\overline{W_i} \mid W_i \in vars(\varphi)\}$ the set of negative literals for the variables occurring in $\varphi$. Literals are associated with game players in most proofs. For a set of players $S$, define $\sigma(S)$ to be the truth assignment where $W_i \in vars(\varphi)$ is true if $W_i$ occurs in $S$, and false, otherwise. The fact that $\sigma(S)$ satisfies $\varphi$ is denoted by $\sigma(S) \models \varphi$. Moreover, we say that a coalition $S \subseteq vars(\varphi) \cup \overline{vars}(\varphi)$ is *consistent* w.r.t. a set of variables $Y \subseteq vars(\varphi)$ if, for each $W_i \in Y$, $|\{W_i, \overline{W_i}\} \cap S| = 1$ holds. In the case where $Y = vars(\varphi)$, we simply say that $S$ is consistent.

We start by demonstrating hardness results for the various membership-checking problems. The first result is the co-**NP**-hardness of CORE-CHECK, which is established on the basis of rather standard arguments reported below just for the sake of completeness. In particular, the reader may find it useful to check that the reduction exploited in the proof is based on games that are cohesive, which makes it different from earlier complexity results given in the literature for specific kinds of compactly specified games.

**Theorem 5.1.** CORE-CHECK *is co-**NP***-hard, even for cohesive TU games with polynomial-time oracles and if the input vector is an imputation.*

*Proof.* Recall that deciding whether a Boolean formula $\Phi$ over the variables $X_1, \ldots, X_n$ is not satisfiable, i.e., deciding whether there exists no truth assignment to the variables making $\Phi$ true, is a co-**NP**-complete problem (Johnson, 1990).

Given such a formula $\Phi$, we build in polynomial time the TU game $\mathcal{G}(\Phi) = \langle N, v \rangle$, where $N = vars(\Phi) \cup \{w, e\}$ and where, for each set of players $S$, $v$ is such that:

$$v(S) = \begin{cases} 1 & \text{if } S = N, \\ 1 & \text{if } e \notin S \wedge w \in S \wedge \sigma(S) \models \Phi, \text{ and} \\ 0 & \text{otherwise.} \end{cases}$$

Consider, now, the vector $\hat{x}$ where $\hat{x}_e = 1$ and $\hat{x}_p = 0$ for each other player $p$, and note that $\hat{x}$ is an imputation. We claim that: $\hat{x} \in \mathscr{C}(\mathcal{G}(\Phi))$ *if and only if $\Phi$ is not satisfiable.*

($\Rightarrow$) $\hat{x} \in \mathscr{C}(\mathcal{G}(\Phi))$ implies that there is no coalition $S$ and $S$-feasible payoff vector $y$ with $y_i > x_i$, for each $i \in S$. Consider any coalition/assignment $\hat{S}$ such that $e \notin \hat{S}$ and $w \in \hat{S}$, and observe that $\hat{x}(\hat{S}) = 0$. Since $\hat{x} \in \mathscr{C}(\mathcal{G}(\Phi))$, we must have $v(\hat{S}) = 0$, which entails that $\sigma(\hat{S})$ does not satisfy $\Phi$, by definition of the worth function. Given that there is a one-to-one correspondence between coalitions $\hat{S}$ (with $e \notin \hat{S}$ and $w \in \hat{S}$) and truth assignments for $\Phi$, we conclude that $\Phi$ is not satisfiable.

($\Leftarrow$) If $\hat{x} \notin \mathscr{C}(\mathcal{G}(\Phi))$, there must exist a coalition $\hat{S} \subseteq N$ such that $\hat{x}(\hat{S}) < v(\hat{S})$, which is only possible if $\hat{x}(\hat{S}) = 0$ and $v(\hat{S}) = 1$. By construction of the worth function, it follows that $\hat{S} \subset N$, $e \notin \hat{S}$, $w \in \hat{S}$ and $\sigma(\hat{S}) \models \Phi$. That is, $\Phi$ is satisfiable.

Finally, observe that the role of player $w$ is to guarantee that the game is cohesive. Indeed, for any partition $\mathcal{S}$ of $N$, there is at most one set $S \in \mathcal{S}$ that contains $w$, and hence that may get 1 as its coalition worth. □

When considering constrained games and arbitrary input vectors (i.e., not necessarily imputations), CORE-CHECK turns out to be slightly more difficult than in the previous case. In fact, we stress here that the use of auxiliary integer variables is crucial in order to establish the result illustrated next.

**Theorem 5.2.** CORE-CHECK *is* $\mathbf{D^P}$*-hard, even for cohesive constrained games with polynomial-time oracles.*

*Proof.* Given a pair of Boolean formulae $(\Phi, \Phi')$, deciding whether $\Phi$ is not satisfiable *and* $\Phi'$ is satisfiable is a prototypical $\mathbf{D^P}$-complete problem (Johnson, 1990). Assume, w.l.o.g., that $\Phi' = c_1 \wedge \ldots \wedge c_m$, with $c_i = t_{i,1} \vee t_{i,2} \vee t_{i,3}$, for each $i \in \{1, \ldots, m\}$. That is, $\Phi'$ is in conjunctive normal form and every clause contains exactly three literals. Moreover, let $vars(\Phi') = \{Y_1, \ldots, Y_\ell\}$ and $vars(\Phi) = \{X_1, \ldots, X_n\}$, and assume w.l.o.g. that $vars(\Phi) \cap vars(\Phi') = \varnothing$.

Consider the TU game $\mathcal{G}(\Phi) = \langle N, v \rangle$ built in the proof of Theorem 5.1, and recall that $N = vars(\Phi) \cup \{w, e\}$ and that the vector $\hat{x}$ (where $\hat{x}_e = 1$ and $\hat{x}_p = 0$, for each player $p \in N$ with $p \neq e$) belongs to $\mathscr{C}(\mathcal{G}(\Phi))$ if and only if $\Phi$ is not satisfiable.

Consider then the following set of constraints:

$$\mathtt{LC} = \left\{ \begin{array}{l} 1 \geq T_{Y_j} \geq 0, \forall j \in \{1, \ldots, \ell\} \\ \rho(t_{i,1}) + \rho(t_{i,2}) + \rho(t_{i,3}) \geq 1, \forall i \in \{1, \ldots, m\} \\ x_p \in \mathbb{R}, \forall p \in N \\ T_{Y_j} \in \mathbb{Z}, \forall j \in \{1, \ldots, \ell\} \end{array} \right.$$

where $\rho(t_{i,h})$ denotes the expression $1 - T_{t_{i,h}}$ if $t_{i,h}$ is a negative literal, and the expression $T_{t_{i,h}}$ if $t_{i,h}$ is a positive literal. Note that players in $N$ are actually not constrained in $\mathtt{LC}$. Therefore, if $\Omega(\mathtt{LC}) = \varnothing$, then $\Omega(\mathtt{LC})[N] = \varnothing$ trivially holds (since $\Omega(\mathtt{LC})[N]$ is the restriction of an empty set over $\mathbb{R}^N$). Otherwise, i.e., if $\Omega(\mathtt{LC}) \neq \varnothing$, then $\Omega(\mathtt{LC})[N] = \mathbb{R}^N$ and therefore these constraints are immaterial. Of course, if $\Omega(\mathtt{LC}) = \varnothing$, then there is no imputation of $\mathcal{G}(\Phi)|_{\mathtt{LC}}$; otherwise, all the solution concepts for $\mathcal{G}(\Phi)$ are preserved in $\mathcal{G}(\Phi)|_{\mathtt{LC}}$, since constraints do not play any role in this case.

Observe now that, for each $j \in \{1, \ldots, \ell\}$, $T_{Y_j}$ is constrained over the domain $\{0, 1\}$ as to encode the truth value of the Boolean variable $Y_j$. Clearly, $\mathtt{LC}$ can be computed in polynomial time from $\Phi$, and it is immediate to check that $\Omega(\mathtt{LC}) \neq \varnothing$ if and only if $\Phi'$ is satisfiable. It follows that the vector $\hat{x}$ is in the core of $\mathcal{G}(\Phi)|_{\mathtt{LC}}$ if and only if $\hat{x}$ belongs to the core of $\mathcal{G}(\Phi)$ (i.e., $\Phi$ is not satisfiable) *and* $\Omega(\mathtt{LC}) \neq \varnothing$ (i.e., $\Phi'$ is satisfiable). $\qquad\blacksquare$

We now turn to the study of the bargaining set. Notice that for the class of graph games (which is an instance of the more general framework we are considering here) completeness for BARGAININGSET-CHECK in $\mathbf{\Pi_2^P}$ has recently been established by Greco et al. (2009b). Clearly enough, this result already implies that BARGAININGSET-CHECK is $\mathbf{\Pi_2^P}$-hard on TU games with polynomial-time oracles. Below, we show that this hardness result still holds on games with polynomial-time oracles which are moreover cohesive.

**Theorem 5.3.** BARGAININGSET-CHECK *is* $\mathbf{\Pi_2^P}$*-hard, even for cohesive TU games with polynomial-time oracles and if the input vector is an imputation.*

*Proof.* We show a polynomial-time reduction from the problem of deciding whether a quantified Boolean formula $H = \forall Y_1, \ldots, Y_n \exists Z_1, \ldots, Z_q \Phi$ is valid, which is a well-known $\mathbf{\Pi_2^P}$-complete problem (Johnson, 1990). Let $\mathbf{Y} = \{Y_1, \ldots, Y_n\}$ and $\mathbf{Z} = \{Z_1, \ldots, Z_q\}$ denote the sets of universally and existentially quantified variables, respectively.

Based on $H$, we build a game $\mathcal{G}(H) = \langle N, v \rangle$, where $N = vars(\Phi) \cup \overline{vars}(\Phi) \cup \{a, a'\}$ and where, for each set of players $S$, $v$ is such that:

$$v(S) = \begin{cases} 2 & \text{if } S = N, \\ 1 & \text{if } |S| = n \text{ and } S \text{ is consistent w.r.t. } \{Y_1, \ldots, Y_n\}, \\ 1 & \text{if } S \text{ is consistent, } |\{a, a'\} \cap S| = 1, \text{ and } \sigma(S) \models \Phi, \\ 0 & \text{otherwise.} \end{cases}$$

Let $\hat{x}$ be the imputation with $\hat{x}_a = \hat{x}_{a'} = 1$ and $\hat{x}_p = 0$, for each other player $p$. The construction of $\mathcal{G}(H)$ and $\hat{x}$ is defined as to guarantee two basic properties, which are intuitively illustrated next:

(1) Recall that an objection $(y, S)$ of player $i$ against player $j$ to $\hat{x}$ is such that $i \in S$, $j \notin S$, $y(S) \leq v(S)$ and $y_k > \hat{x}_k$, for each $k \in S$. Since $v(S) > \hat{x}(S)$ must hold for any objection $(y, S)$, it is the case that objections are one-to-one associated with truth assignments for the variables in $\mathbf{Y}$; indeed, this is to have $v(S) = 1$ (and $\hat{x}(S) = 0$). Let $\sigma(\mathbf{Y} \setminus S)$ be the truth assignment associated with coalition $S$.

(2) Recall that a counterobjection $(z, T)$ to an objection $(y, S)$ of player $i$ against player $j$ to $\hat{x}$ is such that $i \notin T$, $j \in T$, $z(T) \leq v(T)$, $z_k \geq y_k$, for each $k \in T \cap S$, and $z_k \geq \hat{x}_k$, for each $k \in T \setminus S$. If $(y, S)$ is an objection against a player $j \notin \{a, a'\}$, then $(z, \{j\})$ with $z_j = 0$ is a trivial counterobjection. On the other hand, counterobjections $(z, T)$ to objections $(y, S)$ against $a$ or $a'$ are necessarily such that $T \cap S = \varnothing$, because $z(T) \leq 1$ and $\hat{x}_a = \hat{x}_{a'} = 1$. In particular, $z(T) = 1$ must hold. Thus, these counterobjections are one-to-one associated with all the possible *satisfying* truth assignments for the variables in $H$, which are moreover obtained as extensions of the assignment $\sigma(\mathbf{Y} \setminus S)$.

By Definition 2.3, $\hat{x}$ is in the bargaining set of $\mathcal{G}(H)$ if and only if for each objection (i.e., assignment $\sigma(\mathbf{Y} \setminus S)$ to the variables in $\mathbf{Y}$), there is a counterobjection (i.e., satisfying assignment obtained by extending $\sigma(\mathbf{Y} \setminus S)$). Therefore, the following claim holds, whose formal proof is reported in the Appendix:

**Claim A.** $\hat{x} \in \mathscr{B}(\mathcal{G}(H))$ *if and only if $H$ is valid.*

To conclude the proof, note that the game is cohesive. Indeed, for each coalition $S$ where $v(S) = 1$, it is the case that $|S \cap \{Y_1, \ldots, Y_n, \bar{Y}_1, \ldots, \bar{Y}_n\}| = n$. Thus, given any three coalitions $S_1$, $S_2$ and $S_3$ with $v(S_1) = v(S_2) = v(S_3) = 1$, it must be the case that two of them overlap over some players. Therefore, any partition $\mathcal{S}$ of $N$ contains at most two coalitions getting a worth greater than 0, and the result follows since $v(N) = 2$. $\qquad\square$

In the remainder of the section we prove our hardness results for non-emptiness problems. We start by showing that adding constraints to the game causes the complexity of the non-emptiness problem for the core to raise one level up in the polynomial hierarchy—from co-**NP** in absence of constraints (Malizia et al., 2007) to $\mathbf{\Sigma_2^P}$. Note that in the proof below, integer and auxiliary variables do not play any role.

**Theorem 5.4.** CORE-NONEMPTINESS *is $\mathbf{\Sigma_2^P}$-hard, even for cohesive constrained games with polynomial-time oracles, and where integer and auxiliary variables are not allowed.*

*Proof.* Deciding whether a quantified Boolean formula $F = \exists X_1, \ldots, X_n \forall Y_1, \ldots, Y_q \Phi$ is valid is a well-known $\mathbf{\Sigma_2^P}$-complete problem (Johnson, 1990).

Based on $F$, we build in polynomial time the game $\mathcal{G}(F) = \langle N, v \rangle$, where $N = vars(\Phi) \cup \overline{vars}(\Phi) \cup \{a\}$ and where, for each set of players $S$, $v$ is such that:

$$v(S) = \begin{cases} 3 \times n & \text{if } S = N, \\ n & \text{if } S \text{ is consistent and } \sigma(S) \not\models \Phi, \\ 0 & \text{otherwise.} \end{cases}$$

In addition, we build in polynomial time a set $\mathtt{LC}$ that, for each $1 \leq i \leq n$, contains the following constraints:

$$\mathtt{LC} = \begin{cases} x_{X_i} + x_{\bar{X}_i} = 1 \\ x_{X_i} \geq 0 \\ x_{\bar{X}_i} \geq 0 \\ x_a = 2 \times n \\ x_{X_i}, x_{\bar{X}_i} \in \mathbb{R} \\ x_a \in \mathbb{R} \end{cases}$$

First, note that $\mathtt{LC}$ forces $x_{X_i} + x_{\bar{X}_i} = 1$, and forces $x_a$ to take value $2 \times n$. Thus, since $v(N) = 3 \times n$, any imputation $x$ for the constrained game $\mathcal{G}(F)|_{\mathtt{LC}}$ does not distribute any worth to the players associated with the variables in $\{Y_1, \ldots, Y_q\}$. An imputation $x$ is then associated with an assignment $\sigma(x)$ to the variables in $\{X_1, \ldots, X_n\}$ such that $X_i$ is true in $\sigma(x)$ if and only if $x_{X_i} < 1$—note that we are associating 1 with false, here.

To understand the salient features of the reduction, recall now that an objection $(y, S)$ to an imputation $x$ is such that $y \in V_{\mathtt{LC}}(S)$ and $y_k > x_k$ for all $k \in S$. Since $y(S) > x(S)$ holds, we have only to take care of coalitions $S \subset N$ such that $S$ is consistent and $\sigma(S)$ is not a satisfying truth assignment. Recall that $V_{\mathtt{LC}}(S) = \{x \in \mathbb{R}^S \mid x(S) \leq v(S)\} \cap \Omega(\mathtt{LC})[S]$;

thus, for any such objection $(y, S)$ with $X_i \in S$ (resp., $\bar{X}_i \in S$), we have $y_{X_i} \leq 1$ (resp., $y_{\bar{X}_i} \leq 1$). Therefore, $S$ cannot include players in $\{X_1, \bar{X}_1, ..., X_n, \bar{X}_n\}$ getting a worth 1 in $x$. It follows that the set of all possible objections $(y, S)$ to any imputation $x$ corresponds to a superset of all truth assignments $\sigma(S)$ which are not satisfying and which are extensions of $\sigma(x)$. This correspondence allows us to establish the following result (whose formal proof is deferred to the Appendix).

**Claim B.** $\mathscr{C}(\mathcal{G}(F)|_{\text{LC}}) \neq \varnothing$ if and only if $F$ is valid.

To conclude the proof, we notice that $\mathcal{G}(F)|_{\text{LC}}$ is cohesive. Indeed, each coalition $S$ with $v(S) = n$ must be consistent, and thus $|S \cap (vars(\Phi) \cup \overline{vars}(\Phi))| = n + q$. Therefore, given any three coalitions $S_1$, $S_2$ and $S_3$ with $v(S_1) = v(S_2) = v(S_3) = n$, it must be the case that two of them overlap over some players. It follows that any partition $\mathcal{S}$ of $N$ contains at most two coalitions getting worth $n$. $\qquad \square$

The non-emptiness problems for the bargaining set is trivial over TU games, since this concept is always non-empty there. This is no longer the case over constrained games, where this problem turns out to be quite difficult. As in the proof of Theorem 5.4, integer and auxiliary variables play no role in the result shown below.

**Theorem 5.5.** BARGAININGSET-NONEMPTINESS *is* $\boldsymbol{\Sigma_3^P}$*-hard, even for cohesive constrained games with polynomial-time oracles, and where integer and auxiliary variables are not allowed.*

*Proof.* Deciding the validity of the formula $P = \exists X_1, \ldots, X_m \forall Y_1, \ldots, Y_n \exists Z_1, \ldots, Z_q \Phi$ is a well-known $\boldsymbol{\Sigma_3^P}$-complete problem (Johnson, 1990).

Based on $P$, we build in polynomial time a game $\mathcal{G}(P) = \langle N, v \rangle$, where $N = vars(\Phi) \cup \overline{vars}(\Phi) \cup \{a, w\}$ and where, for each set of players $S$, $v$ is such that:

$$v(S) = \begin{cases} m+1 & \text{if } S = N \\ 1 & \text{if } w \in S \wedge |S| = n+1 \wedge \\ & \quad S \text{ is consistent w.r.t. } \{Y_1, \ldots, Y_n\}, \\ 1 & \text{if } S = \{X_i, \bar{X}_i\}, \text{ for some } i \\ 1 & \text{if } a \in S \wedge S \setminus \{a\} \text{ is consistent } \wedge \\ & \quad \sigma(S) \models \Phi, \text{ and} \\ 0 & \text{otherwise.} \end{cases}$$

We also build in polynomial time a set $\text{LC}$ that, for each $1 \leq i \leq m$, contains the following constraints:

$$\begin{cases} x_{X_i} + x_{\bar{X}_i} = 1 \\ x_{X_i} \geq 0 \\ x_{\bar{X}_i} \geq 0 \\ x_a = 1 \\ x_{X_i}, x_{\bar{X}_i} \in \mathbb{R} \\ x_a \in \mathbb{R} \end{cases}$$

First, we observe that, because of the above constraints and of the fact that $v(N) = m+1$, in any imputation of this game all players $Y_j$, $1 \leq j \leq n$, and all players $Z_r$, $1 \leq r \leq q$, get payoff 0. Moreover, any imputation $x$ for which there is an index $\bar{i}$, $1 \leq \bar{i} \leq m$, such that $x_{X_{\bar{i}}} > 0$ and $x_{\bar{X}_{\bar{i}}} > 0$ cannot belong to the bargaining set of $\mathcal{G}(P)|_{\text{LC}}$, for the objection $(y, \{w, Y_1, ..., Y_n\})$ against player $a$ such that $y_{Y_j} = \frac{1}{n+1}$ is justified. Indeed, if $(z, T)$ were a counterobjection with $a \in T$, we would have $z_a \geq x_a = 1$ (indeed, $x_a = 1$ is prescribed by LC). Moreover, because of the definition of the worth function, $T$ would be such that $T \setminus \{a\}$ is consistent, i.e., for each $i \in \{1..., m\}$, $|T \cap \{X_i, \bar{X}_i\}| = 1$. Assume that $X_{\bar{i}} \in T$ (the same line of reasoning applies if $\bar{X}_{\bar{i}} \in T$). Then, $z_{X_{\bar{i}}} \geq x_{X_{\bar{i}}} > 0$ must hold and we would have $z(T) > 1$, which is impossible since $v(T) \leq 1$ and since $v(T) \geq z(T)$ holds for any counterobjection. Thus, the set of imputations $x$ that might possibly belong to the bargaining set are restricted to those where variables $x_{X_i}$ and $x_{\bar{X}_i}$ take distinct values from the set $\{0, 1\}$. As a result, we can associate any such imputation $x$ of the constrained game $\mathcal{G}(P)|_{\text{LC}}$ with an assignment $\sigma(x)$ to the variables in $\{X_1, ..., X_m\}$ such that $X_i$ is true in $\sigma(x)$ if and only if $x_{X_i} = 0$. Note that we are associating 0 with true here.

In fact, in order to show the correctness of the reduction, we may basically follow the spirit of the proof of Theorem 5.4. For any imputation $x$ (with the properties illustrated above), the set of the possible objections $(y, S)$ corresponds to the set of all possible truth assignments $\sigma(\mathbf{Y} \setminus S)$ for the variables in $\mathbf{Y} = \{Y_1, ..., Y_n\}$. Objections that might be possibly justified are then restricted to those against player $a$, for which counterobjections correspond to satisfying assignments extending $\sigma(x)$ and $\sigma(\mathbf{Y} \setminus S)$. Thus, the following can be shown, whose detailed proof is reported in the Appendix.

**Claim C.** $\mathscr{B}(\mathcal{G}(P)|_{\text{LC}}) \neq \varnothing$ *if and only if $P$ is valid.*

Finally, note that the game is cohesive. Indeed, consider any partition $\mathcal{S}$ of the players in $N$, and a coalition $S \in \mathcal{S}$ where $v(S) = 1$. In the case where $a \in S$ and $S$ is consistent w.r.t. $vars(\Phi)$, then there cannot exist any other coalition $S' \in \mathcal{S}$ with $v(S') = 1$ and with $S' = \{X_i, \bar{X}_i\}$ for some $i$. In addition, there can exist at most one further coalition $S'' \in \mathcal{S}$ with $v(S'') = 1$ (for $|S''| = n + 1$, $w \in S''$, and $S''$ is consistent w.r.t. $\{Y_1, ..., Y_n\}$). Thus, $\sum_{S \in \mathcal{S}} v(S) \leq 2$. Similarly, if there is no coalition $S \in \mathcal{S}$ such that $a \in S$ and $S$ is consistent w.r.t. $vars(\Phi)$, then $\sum_{S \in \mathcal{S}} v(S) \leq m + 1$. Indeed, $\mathcal{S}$ might contain the coalitions $\{X_1, \bar{X}_1\}, ..., \{X_m, \bar{X}_m\}$, plus at most one coalition that is consistent w.r.t. $\{Y_1, ..., Y_n\}$ and which gets worth 1. In particular, $\mathcal{S}$ cannot contain two coalitions $S'$ and $S''$ consistent w.r.t. $\{Y_1, ..., Y_n\}$ and with $v(S') = v(S'') = 1$, as $w$ should be contained in both. □

## 5.3 Membership Results

We now complete the picture of the complexity arising in the context of constrained games by proving membership results that, together with the proofs in the previous section, provide the completeness results reported in Figure 2. In particular, we shall consider the case where the worth function $v$ is an oracle that can be computed in deterministic polynomial time in the size $||\mathcal{G}|_{\text{LC}}||$ of the constrained game, while deferring a discussion about how these results can be extended to the case where $v$ is an oracle computable in non-deterministic polynomial time to Section 5.3.1.

We start our analysis by stating the complexity of checking whether a vector is an imputation.

**Lemma 5.6.** *Deciding whether a vector is an imputation is in $\mathbf{D}^{\mathbf{P}}$ for constrained games. In particular, the problem is in* co-**NP** *for constrained games without auxiliary integer variables, and in $\mathbf{P}$ for constrained games without integer variables.*

*Proof.* Let $\mathcal{G} = \langle N, v \rangle$ be a TU game and let LC be a set of constraints. Let $\hat{x}$ be a vector assigning a payoff value to each player in $N$. Recall that $\hat{x}$ is an imputation in $X(\mathcal{G}|_{\mathtt{LC}})$ if: (1) $\hat{x} \in V_{\mathtt{LC}}(N) = \{x \in \mathbb{R}^N \mid x(N) \leq v(N)\} \cap \Omega(\mathtt{LC})[N]$; (2) $\hat{x}$ is efficient; and (3) $\hat{x}$ is individually rational.

(1) Checking whether $\hat{x}(N) \leq v(N)$ is feasible in polynomial time. Moreover, checking whether $\hat{x} \in \Omega(\mathtt{LC})[N]$ is feasible in **NP**. Indeed, we can consider the set of linear inequalities $\mathtt{LC}'$ derived from LC by replacing all player variables by their values according to $\hat{x}$. Note that $\mathtt{LC}'$ is a mixed integer linear program defined over the variables (if any) in $real(\mathtt{LC}) \cup int(\mathtt{LC}) \setminus \{x_i \mid i \in N\}$, and that $\hat{x} \in \Omega(\mathtt{LC})[N]$ if and only if $\mathtt{LC}'$ is satisfiable. By well-known results on mixed integer linear programming (see, e.g., Nemhauser & Wolsey, 1988), $\mathtt{LC}'$ admits a solution if and only if it admits a solution that can be represented with polynomially many bits (in the size of $\mathtt{LC}'$). Thus, the problem can be solved by first guessing in **NP** a vector $\hat{x}'$ assigning a value to each variable in $real(\mathtt{LC}) \cup int(\mathtt{LC}) \setminus \{x_i \mid i \in N\}$, and by subsequently checking whether $\hat{x}'$ satisfies all the constraints in $\mathtt{LC}'$ (which is feasible in polynomial time). Of course, if $int(\mathtt{LC}) \subseteq \{x_i \mid i \in N\}$, then $\mathtt{LC}'$ is a linear program without integer variables. In this special case, the satisfiability of $\mathtt{LC}'$ can be checked in **P** (see, e.g., Papadimitriou & Steiglitz, 1998).

(2) Recall that $\hat{x}$ is efficient if for each $x \in V_{\mathtt{LC}}(N)$, there is a player $i \in N$ such that $\hat{x}_i \geq x_i$. Consider the set of linear inequalities $\mathtt{LC}''$ derived from LC by adding the $|N| + 1$ inequalities: $x(N) \leq v(N)$, and $x_i > \hat{x}_i$ for each $i \in N$. Then, $\hat{x}$ is efficient if and only if $\mathtt{LC}''$ is not satisfiable. This latter task is feasible in co-**NP**, since $\mathtt{LC}''$ is a mixed integer linear program whose satisfiability can be checked in **NP**—see (1). In the special case where $int(\mathtt{LC}) = \varnothing$, $\mathtt{LC}''$ does not contain integer variables and, hence, its (un)satisfiability can be checked in polynomial time.

(3) Recall that $\hat{x}$ is individually rational if for each player $i \in N$, $\hat{x}_i \geq \max\{x_i \mid x_i \in V_{\mathtt{LC}}(\{i\})\}$. Consider the set of linear inequalities $\mathtt{LC}'''_i$ derived from LC by adding the two inequalities $x_i \leq v(\{i\})$ and $x_i > \hat{x}_i$. The individual rationality holds if and only if $\mathtt{LC}'''_i$ is not satisfiable, for each $i \in N$. As in the point (2) above, this task is feasible in co-**NP** in general, and in polynomial time whenever $int(\mathtt{LC}) = \varnothing$.

We can now conclude that deciding whether $\hat{x}$ is an imputation is the conjunction of problem (1), which is feasible in **NP**, and of problems (2) and (3), which are feasible in co-**NP**. Thus, the problem is in $\mathbf{D}^P$.

In the case where $int(\mathtt{LC}) \subseteq \{x_i \mid i \in N\}$ holds, (1) is feasible in polynomial time and, hence, deciding whether $\hat{x}$ is an imputation is in co-**NP**.

Finally, if $int(\mathtt{LC}) = \varnothing$, problems (1), (2), and (3) are feasible in polynomial time. ◻

Let us now consider the membership of CORE-CHECK. This proof is routine and is reported for the sake of completeness only.

**Theorem 5.7.** CORE-CHECK *is in* $\mathbf{D}^P$. *In particular, it is in* co-**NP** *for constrained games without auxiliary integer variables.*

*Proof.* Let $\hat{x}$ be the input vector for the game $\mathcal{G}|_{\mathtt{LC}}$, where $\mathcal{G} = \langle N, v \rangle$. We have to check that $\hat{x}$ satisfies the conditions of the core and that $\hat{x}$ is indeed an imputation.

Concerning the former task, recall that the complementary problem of deciding whether $\hat{x}$ is not in the core amounts to finding a coalition $S$ and a vector $x \in V_{\mathtt{LC}}(S)$ such that $x_i > \hat{x}_i, \forall i \in S$. Consider the set of linear inequalities $\mathtt{LC}_S$ derived from $\mathtt{LC}$ by adding the $|S| + 1$ inequalities $x(S) \leq v(S)$, and $x_i > \hat{x}_i \ \forall i \in S$. Then, $\hat{x}$ is not in the core if there is a coalition $S$ such that $\mathtt{LC}_S$ is satisfiable. This task can be therefore solved by guessing in **NP** a coalition $S$ together with a vector $\hat{x}'$ assigning a value to each variable in $\mathtt{LC}_S$, and by subsequently checking that $\hat{x}'$ does indeed satisfy all the constraints in $\mathtt{LC}_S$. It follows that deciding whether $\hat{x}$ satisfies the conditions of the core is feasible in co-**NP**.

Concerning the task of checking whether $\hat{x}$ is an imputation, we use the results in Lemma 5.6. Thus, for general games, CORE-CHECK can be solved by the conjunction of a problem in co-**NP** and a problem in $\mathbf{D^P}$. Of course, this is again a problem feasible in $\mathbf{D^P}$. Moreover, if $int(\mathtt{LC}) \subseteq \{x_i \mid i \in N\}$ holds, then CORE-CHECK is feasible in co-**NP**. □

Deriving the membership result for BARGAININGSET-CHECK on constrained games requires a more sophisticated line of reasoning. We start by recalling that, for TU games, it is has been shown that BARGAININGSET-CHECK is in $\mathbf{\Pi_2^P}$ (Greco et al., 2009b). In fact, this result has been established by exploiting a characterization for the bargaining set that does not hold in the presence of constraints. Below, by exploiting a completely different proof technique, we shall show that, surprisingly, the presence of the constraints does not alter the computational properties of this problem.

For the following proofs, we recall that given a set $\mathtt{LC}$ of linear (in)equalities over $n$ real variables, the set $\Omega(\mathtt{LC})$ is a *polyhedron* in $\mathbb{R}^n$, whose faces are given by the halfspaces associated with the (in)equalities in $\mathtt{LC}$, and whose vertices are given by the intersection of $n$ inequalities from $\mathtt{LC}$, and hence can be represented with polynomially many bits in the size of $\mathtt{LC}$ (see, e.g., Papadimitriou & Steiglitz, 1998; Nemhauser & Wolsey, 1988). A bounded polyhedron is called a *polytope*. Moreover, we use the following notation. Let $S \subseteq N$ be a set of players and let $y_S$ be the set of variables $\{y_k \mid k \in S\}$. We denote by $\mathtt{LC}_{y_S}$ the copy of the system of mixed-integer linear inequalities $\mathtt{LC}$ where every player variable $x_i$, with $i \in S$, is renamed as $y_i$, and every other variable $v$ in $\mathtt{LC}$ is renamed as $v_{y_S}$.

**Lemma 5.8.** *Let $\mathcal{G} = \langle N, v \rangle$ be a TU game, $\mathtt{LC}$ be a set of constraints, and $\hat{x}$ be an imputation for $\mathcal{G}|_{\mathtt{LC}}$ that does not belong to $\mathscr{B}(\mathcal{G}|_{\mathtt{LC}})$. Then, there exists a justified objection to $\hat{x}$ that is representable with polynomially many bits.*

*Proof.* Since $\hat{x} \notin \mathscr{B}(\mathcal{G}|_{\mathtt{LC}})$, there are two players $i$ and $j$, a coalition $S$ with $i \in S$ and $j \notin S$, and an $S$-feasible vector $y$ such that $(y, S)$ is a justified objection of $i$ against $j$ to $\hat{x}$. Let $\mathtt{LC}^{i,j,S}$ be the system consisting of the (in)equalities in $\mathtt{LC}_{y_S}$ plus the $|S| + 1$ inequalities: $y(S) \leq v(S)$ and $y_k > \hat{x}_k, \ \forall k \in S$. Then, the set $\Omega(\mathtt{LC}^{i,j,S})[y_S]$ consists of all the $S$-feasible vectors $y$ such that $(y, S)$ is an objection of $i$ against $j$ to $\hat{x}$.

Let us now consider possible candidate counterobjections. For any $T \subseteq N$ with $j \in T$ and $i \notin T$, let $\mathtt{LC}^{i,j,S,T}$ be the system including the (in)equalities in $\mathtt{LC}_{y_S}$ and in $\mathtt{LC}_{z_T}$, plus the inequalities $y(S) \leq v(S)$, $y_k > \hat{x}_k, \ \forall k \in S$, $z(T) \leq v(T)$, $z_k \geq y_k, \ \forall k \in T \cap S$, and $z_k \geq \hat{x}_k, \ \forall k \in T \setminus S$. Note that $\Omega(\mathtt{LC}^{i,j,S,T})[y_S]$ contains all $y$ over the index set $S$ such that there exists a counterobjection through $T$, and hence of the form $(z, T)$, to the objection $(y, S)$ of $i$ against $j$ to $\hat{x}$. It follows that the set of all vectors $y$ such that $(y, S)$ is a justified objection of $i$ against $j$ to $\hat{x}$ is the set:

$$\Omega(\mathtt{LC}^{i,j,S})[y_S] \setminus \bigcup_{T \mid i \notin T \wedge j \in T} \Omega(\mathtt{LC}^{i,j,S,T})[y_S].$$

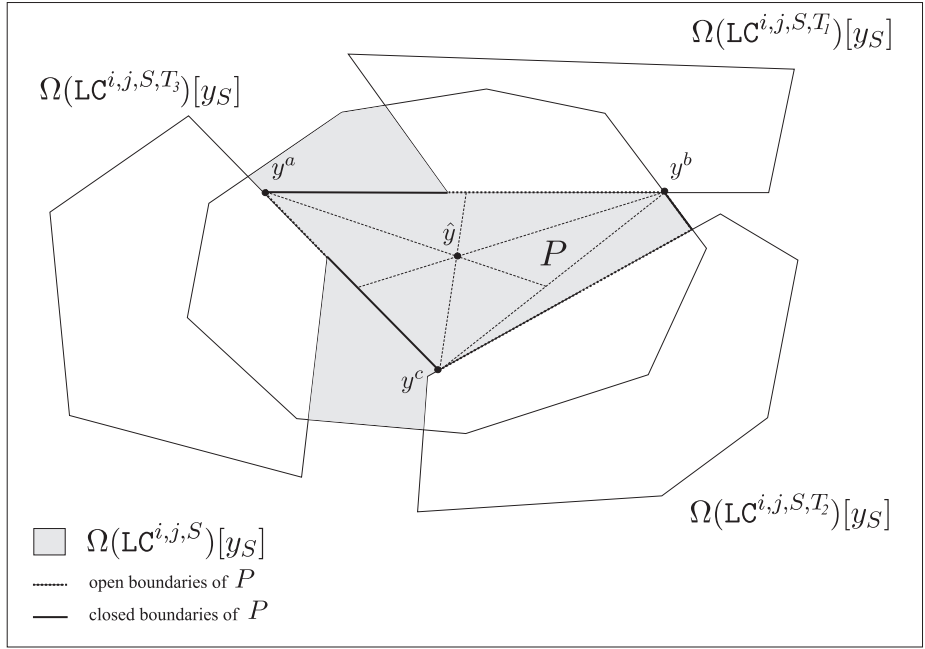

Figure 3: Illustration of Claim D, with coalitions $T_1$, $T_2$, and $T_3$.

To conclude the proof we claim the following.

**Claim D.** $\Omega(\mathtt{LC}^{i,j,S})[y_S] \setminus \bigcup_{T \mid i \notin T \wedge j \in T} \Omega(\mathtt{LC}^{i,j,S,T})[y_S]$ *contains a point (i.e., a justified objection to $\hat{x}$) that can be represented with polynomially many bits.*

To prove the claim, let us consider the following geometrical arguments: Consider first the case where $\mathtt{LC}^{i,j,S}$ and $\mathtt{LC}^{i,j,S,T}$ (for each $T \mid i \notin T \wedge j \in T$) contain no integer variables, and let $P$ be a maximal convex subset of $\Omega(\mathtt{LC}^{i,j,S})[y_S] \setminus \bigcup_{T \mid i \notin T \wedge j \in T} \Omega(\mathtt{LC}^{i,j,S,T})[y_S]$. The vertices of $P$, which are points of $\mathbb{R}^S$, are given by the intersection of at most $|S|$ (independent) halfspaces that are facets of $\Omega(\mathtt{LC}^{i,j,S})[y_S]$ or of some $\Omega(\mathtt{LC}^{i,j,S,T})[y_S]$, and thus they can be represented with polynomially many bits. In fact, $P$ might not contain its own boundaries. Thus, if some of its vertices actually belongs to $P$, then the result straightforwardly holds. On the other hand, if $P$ is a possibly open segment with endpoints $y^a$ and $y^b$ (representable with polynomially many bits), then the middle point $y_m$ necessarily belongs to $P$ (since $P$ is convex) and can be represented with polynomially many bits. Finally, if the polytope $P$ has more than two vertices (as shown in Figure 3), then it must have at least three vertices $y^a$, $y^b$, and $y^c$ that do not belong to the same face of $P$. Therefore, the barycenter $\hat{y}$ of the triangle with vertices $y^a$, $y^b$, and $y^c$ belongs to $P$, and it can be represented with polynomially many bits, because this is the case for $y^a$, $y^b$, and $y^c$.

To conclude the proof, we observe that integer variables in $\mathtt{LC}^{i,j,S}$ and $\mathtt{LC}^{i,j,S,T}$ (for each $T \mid i \notin T \wedge j \in T$) can be easily preprocessed. Roughly—the technical details are reported in the Appendix—, since $\Omega(\mathtt{LC}^{i,j,S})[y_S]$ is a polytope by construction of $\mathtt{LC}^{i,j,S}$ and since, therefore, its vertices can be represented with polynomially many bits, all integer components of interest (basically those falling within $\Omega(\mathtt{LC}^{i,j,S})[y_S]$) can be represented with polynomially many bits, as well. Thus, to find a point with polynomially many bits as asked for in Claim D, we can iterate over all the possible combinations of integer values and, at each step, evaluate the expression $\Omega(\mathtt{LC}^{i,j,S})[y_S] \setminus \bigcup_{T \mid i \notin T \wedge j \in T} \Omega(\mathtt{LC}^{i,j,S,T})[y_S]$ by replacing all integer values with the combination of values at hand. Of course, the resulting expression does not involve integer variables, all the inequalities in it are still representable with polynomially many bits, and therefore the above line of reasoning applies. □

Armed with the above lemma, we can state the complexity of BARGAININGSET-CHECK.

**Theorem 5.9.** BARGAININGSET-CHECK *is in* $\mathbf{\Pi}_2^P$.

*Proof.* We show that the complementary problem of deciding whether a vector $\hat{x}$ is not in the bargaining set of a given constrained game $\mathcal{G}|_{\mathtt{LC}}$ is in $\mathbf{\Sigma}_2^P$. We start by checking whether $\hat{x}$ is an imputation in $\mathbf{D}^P$ (cf. Lemma 5.6)—recall that $\mathbf{D}^P$ is contained in $\mathbf{\Sigma}_2^P$. If this is the case, by Lemma 5.8 we guess in non-deterministic polynomial time a justified objection to $\hat{x}$, that is, a coalition $S$, two players $i \in S$ and $j \notin S$, and a vector $\hat{y}$ such that $(\hat{y}, S)$ is an objection of $i$ against $j$ to $\hat{x}$. Consider the system $\mathtt{LC}'$ of the (in)equalities obtained from $\mathtt{LC}^{i,j,S}$ (recall its definition in the proof of Lemma 5.8) by replacing player variables associated with the coalition $S$ by the respective values in $\hat{y}$. Of course, $(\hat{y}, S)$ is an objection if and only if $\mathtt{LC}'$ is satisfiable. By well-known results on mixed integer linear programming (see, e.g., Nemhauser & Wolsey, 1988), $\mathtt{LC}'$ admits a solution if and only if it admits a solution that can be represented with polynomially many bits. Therefore, within the same non-deterministic step, we can also guess an assignment (call it $\bar{w}'$) to all the variables in $\mathtt{LC}'$, and then check in polynomial time that $\bar{w}'$ actually satisfies all the constraints (i.e., that $\hat{y}$ is actually an objection).

To conclude the algorithm for solving BARGAININGSET-CHECK, we now have to check that there is no counterobjection $(z, T)$ to the objection $(\hat{y}, S)$ of $i$ against $j$ to $\hat{x}$. This task requires a co-**NP** oracle call. In particular, the oracle works by checking the complementary condition in **NP**. To this end, in a non-deterministic step, it first guesses a coalition $T$ with $j \in T$ and $i \notin T$. Consider now the system $\mathtt{LC}''$ including the (in)equalities in $\mathtt{LC}_{z_T}$, plus the inequalities $z(T) \leq v(T)$, $z_k \geq \hat{y}_k$, $\forall k \in T \cap S$, and $z_k \geq \hat{x}_k$, $\forall k \in T \setminus S$. Then, there is a counterobjection $(z, T)$ to $(\hat{y}, S)$ for some vector $z$ if and only if $\mathtt{LC}''$ is satisfiable. As in the case above, a solution to $\mathtt{LC}''$ is guaranteed to exist which can be represented with polynomially many bits, so that this solution (call it $\bar{w}''$) can be guessed within the same non-deterministic step by the oracle. In fact, to check that $\bar{w}''$ actually satisfies $\mathtt{LC}''$ is trivially feasible in polynomial time. □

We now turn to analyze non-emptiness problems. We start with the non-emptiness of the core, which is a co-**NP**-complete problem for TU games (Malizia et al., 2007). Constraints do play a role here, since we have shown that CORE-NONEMPTINESS is $\mathbf{\Sigma}_2^P$-hard (cf. Theorem 5.4). Below, we confirm that this is the exact complexity of this problem.

**Theorem 5.10.** Core-NonEmptiness *is in* $\Sigma_2^P$.

*Proof.* Let us again adopt the notation used in the proof of Lemma 5.8. Let $S \subseteq N$ be a coalition, and $\mathtt{LC}^S$ be the set of mixed-integer linear (in)equalities including the (in)equalities in $\mathtt{LC}_{x_N}$ and $\mathtt{LC}_{y_S}$, plus the inequalities $y(S) \le v(S)$ and $y_i > x_i$, for each $i \in S$. We get:

$$\mathscr{C}(\mathcal{G}|_{\mathtt{LC}}) = X(\mathcal{G}|_{\mathtt{LC}}) \setminus \bigcup_{S \subseteq N} \Omega(\mathtt{LC}^S)[x_N].$$

Let $\mathtt{LC}^X$ be the set of the (in)equalities in $\mathtt{LC}_{x_N}$ plus the inequality $x(N) \le v(N)$. Moreover, for each player $i$, let $\mathtt{LC}^i$ be the set of the (in)equalities $\mathtt{LC}_{x_N}$ and $\mathtt{LC}_{y'_{\{i\}}}$, plus the inequalities $x(N) \le v(N)$ and $x_i < y'_i \le v(\{i\})$. Then, the set $\Omega(\mathtt{LC}^i)[x_N]$ consists of all the vectors that are not individually rational (w.r.t. player $i$). Thus,

$$\mathscr{C}(\mathcal{G}|_{\mathtt{LC}}) = \left( \Omega(\mathtt{LC}^X)[x_N] \setminus \bigcup_{i \in N} \Omega(\mathtt{LC}^i)[x_N] \right) \setminus \bigcup_{S \subseteq N} \Omega(\mathtt{LC}^S)[x_N].$$

In particular, note that the efficiency condition on imputations is guaranteed. Indeed, the points that are not efficient are removed, because they belong to the set $\Omega(\mathtt{LC}^N)[x_N]$, which is considered above for $S = N$.

By applying the same line of reasoning as in Claim D (in Lemma 5.8) on the above expressions, we have that, if $\mathscr{C}(\mathcal{G}|_{\mathtt{LC}})$ is not empty, then it contains an imputation that is representable with polynomially many bits. Thus, we can decide the non-emptiness of the core by first guessing in **NP** a vector $\hat{x}$. Then, we may call a $\mathbf{D}^P$ oracle (corresponding to the invocation of an **NP** and a co-**NP** oracle) to check that $\hat{x}$ is an imputation (cf. Lemma 5.6), and finally we can verify that $\hat{x}$ is in the core with a further call to a co-**NP** oracle. In particular, this latter oracle works by checking the complementary condition in **NP**, i.e., it checks whether $\hat{x}$ is not in the core. To this end, the oracle guesses in a non-deterministic step a coalition $S$. Consider now the system $\mathtt{LC}'$ formed by the (in)equalities in $\mathtt{LC}_{y_S}$ plus the $|S| + 1$ inequalities: $y(S) \le v(S)$ and $y_k > \hat{x}_k$, $\forall k \in S$. Then, there is an objection $(y, S)$ to $\hat{x}$ for some vector $y$ if and only if $\mathtt{LC}'$ is satisfiable. Again, $\mathtt{LC}'$ admits a solution if and only if it admits a solution that can be represented with polynomially many bits. Therefore, within the same non-deterministic step, we can also guess an assignment (call it $\bar{w}'$) to all the variables in $\mathtt{LC}'$, and then check in polynomial time that $\bar{w}'$ actually satisfies all the constraints. □

Now, we can complete our picture with the bargaining set.

**Theorem 5.11.** BargainingSet-NonEmptiness *is in* $\Sigma_3^P$.

*Proof.* Consider again the setting in the proof of Lemma 5.8. Let $i$ and $j$ be two players in $N$. For any coalition $S$ with $i \in S$ and $j \notin S$, let $\mathtt{LC}^{i,j,S}$ be the system consisting of the (in)equalities in $\mathtt{LC}_{x_N}$ and in $\mathtt{LC}_{y_S}$ plus the $|S| + 1$ inequalities: $y(S) \le v(S)$ and $y_k > x_k$, $\forall k \in S$. Moreover, for each pair of sets of players $S$ and $T$ with $i \in S \setminus T$ and $j \in T \setminus S$, let $\mathtt{LC}^{i,j,S,T}$ be the system of mixed integer inequalities including the inequalities of the systems $\mathtt{LC}_{x_N}$, $\mathtt{LC}_{y_S}$, and $\mathtt{LC}_{z_T}$, plus the inequalities $y(S) \le v(S)$, $y_k > x_k$ for each $k \in S$, $z(T) \leq v(T)$, $z_k \geq y_k$ for each $k \in T \cap S$, and $z_k \geq x_k$ for each $k \in T \setminus S$. That is, we proceed in the same way as in the proof of Lemma 5.8, but here the components of vector $x$ are variables of this linear program, while in the previous lemma they were fixed values.

Observe now that $\Omega(\texttt{LC}^{i,j,S,T})[x_N \cup y_S]$ contains all pairs $\langle x, y \rangle$ such that there exists a counterobjection $(z, T)$ to the objection $(y, S)$ of $i$ against $j$ to $x$, and that $\Omega(\texttt{LC}^{i,j,S})[x_N \cup y_S]$ consists of all pairs $\langle x, y \rangle$ such that $y$ is an $S$-feasible vector with $i \in S$ and $j \notin S$ such that $(y, S)$ is an objection to $x$. Then, the set

$$\bar{\Omega}(i,j,S) = \Omega(\texttt{LC}^{i,j,S})[x_N \cup y_S] \setminus \bigcup_{T \mid i \notin T \wedge j \in T} \Omega(\texttt{LC}^{i,j,S,T})[x_N \cup y_S]$$

is the set of all pairs $\langle x, y \rangle$ such that $(y, S)$ is a justified objection of $i$ against $j$ to $x$. Therefore,

$$\mathscr{B}(\mathcal{G}|_{\texttt{LC}}) = X(\mathcal{G}|_{\texttt{LC}}) \setminus \bigcup_{S \subseteq N \wedge i \in S \wedge j \notin S} \bar{\Omega}(i,j,S)[x_N],$$

where, by considering efficiency and individual rationality (see the notation in the proof of Theorem 5.10), we have

$$X(\mathcal{G}|_{\texttt{LC}}) = \left( \Omega(\texttt{LC}^X)[x_N] \setminus \Omega(\texttt{LC}^N)[x_N] \right) \setminus \bigcup_{i \in N} \Omega(\texttt{LC}^i)[x_N].$$

By slightly adapting the proof of Claim D (in Lemma 5.8), one may show that if the bargaining set is not empty, there exists a vector $\hat{x} \in \mathscr{B}(\mathcal{G}|_{\texttt{LC}})$ which can be represented with polynomially many bits. Therefore, BARGAININGSET-NONEMPTINESS can be solved by first guessing in non-deterministic polynomial time such a vector $\hat{x}$. Then, we may call a $\mathbf{D^P}$ oracle to check that $\hat{x}$ is an imputation (cf. Lemma 5.6), and finally we can verify that $\hat{x}$ is indeed in the bargaining set with a further call to a $\mathbf{\Pi_2^P}$ oracle, in order to solve BARGAININGSET-CHECK on input $\hat{x}$ (cf. Theorem 5.9). □

### 5.3.1 EXTENSION TO MORE GENERAL WORTH FUNCTIONS

In all the membership results above, we have assumed that worth functions are polynomial-time computable and, within this setting, we have shown that various hardness results are indeed tight. Thus, the reader might be inclined to believe that, by considering more powerful worth functions, the complexity for these problems may consistently increase. Surprisingly, this is not the case. Indeed, we can show that nothing has to be paid if more powerful worth functions that encode **NP**-complete problems are considered.

To this end, let $v_{\mathcal{G}|_{\texttt{LC}}}$ denote the worth function of a game $\mathcal{G}|_{\texttt{LC}}$, and define the *worth-function graph* for any class $\mathcal{C}$ of constrained games as the set of tuples $W_{\mathcal{C}} = \{ \langle (\mathcal{G}|_{\texttt{LC}}, S), w \rangle \mid \mathcal{G}|_{\texttt{LC}} \in \mathcal{C} \wedge v_{\mathcal{G}|_{\texttt{LC}}}(S) = w \}$. Recall, e.g., from the work by Johnson (1990), that such a function is computable in non-deterministic polynomial time if there is an integer $k$ such that $W_{\mathcal{C}}$ is *(i) k*-balanced, i.e., $\|w\| \leq (\|\mathcal{G}|_{\texttt{LC}}\| + \|S\|)^k$, and *(ii) k*-decidable, i.e., there is a non-deterministic Turing machine that decides whether a given tuple $t$ belongs to $W_{\mathcal{C}}$ in $O(\|t\|^k)$ time. More precisely, since $v_{\mathcal{G}|_{\texttt{LC}}}$ is a partial (standard) single-valued function (multi-valued functions are also considered in the literature), the class of functions that we consider is called **NPSV** (see, e.g., Selman, 1994).

The complexity of various solution concepts for TU games within a setting where worth functions are given as oracles computable in **NPSV** has been analyzed in an extended version of the work by Greco et al. (2009b). There, it emerged that all the membership results in Figure 2 hold on any class $\mathcal{C}$ of games having such worth functions. Very roughly, the basic observation is that if we consider **NPSV** worth-functions, any non-deterministic algorithm $M$ that guesses in polynomial-time some coalition $S$ for a game $\mathcal{G}|_{\text{LC}} \in \mathcal{C}$, can at the same time (with just a polynomial-time delay) guess a worth $w$ and an additional string $c$ (of polynomial size w.r.t. $||\mathcal{G}|_{\text{LC}}|| + ||S||$), which acts as a "certificate" to decide whether the tuple $\langle (\mathcal{G}|_{\text{LC}}, S), w \rangle$ belongs to the **NP** set $W_{\mathcal{C}}$. Thus, the complexity of any (non-deterministic) algorithm that uses the value $v_{\mathcal{G}|_{\text{LC}}}(S)$ after a guess of the coalition $S$ is not affected by replacing a polynomial-time worth-functions with **NPSV** worth-functions. By exploiting this line of reasoning, it is easy to adapt proofs of membership results in order to deal with such more general worth functions.

**Theorem 5.12.** *The membership results in Figure 2 hold on any class $\mathcal{C}$ of games whose worth functions are in* **NPSV**.

## 6. Discussion and Conclusion

Imposing linear constraints on the outcomes of games is an approach that has been explored by several authors in the context of non-cooperative strategic games (e.g., Charnes, 1953; Semple, 1997; Ryan, 1998). However, in the context of cooperative games this approach has received considerably less attention and, indeed, no general framework was proposed in the literature and no analysis of its properties was conducted so far.

In this paper, we have faced the issue by conducting a systematic study of constrained games within a framework where constraints are defined as mixed-integer linear (in)equalities imposed over an underlying TU game. Seemingly close to the class of constrained games is the class of *linear programming games* (see, e.g., Owen, 1975), where the worth $v(S)$ of a coalition $S$ is implicitly given as a linear program (e.g., as the maximum of a given objective function over a feasible region $\Omega(\text{LC})$ defined in terms of a set of linear (in)equalities $\text{LC}$). Of course, this approach differs from the setting of constrained games where the role of $\text{LC}$ is, instead, to govern the distribution of the worths within an NTU perspective. Moreover, differently from classical NTU formalizations, constrained games allow to define non-convex and non-comprehensive sets of worth distributions, which is an appealing modeling capability that emerged to be useful in several application domains. Finally, the resulting game framework has been analyzed with respect to the preservation and the computational properties of some relevant solution concepts.

It is worthwhile noticing that the framework we have discussed in this paper shares the spirit of the recent arguments by Shoham (2008), who advocated the use of a broader vocabulary than the fairly terse one characterizing the early foundations of the game theory. Also relevant are those proposals that reconsidered basic concepts of cooperative games in the light of a modeling perspective that is closer to the requirements of computer science applications: seminal and influential directions of this type give rise, in particular, to coalitional skill games (Bachrach & Rosenschein, 2008), qualitative coalitional games (Wooldridge & Dunne, 2004), coalitional resource games (Wooldridge & Dunne, 2006), Bayesian coalitional games (Ieong & Shoham, 2008), multi-attribute coalitional games (Ieong & Shoham, 2006), temporal qualitative coalitional games (Ågotnes, van der Hoek, & Wooldridge, 2006), and cooperative Boolean games (Dunne, van der Hoek, Kraus, & Wooldridge, 2008). In the light of the above approaches, an interesting avenue of further research may be to consider more expressive kinds of constraints, formulated for instance via logic-based languages, and where preference criteria can be adopted in place of hard constraints.

Other avenues of research are related to some technical questions that were not explored in the paper. First, our complexity analysis focused on the notions of the core and the bargaining set, which are founded on the concepts of objections and counterobjections. Of course, it would be interesting to complement our results with the analysis of the kernel, the nucleolus, and the Shapely value. Actually, hardness results for the kernel and the nucleolus of TU (graphical) games have recently been illustrated by Greco et al. (2009b), and indeed they trivially provide lower bounds for the complexity of such solution concepts in the setting of constrained games. However, providing tighter computational bounds requires a deeper understanding of the computational aspects underlying Kalai's axiomatization, which is outside the scope of this paper. Furthermore, for the Shapley value, it is of interest to study other extensions that have been provided in the literature for NTU games and to assess their behavior when applied to constrained games.

Moreover, our hardness results have been shown to hold even by restricting the underlying TU games to use cohesive worth functions only. It might be of interest to study the complexity when different specific kinds of functions are considered (for instance, *monotone*, *superadditive*, *weakly superadditive*, or *convex* ones[6]). Similarly, assessing to which extent considering such specific kinds of worth functions affects the analytical properties studied in Section 4 is an interesting question which we leave for further research.

Finally, from a modeling viewpoint, we recall that the framework proposed in the paper exploits *one* set of linear (in)equalities to constrain the outcomes of *all* coalitions. Thus, in the light of adding modeling power to this framework, it might be of interest to study a natural generalization where each coalition is equipped with its specific set of linear (in)equalities. In particular, this setting would call for conceiving suitable mechanisms to compactly represent (exponentially many) different sets of constraints, and for defining formal measures for the "expressivity" of such compact representations for constraint-based NTU games.

### Acknowledgments

A coalitional game framework for dealing with linear constraints imposed on TU games was first illustrated by the authors in an extended abstract published in the proceedings of the 8th International Conference on Autonomous Agents and Multiagent Systems (Greco, Malizia, Palopoli, & Scarcello, 2009a). There, some solution concepts have been defined and studied, which were based on proposing ad-hoc adaptations of solution concepts for TU games. Following the suggestions of the anonymous referees, the constrained game framework proposed in the present paper fits instead the framework of NTU games, in its most general form. Thus, the solution concepts studied in this paper are just given as suitable specializations of the standard solution concepts defined for NTU games.

---

6. The worth function $v : 2^N \mapsto \mathbb{R}$ is monotone if $v(S) \geq v(T)$ holds, for each pair of coalitions $S, T \subseteq N$ such that $S \supseteq T$; $v$ is superadditive if $v(S \cup T) \geq v(S) + v(T)$ holds, for each pair of coalitions $S, T \subseteq N$ such that $S \cap T = \varnothing$; $v$ is weakly superadditive if $v(S \cup \{i\}) \geq v(S) + v(\{i\})$, $\forall S \subseteq N$ and $\forall i \in N \setminus S$; $v$ is convex if $v(S \cup T) + v(S \cap T) \geq v(S) + v(T)$, $\forall S, T \subseteq N$ (see, e.g., Peleg & Sudhölter, 2007).

# Appendix A. Computational Complexity

In this appendix we recall some basic definitions of complexity theory, by referring the reader to the work by Johnson (1990) for more on this topic.

## A.1 The Complexity of Decision Problems: P, NP, and co-NP

*Decision* problems are maps from strings (encoding the input instance over a fixed alphabet, e.g., the binary alphabet $\{0, 1\}$) to the set $\{$"*yes*", "*no*"$\}$. The class **P** is the set of decision problems that can be solved by a deterministic Turing machine in polynomial time with respect to the input size, that is, with respect to the length of the string that encodes the input instance. For a given input $x$, its size is usually denoted by $||x||$.

Throughout the paper, we often refer to computations carried out by *non-deterministic* Turing machines, too. Recall that these are Turing machines that, at some points of the computation, may not have one single next action to perform, but a *choice* between several possible next actions. A non-deterministic Turing machine answers a decision problem if on any given input $x$: (*i*) there is at least one sequence of choices leading to halt in an accepting state if $x$ is a "yes" instance; and (*ii*) all possible sequences of choices lead to some rejecting state if $x$ is a "no" instance.

The class of decision problems that can be solved by non-deterministic Turing machines in polynomial time is denoted by **NP**. Problems in **NP** enjoy a remarkable property: any "yes" instance $x$ has a *certificate* of its being a "yes" instance, which has polynomial length and which can be checked in polynomial time (in the size $||x||$). As an example, the problem of deciding whether a Boolean formula $\Phi$ over the variables $X_1, \ldots, X_n$ is satisfiable, i.e., deciding whether there exists some truth assignment to the variables making $\Phi$ true, is a well-known problem in **NP**; in fact, any satisfying truth assignment for $\Phi$ is obviously a certificate that $\Phi$ is a "yes" instance, i.e., that $\Phi$ is satisfiable.

The class of problems whose complementary problems are in **NP** is denoted by co-**NP**. As an example, the problem of deciding whether a Boolean formula $\Phi$ is *not* satisfiable is in co-**NP**. Of course, the class **P** is contained both in **NP** and co-**NP**.

The class $\mathbf{D}^P$ is the class of problems that can be defined as a conjunction of two problems, one from **NP** and one from co-**NP**, respectively. For instance, it is in $\mathbf{D}^P$ to decide whether, for a given pair of Boolean formulae $(\Phi, \Phi')$, $\Phi'$ is satisfiable *and* $\Phi$ is not.

## A.2 Further Complexity Classes: The Polynomial Hierarchy

Throughout the paper, we also refer to a particular type of computation called computation with *oracles*. Intuitively, oracles are subroutines that have unary cost.

The classes $\mathbf{\Sigma}_k^P$, $\mathbf{\Pi}_k^P$, and $\mathbf{\Delta}_k^P$, forming the *polynomial hierarchy*, are defined as follows: $\mathbf{\Sigma}_0^P = \mathbf{\Pi}_0^P = \mathbf{P}$ and for all $k \geq 1$, $\mathbf{\Sigma}_k^P = \mathbf{NP}^{\Sigma_{k-1}^P}$, $\mathbf{\Delta}_k^P = \mathbf{P}^{\Sigma_{k-1}^P}$, and $\mathbf{\Pi}_k^P = \text{co-}\mathbf{\Sigma}_k^P$ where co-$\mathbf{\Sigma}_k^P$ denotes the class of problems whose complementary problem is solvable in $\mathbf{\Sigma}_k^P$. Here, $\mathbf{\Sigma}_k^P$ (resp. $\mathbf{\Delta}_k^P$) models computability by a non-deterministic (resp., deterministic) polynomial-time Turing machine that may use an oracle in $\mathbf{\Sigma}_{k-1}^P$. Note that $\mathbf{\Sigma}_1^P$ coincides with **NP**, and that $\mathbf{\Pi}_1^P$ coincides with co-**NP**.

A well-known problem at the $k$-th level of the polynomial hierarchy is deciding the validity of a quantified Boolean formula with $k$ quantifier alternations. A quantified Boolean formula (short: QBF) with $k$ quantifier alternations has the form $Q_1 \bar{X}_1 Q_2 \bar{X}_2 ... Q_k \bar{X}_k \Phi$, where $k \geq 1$, $\bar{X}_i$ $(1 \leq i \leq k)$ is a set of variables, $Q_i \in \{\exists, \forall\}$ $(1 \leq i \leq k)$, $Q_i \neq Q_{i+1}$ $(1 \leq i < k)$, and $\Phi$ is Boolean formula over the variables in $\bigcup_{i=1}^{k} \bar{X}_i$. The set of all quantified Boolean formulae with $k$ quantifier alternations and $Q_1 = \exists$ (resp., $Q_1 = \forall$) is denoted by $\text{QBF}_{k,\exists}$ (resp., $\text{QBF}_{k,\forall}$). Deciding the validity of a quantified Boolean formula in $\text{QBF}_{k,\exists}$ (resp., $\text{QBF}_{k,\forall}$) is a well-known problem in $\mathbf{\Sigma_k^P}$ (resp., $\mathbf{\Pi_k^P}$). Note that for $k = 1$, this problem coincides with the problem of deciding whether the Boolean formula $\Phi$ is satisfiable (resp., not satisfiable), which is indeed in $\mathbf{NP}$ (resp., co-$\mathbf{NP}$).

### A.3 Reductions Among Decision Problems

A decision problem $A_1$ is *polynomially reducible* to a decision problem $A_2$, denoted by $A_1 \leq_p A_2$, if there is a polynomial time computable function $h$ such that, for every $x$, $h(x)$ is defined and $x$ is a "yes" instance of $A_1$ if and only if $h(x)$ is a "yes" instance of $A_2$. A decision problem $A$ is *complete* for the class $\mathcal{C}$ of the polynomial hierarchy (beyond $\mathbf{P}$) if $A$ belongs to $\mathcal{C}$ and every problem in $\mathcal{C}$ is polynomially reducible to $A$. Thus, problems that are complete for the class $\mathcal{C}$ are the most difficult of the problems in $\mathcal{C}$.

It is worthwhile observing that the problems discussed in this section are known to be complete for the classes in which the membership has been pointed out. In particular, deciding the validity of a $\text{QBF}_{k,\exists}$ (resp., $\text{QBF}_{k,\forall}$) formula is the prototypical $\mathbf{\Sigma_k^P}$-complete (resp., $\mathbf{\Pi_k^P}$-complete) problem.

## Appendix B. Proofs in Section 3

**Proposition 3.2** *Let $\mathcal{G} = \langle N, v \rangle$ be a TU game and let $\mathcal{X} \subseteq X(\mathcal{G})$ be an arbitrary finite set of imputations. Then, there is a finite set of constraints $\texttt{LC}$ such that $X(\mathcal{G}|_{\texttt{LC}}) = \mathcal{X}$.*

*Proof.* Consider the game $\mathcal{G} = \langle N, v \rangle$, the set $\mathcal{X} = \{\hat{x}^1, \ldots, \hat{x}^k\}$ of imputations of $\mathcal{G}$, and the set of constraints:

$$\texttt{LC} = \begin{cases} x_i = \hat{x}_i^1 \times y^1 + \cdots + \hat{x}_i^k \times y^k, \ \forall 1 \leq i \leq |N| \\ 0 \leq y_j \leq 1, \ \forall 1 \leq j \leq k \\ y^1 + \cdots + y^k = 1 \\ x_1, \ldots, x_n \in \mathbb{R} \\ y^1, \ldots, y^k \in \mathbb{Z} \end{cases}$$

where $\hat{x}_i^1, \ldots, \hat{x}_i^k$ (for $1 \leq i \leq |N|$) are constants.

It is immediate to check that $V_{\texttt{LC}}(N) = \mathcal{X}$. Moreover, we observe that each vector $\hat{x}^j \in V_{\texttt{LC}}(N)$ $(1 \leq j \leq k)$ is efficient, since it cannot be dominated by any other vector in $\mathcal{X}$ (just notice that $\mathcal{X}$ is a set of imputations for $\mathcal{G}$). Finally, notice that each imputation $x \in \mathcal{X}$ is individually rational w.r.t. $\mathcal{G}$. Thus, by Proposition 3.7, $x$ is individually rational w.r.t. the constrained game too. It follows that $V_{\texttt{LC}}(N) = X(\mathcal{G}|_{\texttt{LC}})$. $\qquad\square$

**Proposition 3.3** *There exists a class $\mathcal{C} = \{\mathcal{G}|_{\texttt{LC}}^n\}_{n>0}$ of constrained games such that each game $\mathcal{G}|_{\texttt{LC}}^n$ is over $n+1$ players, $\texttt{LC}$ consists of $2 \times n + 1$ inequalities, and $|X(\mathcal{G}|_{\texttt{LC}})| = 2^n$.*

*Proof.* Consider the class $\mathcal{C} = \{\mathcal{G}|_{\text{LC}}^n\}_{n>0}$ where the game $\mathcal{G}^n = \langle N, v \rangle$ is such that $N = \{1, \ldots, n, n+1\}$, $v(N) = n$, $v(S) = 0$, for each coalition $S \subset N$, and $\text{LC} = \{0 \le x_i \le 1, x_i \in \mathbb{Z}, \forall 1 \le i \le n\} \cup \{x_{n+1} \ge n - \sum_{i=1}^n x_i\}$. It can be easily checked that $|X(\mathcal{G}|_{\text{LC}}^n)| = 2^n$. $\quad\square$

**Proposition 3.7** *Let $\mathcal{G} = \langle N, v \rangle$ be a TU game and let $x$ be a payoff vector that is individually rational w.r.t. $\mathcal{G}$ (i.e., $x_i \ge v(\{i\})$, for each player $i \in N$). Then, for each set $\text{LC}$ of constraints, $x$ is individually rational w.r.t. the constrained game $\mathcal{G}|_{\text{LC}}$.*

*Proof.* Let $x$ be a payoff vector such that $x_i \ge v(\{i\})$, for each player $i \in N$. Consider the constrained game $\mathcal{G}|_{\text{LC}}$ and a player $i \in N$. If $V_{\text{LC}}(\{i\}) = V_v(\{i\}) \cap \Omega(\text{LC})[\{i\}] = \varnothing$, we trivially have that $x_i > -\infty$. Otherwise, i.e., if $V_{\text{LC}}(\{i\}) \ne \varnothing$, we can just notice that $\max\{y_i \mid y_i \in V_{\text{LC}}(\{i\})\} \le v(\{i\})$. Thus, we have that $x_i \ge \max\{y_i \mid y_i \in V_{\text{LC}}(\{i\})\}$. $\quad\square$

## Appendix C. Proofs of Claims in Section 5

**Claim A.** $\hat{x} \in \mathscr{B}(\mathcal{G}(H))$ *if and only if $H$ is valid.*

*Proof.* Let us study the structure of a possible objection to $\hat{x}$. Recall that $(y, S)$ is an objection of player $i$ against player $j$ to $\hat{x}$ if and only if: $i \in S$, $j \notin S$, $y(S) \le v(S)$ and $y_k > \hat{x}_k$, for each $k \in S$. Thus, we observe that $v(S) = 1$ must hold, in order to improve the payoffs of its members, and because a worth value equal to 2 can only be obtained by the grand-coalition. In addition, since $\hat{x}_a = \hat{x}_{a'} = 1$ and since $y(S) \le v(S) = 1$, it must also be the case that $\{a, a'\} \cap S = \varnothing$, because these players get 1 in the current imputation $\hat{x}$. Due to the definition of the worth function, this entails that $S$ is consistent over the variables $\mathbf{Y}$, i.e., it is a set of $n$ players corresponding to literals over the universally quantified variables. We thus associate with each possible objection $(y, S)$ to $\hat{x}$ (where $y(S) \le 1$ and $y_k > 0$, $\forall k \in S$) a truth-value assignment to the variables in $\mathbf{Y}$ such that, $\forall 1 \le i \le n$, $Y_i$ is assigned *false* if $Y_i \in S$ (and *true* if $\bar{Y}_i \in S$). According to our notation, this means that such an objection is associated with the truth-value assignment $\sigma(\mathbf{Y} \setminus S)$. But we can define the converse, as well. For any truth-value assignment $\sigma_{\mathbf{Y}}$ for the universally quantified variables, its associated objection is the pair $(y, S)$ such that $S = \{Y_k \in \mathbf{Y} \mid \sigma_{\mathbf{Y}}(Y_k) = false\} \cup \{\bar{Y}_k \mid Y_k \in \mathbf{Y}, \sigma_{\mathbf{Y}}(Y_k) = true\}$, and $y_p = \frac{1}{|S|}$, for each $p \in S$. Note that $(y, S)$ is an objection of any player in $S$ against player $a$ or $a'$ to $\hat{x}$, and that $\sigma(\mathbf{Y} \setminus S) = \sigma_{\mathbf{Y}}$.

Indeed, if $(y, S)$ is an objection against a player $j \notin \{a, a'\}$, then $(z, \{j\})$ with $z_j = 0$ is a trivial counterobjection, since $v(\{j\}) = \hat{x}_j = 0$ and $j \notin S$. It follows that the set of objections that are possibly justified has to be restricted to those against player $a$ or player $a'$. Next, we consider the case of an objection against $a$, but exactly the same arguments hold for objections against $a'$. Let $(y, S)$ be an objection of $i$ against player $a$ to $\hat{x}$. Any counterobjection $(z, T)$ must now be with $a \in T$ (and $i \notin T$). Thus, in order to have $z_a \ge \hat{x}_a = 1$, it must be the case that $v(T) \ge 1$ and, actually, that $v(T) = 1$, due to the definition of the worth function. In particular, the latter (with the fact that $z_a = 1$) entails that, for each player $p \ne a$ with $p \in T$, it holds that $z_p = 0$. Thus, $T \cap S$ must be empty, because all members of $S$ get something according to $y$, $T \setminus \{a\}$ is consistent, and $\sigma(T) \models \Phi$. In particular, since $T \cap S = \varnothing$, according to this satisfying assignment, $Y_i$ is true in $\sigma(T)$ if and only if $Y_i \notin S$. It follows that $\sigma(T)$ coincides with $\sigma(\mathbf{Y} \setminus S)$ over the universally quantified variables, and thus it is in fact an extension of this truth-value assignment to the set of all variables occurring in the formula $\Phi$. Conversely, note that every satisfying assignment for $\Phi$ that extends $\sigma(\mathbf{Y} \setminus S)$ corresponds to such a counterobjection to $(y, S)$.

Given the above observations, we show the claim: $\hat{x} \in \mathscr{B}(\mathcal{G}(H)) \iff H$ is valid.

($\Rightarrow$) Assume that $\hat{x} \in \mathscr{B}(\mathcal{G}(H))$. Let $\sigma_{\mathbf{Y}}$ be any truth-value assignment for the universally quantified variables, and let $(y, S)$ be the objection to $\hat{x}$ associated with $\sigma_{\mathbf{Y}}$. In particular, $\sigma(\mathbf{Y} \setminus S) = \sigma_{\mathbf{Y}}$, by construction. Since $\hat{x} \in \mathscr{B}(\mathcal{G}(H))$, there exists a valid counterobjection $(z, T)$ to $(y, S)$, and we have seen above that its corresponding truth-value assignment $\sigma(T)$ is an extension of $\sigma(\mathbf{Y} \setminus S)$ to the set of all variables $vars(\Phi)$, and that $\sigma(T) \models \Phi$. It follows that $H$ is valid.

($\Leftarrow$) Assume that $\hat{x} \notin \mathscr{B}(\mathcal{G}(H))$. Then, there is a justified objection $(y, S)$ against $a$ (or $a'$) to $\hat{x}$. It follows from the discussion above that there is no truth-value assignment that is able to extend the assignment $\sigma(\mathbf{Y} \setminus S)$ to all the variables $vars(\Phi)$, and to satisfy $\Phi$. Indeed, such an extension would be associated with a counterobjection to $(y, S)$. It follows that $H$ is not valid. □

**Claim B.** $\mathscr{C}(\mathcal{G}(F)|_{\mathsf{LC}}) \neq \varnothing$ *if and only if $F$ is valid.*

*Proof.*

($\Rightarrow$) Assume that $\hat{x} \in \mathscr{C}(\mathcal{G}(F)|_{\mathsf{LC}})$, i.e., that $\forall S \subseteq N$, there is no $S$-feasible payoff vector such that $y_i > x_i$ for each $i \in S$. We claim that $\sigma(\hat{x})$ is a truth assignment to the variables in $\{X_1, \ldots, X_n\}$ witnessing the validity of $F$. Indeed, assume, for the sake of contradiction, that there is a truth assignment $\sigma_Y$ to the variables in $\{Y_1, \ldots, Y_q\}$ such that $\sigma(\hat{x}) \cup \sigma_Y \not\models \Phi$. Consider the coalition $\bar{S}$ such that $\bar{S}$ is consistent and $\sigma(\bar{S}) = \sigma(\hat{x}) \cup \sigma_Y$. By definition of the worth function, $v(\bar{S}) = n$. Moreover, observe that $\hat{x}(\bar{S}) < n$ holds, by definition of the assignment $\sigma(\hat{x})$ and given that $\hat{x}_{Y_j} = 0$ and $\hat{x}_{\bar{Y}_j} = 0$, for each imputation $\hat{x}$ and variable $Y_j$ $(1 \leq j \leq q)$. Note, in fact, that if $X_i \in \bar{S}$ then $X_i$ is true in $\sigma(\hat{x})$ and then $\hat{x}_{X_i} < 1$, and if $\bar{X}_i \in \bar{S}$ then $X_i$ is false in $\sigma(\hat{x})$ and then $\hat{x}_{\bar{X}_i} = 0$ (because in this case $\hat{x}_{X_i} = 1$ holds). Now, let $\epsilon = \frac{\min_{i \in \bar{S}}(1 - \hat{x}_i)}{n + q}$ and notice that $\epsilon > 0$, since $\hat{x}_i < 1$ is, in particular, prescribed by the definition of $\sigma(\hat{x})$. Consider the vector $y \in \mathbb{R}^{\bar{S}}$ such that $y_i = \hat{x}_i + \epsilon$ for each $i \in \bar{S}$. Note that $y(\bar{S}) \leq n$ because $|\bar{S}| = n + q$; moreover, $y \in V_{\mathsf{LC}}(\bar{S})$ holds because $v(\bar{S}) = n$ and because constraints are satisfied at $y$ (just notice that $\bar{S}$ contains exactly one player in $\{X_i, \bar{X}_i\}$, for each variable $X_i$, which is associated with a payoff less than or equal to 1 in $y$ and that constraints '$x_{X_i} + x_{\bar{X}_i} = 1$' do not play any role here because only one player $X_i$ or $\bar{X}_i$ is in $\bar{S}$ and hence given a variable $y_{X_i}$ (resp., $y_{\bar{X}_i}$) with $y_{X_i} \leq 1$ (resp., $y_{\bar{X}_i} \leq 1$) always exists a non negative value for $y_{\bar{X}_i}$ (resp., $y_{X_i}$) such that $y_{X_i} + y_{\bar{X}_i} = 1$). Since $y_i > x_i$, for each $i \in \bar{S}$, we then conclude that $\hat{x} \notin \mathscr{C}(\mathcal{G}(F)|_{\mathsf{LC}})$, which is impossible.

($\Leftarrow$) Assume that there is a truth assignment $\sigma$ to the variables in $\{X_1, \ldots, X_n\}$ witnessing the validity of $F$, and let $\hat{x}$ be an imputation such that $\sigma(\hat{x})$ coincides with $\sigma$, where in particular $\hat{x}_{X_i} = 1$ (resp., $\hat{x}_{X_i} = 0$) if $X_i$ is false (resp., true) in $\sigma$. We claim that $\hat{x} \in \mathscr{C}(\mathcal{G}(F)|_{\mathsf{LC}})$. Indeed, assume for the sake of contradiction, that there is a coalition $\bar{S}$ and an $\bar{S}$-feasible payoff vector $y$ such that $y_i > \hat{x}_i$ for each $i \in S$. Since $v(\bar{S}) \geq y(\bar{S}) > \hat{x}(\bar{S})$ and since $\hat{x}(\bar{S}) \geq 0$, given the definition of the worth function, it is actually the case that $\bar{S}$ is consistent (and, thus, $\sigma(\bar{S})$ is a truth assignment) and that $\sigma(\bar{S})$ is not satisfying. In particular, recall that $V_{\mathsf{LC}}(\bar{S}) = \{x \in \mathbb{R}^{\bar{S}} \mid x(\bar{S}) \leq v(\bar{S})\} \cap \Omega(\mathsf{LC})[\bar{S}]$; thus, for each player $X_i \in \bar{S}$ (resp., $\bar{X}_i \in \bar{S}$), we have $y_{X_i} \leq 1$ (resp., $y_{\bar{X}_i} \leq 1$). Therefore, $\bar{S}$ cannot include any player in $\{X_1, \bar{X}_1, ..., X_n, \bar{X}_n\}$ getting a worth 1 in $\hat{x}$. It follows that $\sigma(\bar{S})$ is an extension of $\sigma(\hat{x})$, which is moreover not satisfying. Thus, $\sigma(\hat{x})$ would not witness the validity of $F$, which is impossible. $\square$

**Claim C.** $\mathscr{B}(\mathcal{G}(P)|_{\mathsf{LC}}) \neq \varnothing$ if and only if $P$ is valid.

*Proof.* Consider any imputation $\hat{x}$ where $\hat{x}_a = 1$, and where $\hat{x}_{X_i}$ and $\hat{x}_{\bar{X}_i}$ take distinct values from the set $\{0, 1\}$, for each variable $X_i \in \{X_1, ..., X_m\}$. Any objection $(y, S)$ to $\hat{x}$ must be such that $v(S) = 1$ (which is indeed the maximum available worth over each coalition $S \subset N$) and there is no player $i \in S$ with $\hat{x}_i = 1$. It follows that objections are necessarily of the form $(y, S)$ where $w \in S$, $|S| = n+1$, and $S$ is consistent w.r.t. $\{Y_1, ..., Y_n\}$. In other words, any objection $(y, S)$ to $\hat{x}$ is such that $S$ contains $w$ plus one "universal player" per universally quantified variable, and thus it is uniquely associated with a truth-value assignment to the universally quantified variables $\mathbf{Y} = \{Y_1, ..., Y_n\}$. Let $\sigma(\mathbf{Y} \setminus S)$ denote this assignment, where we set $Y_j = true$ if and only if $Y_j \notin S$, for any $1 \leq j \leq n$. We define also the converse: given any truth value assignment $\sigma_{\mathbf{Y}}$ to the universally quantified variables, its associated objection is the pair $(y, S)$ such that $S = \{Y_j \mid \sigma_{\mathbf{Y}}(Y_j) = false\} \cup \{\bar{Y}_j \mid \sigma_{\mathbf{Y}}(Y_j) = true\} \cup \{w\}$, and $y_k = \frac{1}{|S|}$, for every $k \in S$.

Now, if $(y, S)$ is an objection against a player $j \notin \{a, X_1, ..., X_m, \bar{X}_1, ..., \bar{X}_m\}$, then $(z, \{j\})$ with $z_j = 0$ is a trivial counterobjection. Indeed, such a player $j$ does not belong to $S$ and may be either an element of $\{Y_1, ..., Y_n\}$ or an element of $\{Z_1, ..., Z_q\}$. In either cases, $v(\{j\}) = \hat{x}_j = 0$. On the other hand, if $(y, S)$ is an objection against a player $X_i$ (or, $\bar{X}_i$), then $(z, \{X_i, \bar{X}_i\})$ with $z_{X_i} = \hat{x}_{X_i}$ and $z_{\bar{X}_i} = \hat{x}_{\bar{X}_i}$ is again a counterobjection, because $v(\{X_i, \bar{X}_i\}) = z(\{X_i, \bar{X}_i\}) = 1$ and $\{X_i, \bar{X}_i\} \cap S = \varnothing$. It follows that the set of objections that are possibly justified has to be restricted to the objections against player $a$. Let $(y, S)$ be an objection of $i \in S$ against player $a$ to $\hat{x}$. Any counterobjection $(z, T)$ must have $a \in T$. Thus, in order to have $z_a \geq \hat{x}_a = 1$, it must be the case that $v(T) \geq 1$ and, actually, that $v(T) = 1$, due to the definition of the worth function. In particular, the latter entails that for each player $p \neq a$ with $p \in T$, it holds that $z_p = 0$. Thus, $T \cap S$ must be empty and, in particular, the only possibility is that $T \setminus \{a\}$ is consistent, and $\sigma(T) \models \Phi$. Finally, since $T \cap S = \varnothing$ and $\hat{x}(p) = 0$ for each $p \in T$ with $p \neq a$, we have that $\sigma(T)$ is a satisfying assignment for $\Phi$ where: $X_i$ is true in $\sigma(T)$ if and only if $X_i$ is true in $\sigma(\hat{x})$; $Y_i$ is true in $\sigma(T)$ if and only if $Y_i \notin S$, and thus $Y_i$ is true in $\sigma(\mathbf{Y} \setminus S)$. That is, $\sigma(T)$ is a complete assignment for $\Phi$ that extends both partial assignments $\sigma(\hat{x})$ and $\sigma(\mathbf{Y} \setminus S)$.

By exploiting the above observations, we can now prove the claim.

$(\Rightarrow)$ Assume that there exists $\hat{x} \in \mathscr{B}(\mathcal{G}(P)|_{\mathsf{LC}})$. We have seen that such an imputation $\hat{x}$ is associated with a truth-value assignment $\sigma(\hat{x})$ to the variables in $\{X_1, ..., X_m\}$—recall that an imputation $x$ for which there is an index $\bar{i}$, $1 \leq \bar{i} \leq m$, such that $x_{X_{\bar{i}}} > 0$ and $x_{\bar{X}_{\bar{i}}} > 0$ cannot belong to the bargaining set of $\mathcal{G}(P)|_{\mathsf{LC}}$. Moreover, we have seen that every assignment $\sigma_{\mathbf{Y}}$ to the universally quantified variables corresponds to a

possible objection $(y, S)$ to $\hat{x}$, and since $\hat{x}$ belongs to the bargaining set there must exist a valid counterobjection $(z, T)$ to $(y, S)$ associated with a satisfying truth-value assignment for $\Phi$ that extends both partial assignments $\sigma(\hat{x})$ and $\sigma_{\mathbf{Y}}$. This means that $\hat{x}$ is a witness of the validity of $P$.

($\Leftarrow$) If $P$ is valid then there is an assignment $\sigma_X$ to the variables in $\{X_1, \ldots, X_m\}$ that witnesses its validity. Consider the imputation $\hat{x}$ such that, $\forall 1 \leq i \leq m$, $\hat{x}_{X_i} = 0$ and $\hat{x}_{\bar{X}_i} = 1$ if $\sigma_X(X_i) = true$, and $\hat{x}_{X_i} = 1$ and $\hat{x}_{\bar{X}_i} = 0$ otherwise. Moreover, $\hat{x}_a = 1$ and all other players get 0. Since, for every extension of $\sigma_X$ to the universally quantified variables (corresponding to a possible objection $(y, S)$ to $\hat{x}$), there exists a further extension to all variables that satisfies $\Phi$ (corresponding to a counterobjection to $(y, S)$), it follows that $\hat{x} \in \mathscr{B}(\mathcal{G}(P)|_{\mathtt{LC}})$. $\qquad\square$

**Claim D.** $\Omega(\mathtt{LC}^{i,j,S})[y_S] \setminus \bigcup_{T | i \notin T \wedge j \in T} \Omega(\mathtt{LC}^{i,j,S,T})[y_S]$ *contains a point (i.e., a justified objection to $\hat{x}$) that can be represented with polynomially many bits.*

*Proof.* The case where $\Omega(\mathtt{LC}^{i,j,S})[y_S]$ and $\Omega(\mathtt{LC}^{i,j,S,T})[y_S]$ (for each $T \mid i \notin T \wedge j \in T$) do not contain integer variables has already been addressed in the proof of Lemma 5.8. We next show how to preprocess such integer variables, if they occur in the programs at hand.

Recall first that $\Omega(\mathtt{LC}^{i,j,S})[y_S]$ is bounded, since $\mathtt{LC}^{i,j,S}$ contains the $|S| + 1$ inequalities: $y(S) \leq v(S)$ and $y_k > \hat{x}_k$, $\forall k \in S$. This implies that we can assume, w.l.o.g., that $\Omega(\mathtt{LC}^{i,j,S})$ is bounded in its turn. Indeed, by standard arguments in linear programming it follows that any point in $\Omega(\mathtt{LC}^{i,j,S})[y_S]$ can be obtained as the projection of a point in $\Omega(\mathtt{LC}^{i,j,S})$ whose auxiliary components (i.e., those not associated with the variables in $y_S$) are bounded by a polynomial in the size of $\mathtt{LC}^{i,j,S}$. Thus, this bound can be made explicit in the definition of $\mathtt{LC}^{i,j,S}$, without altering the projection $\Omega(\mathtt{LC}^{i,j,S})[y_S]$.

Second, we observe that we can also assume, w.l.o.g., that $\Omega(\mathtt{LC}^{i,j,S,T})$ is bounded too, for each $T$ such that $i \notin T \wedge j \in T$. Indeed, in the definition of $\mathtt{LC}^{i,j,S,T}$, we may constrain each variable in $y_S$ to range within the minimum and maximum values it may assume in $\Omega(\mathtt{LC}^{i,j,S})[y_S]$—note that these extreme values can be represented with polynomially many bits, since they are achieved on some vertices of $\Omega(\mathtt{LC}^{i,j,S})[y_S]$. This modification does not alter the set $\Omega(\mathtt{LC}^{i,j,S})[y_S] \setminus \bigcup_{T | i \notin T \wedge j \in T} \Omega(\mathtt{LC}^{i,j,S,T})[y_S]$. Thus, $\Omega(\mathtt{LC}^{i,j,S,T})[y_S]$ is bounded, so that $\Omega(\mathtt{LC}^{i,j,S,T})$ can be assumed to be bounded, too—see above.

We can now resume the main proof and show how integer variables can be easily preprocessed. Let $\mathtt{LC}$ be any program in $\{\mathtt{LC}^{i,j,S}\} \cup \{\mathtt{LC}^{i,j,S,T} \mid i \notin T \wedge j \in T\}$, let $\widetilde{\mathtt{LC}}$ denote the linear relaxation of $\mathtt{LC}$, and recall that each vertex of the polytope $\Omega(\widetilde{\mathtt{LC}})$ can be represented with polynomially many bits. Since $\Omega(\mathtt{LC})$ is contained in $\Omega(\widetilde{\mathtt{LC}})$, such values of the components of the vertices are bounds for every integer component of any vector in $\Omega(\mathtt{LC})$, which can thus be represented with polynomially many bits. Let $U$ be the set of all the admissible values of all such integer components. Let $I(\mathtt{LC})$ denote the set of all the possible assignments of values from the integer variables of $\mathtt{LC}$ to $U$, and for any assignment $\mathbf{z} \in I(\mathtt{LC})$, let $\mathtt{LC}\langle \mathbf{z} \rangle$ denote the linear program where each integer variable in $int(\mathtt{LC})$ is replaced by its corresponding value in $\mathbf{z}$. Now, for any pair of assignments $\mathbf{z}$ and $\mathbf{w}$ belonging to $I(\mathtt{LC}^{i,j,S})$ and to $I(\mathtt{LC}^{i,j,S,T})$, respectively, let us say that $\mathbf{w}$ matches with $\mathbf{z}$ (w.r.t. $y_S$) if $\mathbf{z}$ and $\mathbf{w}$ coincide on their restrictions over $y_S \cap int(\mathtt{LC}^{i,j,S}) \cap int(\mathtt{LC}^{i,j,S,T})$. Furthermore, let $\mathcal{W}$ be the set of all non-integer variables in $y_S$, that is, $y_S \setminus (int(\mathtt{LC}^{i,j,S}) \cup \bigcup_{T | i \notin T \wedge j \in T} int(\mathtt{LC}^{i,j,S,T}))$.

Then, the set $\Omega(\mathtt{LC}^{i,j,S})[y_S] \setminus \bigcup_{T|i \notin T \land j \in T} \Omega(\mathtt{LC}^{i,j,S,T})[y_S]$ contains a point that can be represented with polynomially many bits (resp., is empty) if and only if there is an element $\mathbf{z}$ in $I(\mathtt{LC}^{i,j,S})$ such that (resp., for each element $\mathbf{z}$ in $I(\mathtt{LC}^{i,j,S})$):

$$\Omega(\mathtt{LC}^{i,j,S}\langle\mathbf{z}\rangle)[\mathcal{W}] \setminus \left( \bigcup_{T|i\notin T \land j \in T} \bigcup_{\mathbf{w} \in I(\mathtt{LC}^{i,j,S,T}) \,|\, \mathbf{w} \text{ matches with } \mathbf{z}} \Omega(\mathtt{LC}^{i,j,S,T}\langle\mathbf{w}\rangle)[\mathcal{W}] \right)$$

contains an element that can be represented with polynomially many bits (resp., it is empty).

Note that the above expression has the same form as the original one, but no integer variable occurs in it. To conclude, just observe that $\mathtt{LC}^{i,j,S}\langle\mathbf{z}\rangle$ and $\mathtt{LC}^{i,j,S,T}\langle\mathbf{w}\rangle$ ($T \mid i \notin T \land j \in T$) can be represented with polynomially many bits (w.r.t. the size of the original mixed-integer linear programs), since they are obtained by mapping integer variables into values that are representable with polynomially many bits. □